\pdfoutput=1
\documentclass[a4paper,fleqn,usenatbib]{mnras}

\usepackage{mathptmx}
\usepackage{txfonts}

\usepackage[T1]{fontenc}
\usepackage{ae,aecompl}

\usepackage{graphicx}
\usepackage{longtable}
\usepackage{color}
\usepackage{amssymb}

\newcommand{\HI}{\hbox{{\rm H}{\sc \,i}}}

\newcommand{\Lya}{\hbox{Ly$\alpha$}}

\newcommand{\Lye}{\hbox{Ly$\epsilon$}}

\newcommand{\MgII}{\hbox{{\rm Mg}{\sc \,ii}}}

\newcommand{\kms}{\hbox{km~s$^{-1}$}}

\newcommand{\Om}{\hbox{$\Omega_{m}$}}
\newcommand{\Od}{\hbox{$\Omega_{\Lambda}$}}

\newcommand{\zabs}{\hbox{$z_{\rm abs}$}}
\newcommand{\zem}{\hbox{$z_{\rm em}$}}
\newcommand{\zmin}{\hbox{$z_{\rm min}$}}
\newcommand{\zmax}{\hbox{$z_{\rm max}$}}

\newcommand{\z}{\hbox{$z$}}

\newcommand{\Dz}{\hbox{$\Delta z$}}
\newcommand{\DX}{\hbox{$\Delta X$}}

\newcommand{\gz}{\hbox{$g(z)$}}
\newcommand{\fnx}{\hbox{$f_{\rm HI}(N,X)$}}

\newcommand{\lx}{\hbox{$\ell(X)$}}

\newcommand{\Og}{\hbox{$\Omega_{\rm g}$}}

\newcommand{\Odla}{\hbox{$\Omega_{\rm HI}^{\rm DLA}$}}

\newcommand{\Ohi}{\hbox{$\Omega_{\rm HI}$}}

\newcommand{\NHI}{\hbox{$N(\rm \HI)$}}
\newcommand{\logNHI}{\hbox{$\log \NHI$}}
\newcommand{\elogNHI}{\hbox{$\epsilon[\logNHI]$}}

\title{The evolution of neutral gas in damped Lyman $\alpha$ systems from the XQ-100 survey}
\author[R.~S\'{a}nchez-Ram\'{\i}rez et al.]{R.~S\'{a}nchez-Ram\'{\i}rez$^{1,2,3}$ \thanks{E-mail: ruben@iaa.es},
S.~L.~Ellison$^{4}$,
J.~X.~Prochaska$^{5}$,
T.~A.~M.~Berg$^{4}$, \and
S.~L\'{o}pez$^{6}$,
V.~D'Odorico$^{7}$,
G.~D.~Becker$^{8,9}$,
L.~Christensen$^{10}$,
G. Cupani$^{7}$, \and
K.~D.~Denney$^{11}$,
I.~P\^{a}ris$^{7}$,
G.~Worseck$^{12}$,
and J.~Gorosabel$^{1,2,3,13}$\\
\\
$^{1}$Unidad Asociada Grupo Ciencias Planetarias (UPV/EHU, IAA-CSIC), Departamento de F\'{\i}sica Aplicada I, \\ 
E.T.S. Ingenier\'{\i}a, Universidad del Pa\'{\i}s Vasco (UPV/EHU), Alameda de Urquijo s/n, E-48013 Bilbao, Spain.\\
$^{2}$Ikerbasque, Basque Foundation for Science, Alameda de Urquijo 36-5, E-48008 Bilbao, Spain.\\
$^{3}$Instituto de Astrof\'{\i}sica de Andaluc\'{\i}a (IAA-CSIC), Glorieta de la Astronom\'{\i}a s/n, E-18008, Granada, Spain.\\
$^{4}$Department of Physics and Astronomy, University of Victoria, Victoria, BC V8P 1A1, Canada.\\
$^{5}$Department of Astronomy and Astrophysics, UCO/Lick Observatory, University of California, 1156 High Street, Santa Cruz, \\
CA 95064, USA.\\
$^{6}$Departamento de Astronom\'{\i}a, Universidad de Chile, Casilla 36-D, Santiago, Chile.\\
$^{7}$INAF-Osservatorio Astronomico di Trieste, Via Tiepolo 11, I-34143 Trieste, Italy.\\
$^{8}$Space Telescope Science Institute, 3700 San Martin Drive, Baltimore, MD 21218, USA.\\
$^{9}$Institute of Astronomy and Kavli Institute of Cosmology, Madingley Road, Cambridge CB3 0HA, UK.\\
$^{10}$Dark Cosmology Centre, Niels Bohr Institute, University of Copenhagen, Juliane Maries Vej 30, DK-2100 Copenhagen, Denmark.\\
$^{11}$Department of Astronomy, The Ohio State University, 140 West 18th Avenue, Columbus, OH 43210, USA.\\
$^{12}$Max-Planck-Institut f\"{u}r Astronomie, K\"{o}nigstuhl 17, D-69117 Heidelberg, Germany.\\
$^{13}$Deceased.}

\date{Accepted XXX. Received YYY; in original form ZZZ}

\pubyear{2015}

\begin{document}

\label{firstpage}

\pagerange{\pageref{firstpage}--\pageref{lastpage}}

\maketitle

\begin{abstract}
We present a sample of 38 intervening Damped Lyman $\alpha$ (DLA) systems identified towards 100 $z>3.5$ quasars, observed during the XQ-100 survey. The XQ-100 DLA sample is combined with major DLA surveys in the literature. The final combined sample consists of 742 DLAs over a redshift range approximately $1.6 < z_{\rm abs} < 5.0$.   We develop a novel technique for computing \Odla\ as a continuous function of redshift, and we thoroughly assess and quantify the sources of error therein, including fitting errors and incomplete sampling of the high column density end of the column density distribution function.  There is a statistically significant redshift evolution in \Odla\ ($\geq 3 \sigma$) from $z \sim 2$ to $z \sim$ 5.  In order to make a complete assessment of the redshift evolution of \Ohi, we combine our high redshift DLA sample with absorption surveys at intermediate redshift and 21cm emission line surveys of the local universe.  Although \Odla, and hence its redshift evolution, remains uncertain in the intermediate redshift regime ($0.1 < z_{\rm abs} < 1.6$), we find that the combination of high redshift data with 21cm surveys of the local universe all yield a statistically significant evolution in \Ohi\ from $z \sim 0$ to $z \sim 5$ ($\geq 3 \sigma$).  Despite its statistical significance, the magnitude of the evolution is small: a linear regression fit between \Ohi\ and $z$ yields a typical slope of $\sim$0.17$\times 10^{-3}$, corresponding to a factor of $\sim$ 4 decrease in \Ohi\ between $z=5$ and $z=0$.
\end{abstract}


\section{Introduction}

Measurements of \Ohi, the mass density of atomic hydrogen gas scaled to the critical density, and its evolution with redshift offer cosmological constraints on several aspects of galaxy formation.  The value of \Ohi\ at any epoch characterizes the instantaneous reservoir of cold, neutral gas available for star-formation integrated across the entire galaxy population. This constraint holds independently of the detailed association of individual DLAs to specific galaxy populations, i.e. \Ohi\ is a cosmic quantity \citep[e.g.][]{1995ApJ...454..698W}.  It may serve, therefore, as an input to semi-analytic prescriptions for galaxy formation \cite[e.g.][]{2014arXiv1412.2712S}. The time evolution of \Ohi, meanwhile, tracks the global
balance between the accretion of cold gas onto galaxies against the processes that consume and/or expel that gas \cite[e.g.][]{2012MNRAS.425.2027K, 2013ApJ...772..119L, 2013MNRAS.434.2645D}.  As theorists continue to explore models to capture the complex processes of star-formation and feedback, \Ohi\ offers a cosmic check on their prescriptions. Indeed, there is apparent tension between previous \Ohi\ measurements and galaxy formation models that reproduce other key observables of the galactic population \citep{2015MNRAS.447.1834B, 2014arXiv1412.2712S}. We are hence motivated to assess \Ohi\ and the uncertainties in its estimation across cosmic time.

This cosmic evolution of \Ohi\ can be traced by combining surveys of damped Lyman alpha systems (DLAs) at moderate-to-high redshifts, with 21cm emission surveys at \z$\sim$0. In recent years, there has been significant progress in refining measurements of \Ohi\ with both of these techniques, where large statistical samples have been crucial for addressing biases due to incompleteness.  In the nearby universe the greatest uncertainty for the determination of \Ohi\ in early 21cm surveys was the faint end slope of the \HI\ mass function \citep[e.g.][]{2002ApJ...567..247R, 2003AJ....125.2842Z, 2005MNRAS.364.1467Z}. The Arecibo L-band Fast ALFA (ALFALFA) survey \citep{2005AJ....130.2598G} has now provided the \z$\sim$0 benchmark for \Ohi, based on over 10,000 galaxies in the local universe \citep{2010ApJ...723.1359M}.  Extending the measurement of \Ohi\ to  even $z \sim 0.1$ -- 0.2 is extremely challenging for current 21~cm surveys.  Nonetheless, stacking experiments have produced several estimates of \Ohi\ in this redshift range \citep{2007MNRAS.376.1357L, 2013MNRAS.433.1398D, 2013MNRAS.435.2693R}. Individual detections of 21~cm emission beyond $z=0$ are growing, thanks to surveys such as CHILES \citep{2013ApJ...770L..29F} and HIGHz \citep{2015MNRAS.446.3526C}, although these surveys are not yet large or complete enough to give a statistical perspective on \Ohi.  

Fortunately, \Ohi\ in the higher redshift universe can be effectively measured in absorption, using DLA surveys.  The objective of DLA surveys has largely been to assess the redshift evolution of \Ohi, in comparison to the \z=0 local benchmark. Early surveys of DLAs focused predominantly on the redshift range 2 $<$ \zabs\ $<$ 3.0 \citep[e.g.][]{1986ApJS...61..249W, 1995ApJ...454..698W}. The lower bound of this redshift range was set by the accessibility of the \Lya\ line to ground based spectrographs, and the upper bound by the limited number of bright, high redshift quasars known at the time. Although these early surveys enabled a broad-brush measurement of \Ohi, the limited redshift range and sample sizes were insufficient to study the evolution of the cosmic gas reservoir. Subsequent ground-based surveys were motivated to extend the redshift range to earlier epochs, and reported a tentative peak in \Ohi\ at $z\sim$ 3 \citep{1996MNRAS.282.1330S, 2000ApJ...543..552S}. As the size of DLA samples grew, improved statistics led to an upward revision of \Ohi\ at \z$>$3.5 and evidence for a peak at \z$\sim$3 diminished; the mass density of \HI\ in DLAs appeared to be consistent over the range of redshifts \z$\sim$ 2 -- 5 \citep{2003MNRAS.346.1103P, 2005MNRAS.363..479P}. The apparent down-turn of \Ohi\ seen in earlier surveys seems likely to be caused by poor statistics at the survey limit \citep[e.g.][]{2004PASP..116..622P}.

Despite these early surveys, the error bars on \Ohi\ remained substantial, and it was the advent of the Sloan Digital Sky Survey (SDSS) that led to the first truly robust measure of \Ohi\ redshift evolution.  Several investigations, based on different SDSS data releases, have found a mildly decreasing \Ohi\ from \z$\sim$3.5 to 2 \citep[e.g.][]{2004PASP..116..622P, 2005ApJ...635..123P, 2009ApJ...696.1543P, 2009AA...505.1087N, 2012AA...547L...1N}. All of these works self-consistently show an evolution of at most a factor of two in this redshift range, an effect too subtle to be detectable in previous smaller surveys. Pushing to even higher redshifts, there again seemed to be tentative evidence of a downturn in \Ohi\ above z$\sim$3.5 \citep{2009AA...508..133G, 2010ApJ...721.1448S}. However, with a factor of eight increase in path length over previous compilations, \citet{2015MNRAS.452..217C} have shown that \Ohi\ evolution is statistically consistent (within the observational errors) with a power law of index 0.4 from $z=5$ to the present day. The results of \citet{2015MNRAS.452..217C} therefore support a mild, but steady evolution in the neutral gas content of galaxies since early times.

Despite the uniform decline in \Ohi\ from high \z\ to the present day proposed by \citet{2015MNRAS.452..217C}, the value of \Ohi\ measured at $z=5$ is formally consistent with the value measured by \citet{2006ApJ...636..610R} in the range 0.2 $<$ \zabs\ $<$ 1.5 \citep[see Fig. 12 of][]{2015MNRAS.452..217C}.  A statistically plausible alternative picture to the steady decline of \Ohi\ is therefore one in which gas consumption was almost perfectly balanced by replenishment, with a statistically significant decrease (of a factor of two) only at the most recent epochs. One of the challenges in the interpretation of the  data compilation presented by \citet{2015MNRAS.452..217C} is in the combination of surveys performed at different redshifts, and a homogeneous assessment of the error associated with \Ohi. 

In this paper we present a new survey for DLAs in the range 1.6 $<$ \zabs\ $<$ 4.5 and make a novel assessment of \Ohi\ as a function of redshift.  Our sample combines DLAs from our own survey, with a compilation of literature absorbers that has been carefully checked for duplicates and errors. Rather than showing our new survey results in comparison with previous surveys at different redshifts, we maximize the statistical potential of decades of work by combining previous surveys together.  Moreover, by quantifying \Ohi\ within a sliding redshift window, rather than in contiguous non-overlapping bins, and with a rigorous assessment of error propagation techniques, we are able to determine a holistic perspective of the atomic gas content of galaxies up to $z=5$.

The paper is organized as follows. In Section 2 we describe the XQ-100 survey of 100 $z>3.5$ quasars, and the detection of DLAs therein. The XQ-100 DLA sample is combined with various literature samples and compilations, as described in Section 3. Section 4 presents our statistical analysis, including the description of our technique to determine \Ohi\ evolution `curves', a rigorous assessment of sources of error and the analysis of the column density distribution and line density functions.

In this work, we will use the term \Odla, to mean the neutral hydrogen mass density in Damped Lyman $\alpha$ systems (i.e. the contribution to \Ohi\ from systems above the DLA column density threshold, \logNHI\ $\ge$ 20.3) relative to the critical density.   The total gas mass density is given by \Og, which requires a correction to \Ohi\ by the factor $\mu$=1.3 to account for helium.  Finally, we assume a flat $\Lambda$CDM cosmology with $H_{0}$=70.0~\kms~Mpc$^{-1}$, \Om=0.3 and \Od=0.7.

\section{The XQ-100 sample}
\label{sec:xq100}

\begin{figure*}
\includegraphics[width=0.95\textwidth]{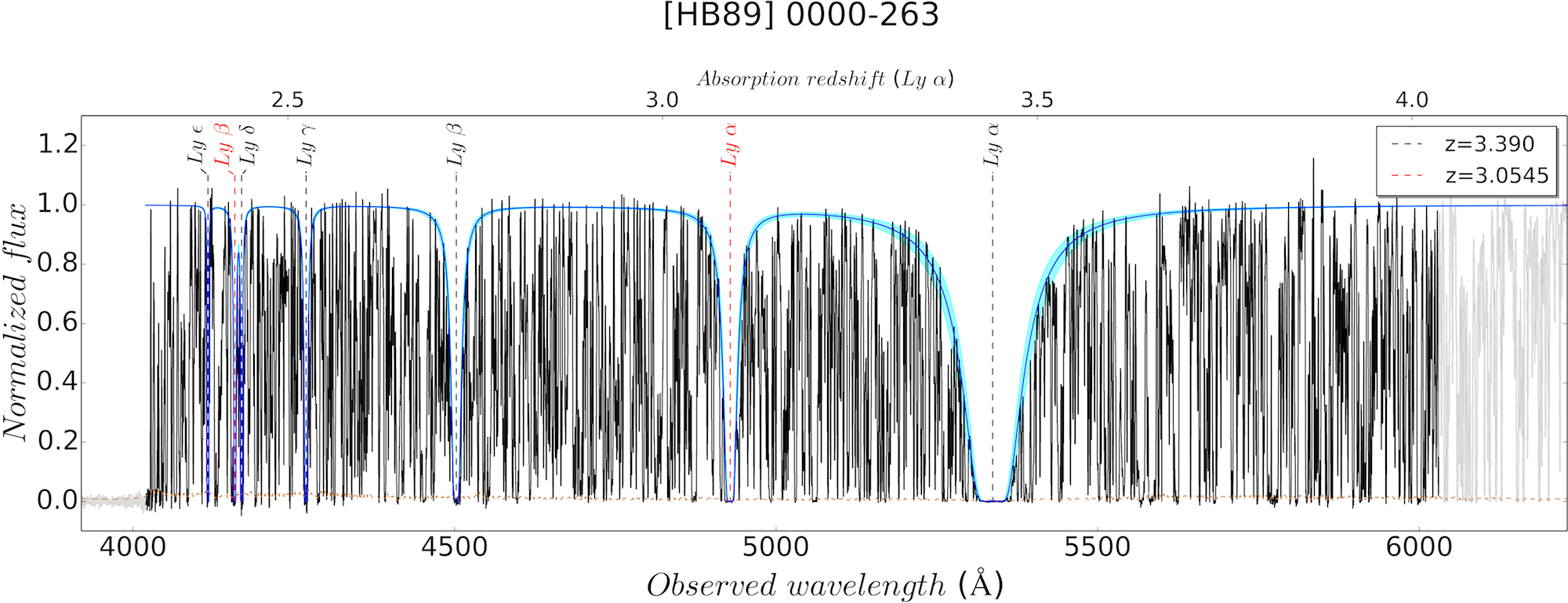}
\includegraphics[width=0.95\textwidth]{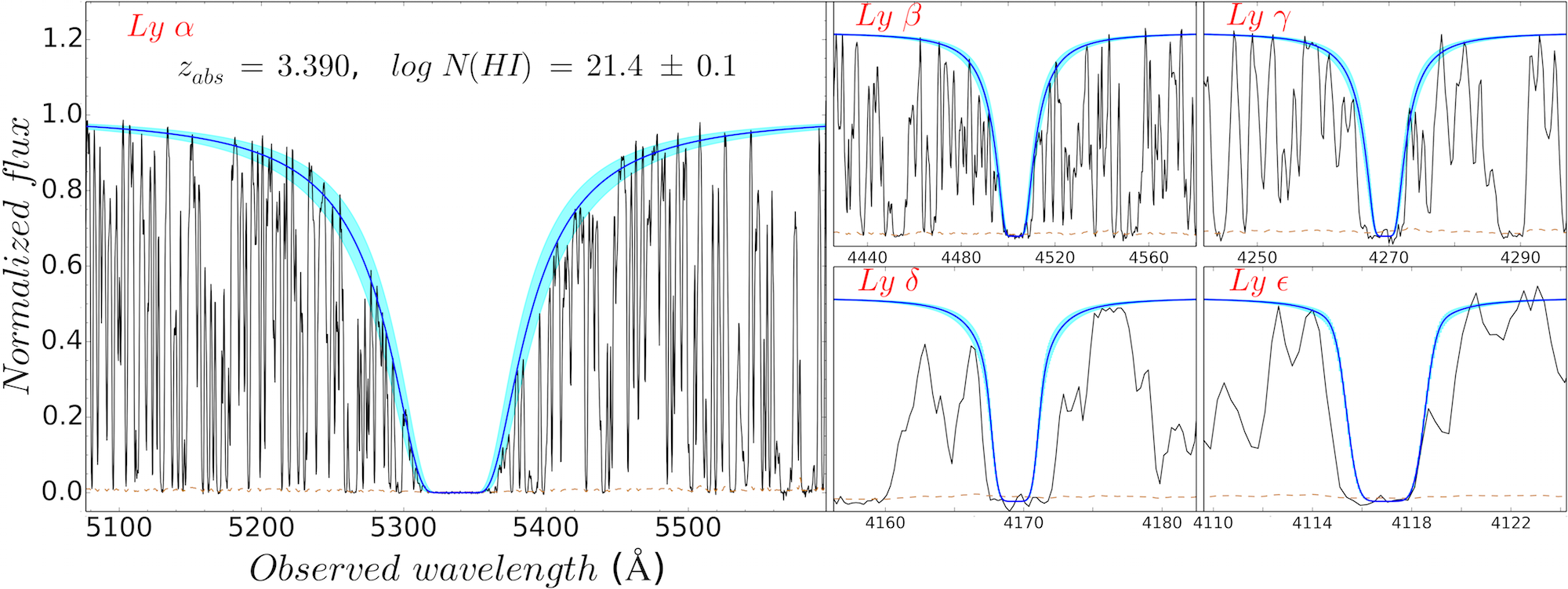}
\caption{Example of normalized \Lya\ forest (upper panel) for the quasar [HB89] 0000-263. The part of the spectrum used for statistics is plotted in black and the rest in gray. The error spectrum is shown in brown. The model of all absorbers with \logNHI$\geq$19.5 is drawn in blue with the 1$\sigma$ error zone shaded in cyan. Each individual system is labeled with a specific color denoted in the legend box.  However, only one of the absorbers, at $z=3.390$, has an \NHI\ above the DLA threshold and is included in our catalog.  The fits to the higher order lines of this DLA are shown in the lower panels.}
\label{fig:fit}
\end{figure*}

The XQ-100 survey is an ESO Large Program (ESO ID 189.A-0424, P.I. S. Lopez) which obtained X-shooter spectra of 100 $3.5<z<4.7$ QSOs in the period between 10-02-2012 and 23-02-2014. X-shooter \citep{2011AA...536A.105V} is a triple-arm spectrograph which obtains moderate resolution spectra with complete wavelength coverage from $\sim$ 320 -- 2500 nm, permitting the simultaneous analysis of QSO absorption lines and emission features from the atmospheric cut-off to the near-IR. A full description of the XQ-100 survey characteristics, sample selection, observational set-up and data reduction is provided in Lopez et al. (in prep). We review only the basic features of the survey design and data here. In brief, the XQ-100 survey adopted slit widths of 0.9\arcsec (UVB arm) and 1.0\arcsec (VIS and NIR arms) to obtain spectra whose resolution ranged from $R \sim 5100 - 8800$. Exposure times ranged from $\sim$ 1700s to 3600s  yielding a median signal-to-noise ratio (SNR) $\sim$ 30 per pixel. The data were reduced using a custom pipeline which provided a notable improvement in the removal of the near-IR sky emission with respect to the ESO provided pipeline.  All of the 1-dimensional spectra (flux and wavelength calibrated) are made publically available (see Lopez et al. in prep for details on the data reduction and public data release). Higher order data products are also made available in the public repository: spectra corrected for telluric absorption and normalized spectra derived from a variety of continuum fitting methods.

In this paper, we make use of the basic 1D products and perform our own normalization of the continuum.  This decision is driven by the sensitivity of the \Lya\ fit in the damping wings to the continuum placement. The nomalization of the spectra and Voigt profile fits were performed using an interactive interface that permits the user to simultaneously identify and fit the absorbers, and iterate on the continuum placement. The iterative procedure is required due to both the complexity of the \Lya\ forest, and the challenges associated with continuum placement over the broad damped profile, particularly in the shallow wing region \citep{2003ApJS..147..227P}. 

The fitting procedure can be summarized as follows. First, we estimated the continuum by manually identifying regions of apparently unabsorbed continuum flux throughout the \Lya\ forest and around the \Lya\ emission. A cubic spline was used to fit these points and make an initial normalization. Next, a Voigt profile with \logNHI=19.0 is moved through the forest range looking for DLA candidates. Although the canonical threshold for DLA classification is \logNHI=20.3, the lower threshold of our scan provides a conservative initial selection for assessment. It is also useful to identify these lower column density systems to aid with fits of blended absorbers. For each potential absorption system, a simultaneous fit of all Lyman series lines up to \Lye\ was performed, with adjustments to the continuum when necessary, in order to determine the HI column density and redshift of the absorber. Metal lines associated with each potential absorber were identified (in Berg et al. in prep. we will present a complete assessment of the abundances in the XQ-100 DLA sample), and in cases of uncertain fits (e.g. due to blends) the redshifts from the metal lines were used to inform the fits (but the metals are not used \textit{a priori} to fix the redshifts for all absorbers). An example of our fits is shown in Figure \ref{fig:fit}; a full montage of all of the fits to our DLA sample is provided in the online version of this paper.  

The conservative initial search threshold of \logNHI=19.0 permits the identification of numerous absorbers whose \NHI\ is below the traditional DLA threshold of \logNHI=20.3, yet still exhibit damping wings that permit the accurate measurement of the HI content. These sub-DLAs have been the subject of targeted research \citep[e.g.][]{2013AA...556A.141Z} and are sometimes included explicitly  \citep[e.g.][]{2005MNRAS.363..479P, 2009AA...508..133G}, or statistically \citep[e.g.][]{2015MNRAS.452..217C} in the calculation of \Ohi. Despite the ability of our dataset to identify absorbers down to at least  \logNHI=19.5, we do not include them in the present work (however, these absorbers are used to compute \fnx\ and \Odla\ uncertainties).  The motivation for this decision is one of homogeneity. Later in this work, we will combine the XQ-100 sample with other available surveys for DLAs at $z>2$, in order to obtain the most statistically robust measure of \Ohi\ from $2 < z < 5.5$.  Since many of the literature samples that we will make use of do not include sub-DLAs (often due to their more limited spectral resolution), we adopt the standard threshold of \logNHI=20.3 for the DLA catalog presented here. However, in a separate future paper, we will present the identification of sub-DLAs, with column density completion functions and assess their contribution to both the neutral gas and metals at $z \sim$ 4. 

The final requirement for the DLA to be included in our XQ-100 statistical sample for the computation of \Odla\ is that its redshift be at least 5000 \kms\ from the background QSO in order to exclude the proximate DLA (PDLA) population.  The PDLAs have been shown to exhibit different clustering properties \citep[they are more prevalent than intervening systems,][]{2002AA...383...91E, 2006MNRAS.367..412R, 2008ApJ...675.1002P} and have also been suggested to manifest different metallicities and ionization conditions \citep{2010MNRAS.406.1435E, 2011MNRAS.412..448E}.  These distinctions justify the exclusion of PDLAs from our statistical study of DLA gas content. These out-of-sample absorbers are flagged with a star in Table \ref{tab:xq}.

Our final XQ-100 DLA sample contains 38 absorbers with absorption redshifts ranging from 2.24 to 4.47. Of these DLAs, 27 are not duplicated in our combined literature sample and 22 are not in the catalog by \citet{2012AA...547L...1N}. Duplicated systems are flagged with a diamond in Table \ref{tab:xq}.

We compare in Figure \ref{fig:comp} our fitted \NHI\ values with previous estimates for some systems, where measurements have been made at a variety of spectral resolutions.  Most of the DLAs in the \citet{2003MNRAS.346.1103P} compilation, as well as those by \citet{2012AA...547L...1N}, are observed at $R<2000$.  The \citep{2009AA...508..133G} sample was observed with ESI, which has a comparable resolution to X-shooter.  The `high resolution' \citep[HR][]{2015MNRAS.452.4326B} data points have typical values of $R \sim 40,000$, observed with either HIRES on Keck or UVES on the VLT. There is a tendency for the XQ-100 fits to exceed those in the literature when the resolution of the latter is low. However, the agreement with high resolution measurements is generally excellent and well within the quoted uncertainties (typically 0.1 -- 0.2 dex).

\begin{figure}
\includegraphics[width=0.45\textwidth]{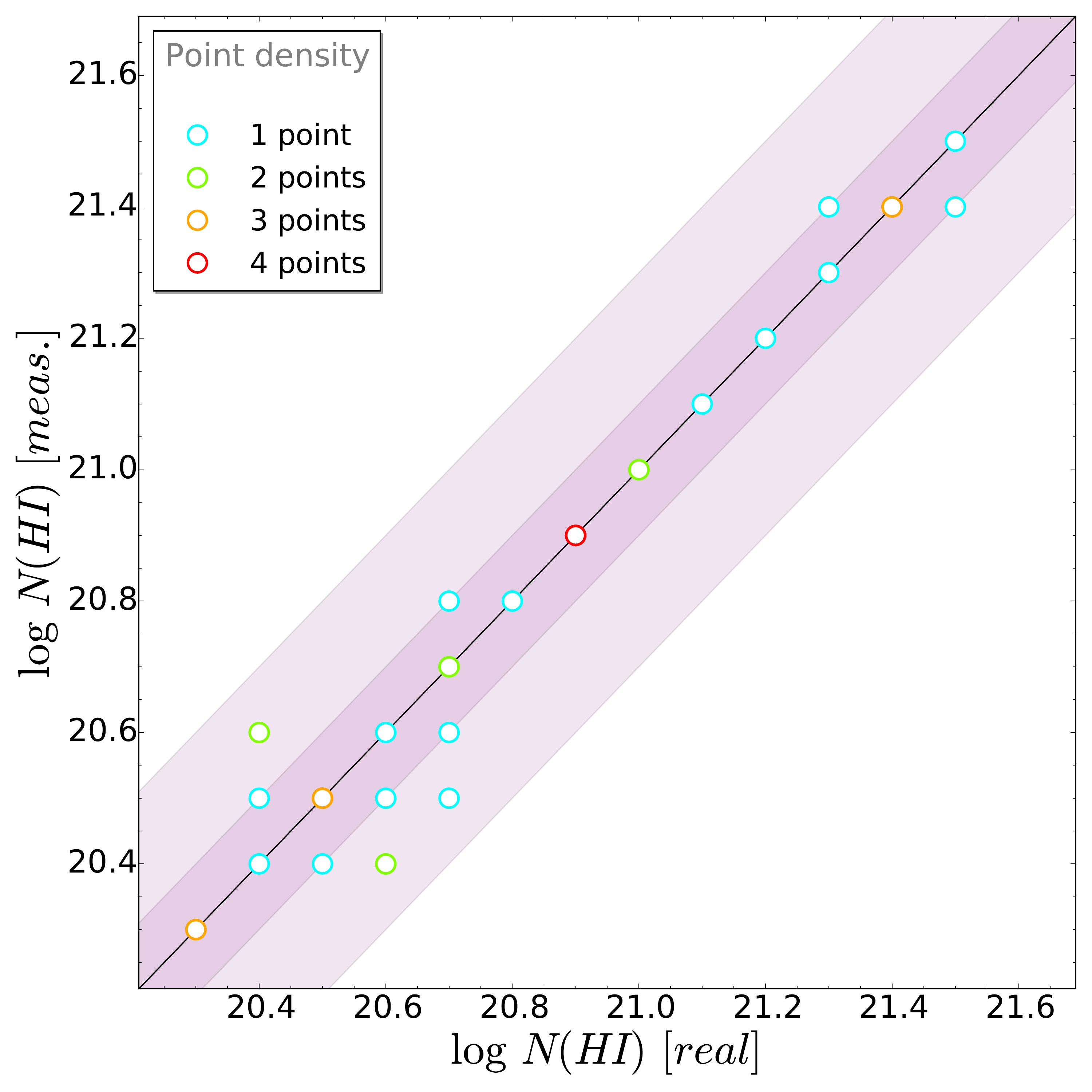}
\caption{Comparison of our measurements of the synthetic DLAs inserted into XQ-100 data. The colour of the points represent the number of points that overlap due to matching \logNHI\ for different redshift tests. Shaded purple regions show 0.1 dex and 0.3 dex intervals.}
\label{fig:logN-test}
\end{figure}

\begin{figure}
\includegraphics[width=0.45\textwidth]{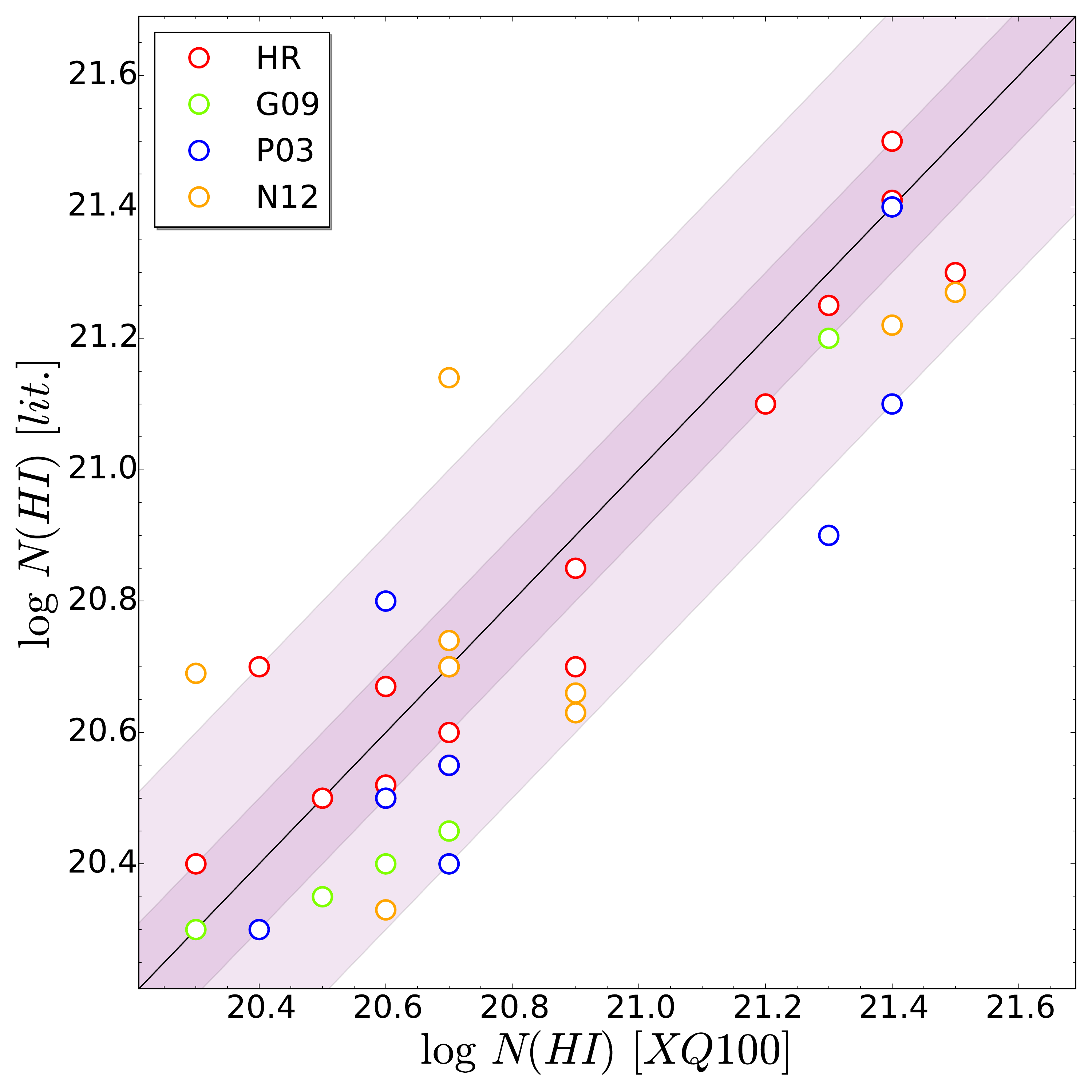}
\caption{Comparison of our XQ-100 measurements with previous estimates. Shaded purple regions show 0.1 dex and 0.3 dex intervals.}
\label{fig:comp}
\end{figure}

We also performed a further test to determine whether our XQ-100 measurements are reliable, or exhibit any systematic bias. Using the pairs of (\z, \logNHI) values of our DLA sample and the number of Lyman series lines used to fit each pair, we injected synthetic aborbers into real XQ-100 spectra.  In this way, we accurately represent the properties of the DLAs in our sample, and our ability to recover them for the noise and resolution properties of the data. We blindly measured the \logNHI\ of the synthetic absorbers using the same procedure as previously employed for real systems. The comparison between the real and measured column densities are shown in Figure \ref{fig:logN-test}.  The agreement at high \logNHI\ is excellent.  At moderate column densities, there is an increased scatter, mainly due to blending with sub-DLAs and Lyman limit systems, but most absorbers are accurately measured within 0.2 dex, and there is no systematic under or over-estimate.

\begin{table*}
\caption{XQ-100 DLA catalog.}
\begin{tabular}{@{}llllllll}
\hline
Name & \zem & \zmin & \zmax & \zabs & \logNHI & \elogNHI & Lines covered \\
\hline
$\Diamond$ $[$HB89$]$ 0000-263 & 4.010 & 2.307 & 4.041 & 3.3900 & 21.40 & 0.10 & Ly-$\alpha$, Ly-$\beta$, Ly-$\gamma$, Ly-$\delta$, Ly-$\epsilon$ \\
$\spadesuit$ $[$HB89$]$ 0053-284 & 3.620 & 2.447 & 3.559 &   &   &   &   \\
$[$HB89$]$ 0055-269 & 3.660 & 1.599 & 3.584 &   &   &   &   \\
BRI 0241-0146 & 4.050 & 2.743 & 3.972 &   &   &   &   \\
BR 0245-0608 & 4.240 & 2.891 & 4.147 &   &   &   &   \\
$\Diamond$ BRI 0952-0115 & 4.430 & 2.907 & 4.329 & 4.0245 & 20.70 & 0.15 & Ly-$\alpha$, Ly-$\beta$, Ly-$\gamma$, Ly-$\delta$ \\
BR 1033-0327 & 4.510 & 2.899 & 4.440 &   &   &   &   \\
BRI 1108-0747 & 3.920 & 2.619 & 3.841 &   &   &   &   \\
$[$HB89$]$ 1159+123 & 3.510 & 1.854 & 3.448 &   &   &   &   \\
BR 2212-1626 & 3.990 & 2.529 & 3.912 &   &   &   &   \\
BR 2213-6729 & 4.470 & 2.768 & 4.389 &   &   &   &   \\
BR 2248-1242 & 4.160 & 2.940 & 4.072 &   &   &   &   \\
$\Diamond$ PKS B1418-064 & 3.690 & 2.356 & 3.611 & 3.4490 & 20.30 & 0.15 & Ly-$\alpha$, Ly-$\beta$, Ly-$\gamma$, Ly-$\delta$, Ly-$\epsilon$ \\
BR J0006-6208 & 4.460 & 2.998 & 4.351 & 3.2030 & 20.90 & 0.15 & Ly-$\alpha$ \\
 &   &   &   & 3.7750 & 21.00 & 0.20 & Ly-$\alpha$, Ly-$\beta$ \\
BR J0030-5129 & 4.170 & 2.529 & 4.088 &   &   &   &   \\
$\bigstar$ PSS J0034+1639 & 4.29 & 2.9813 & 4.2396 & 4.2835 & 21.00 & 0.10 & Ly-$\alpha$, Ly-$\beta$, Ly-$\gamma$, Ly-$\delta$, Ly-$\epsilon$ \\
 $\bigstar$ &   &   &   & 4.2523 & 20.60 & 0.10 & Ly-$\alpha$, Ly-$\beta$, Ly-$\gamma$, Ly-$\delta$, Ly-$\epsilon$ \\
  &   &   &   & 3.7550 & 20.40 & 0.15 & Ly-$\alpha$, Ly-$\beta$ \\
SDSS J004219.74-102009.4 & 3.880 & 2.488 & 3.783 &   &   &   &   \\
BRI J0048-2442 & 4.150 & 2.587 & 4.000 &   &   &   &   \\
$\spadesuit$ PMN J0100-2708 & 3.520 & 2.192 & 3.471 &   &   &   &   \\
$\Diamond$ BRI J0113-2803 & 4.300 & 2.784 & 4.227 & 3.1060 & 21.20 & 0.10 & Ly-$\alpha$ \\
PSS J0117+1552 & 4.240 & 2.364 & 4.157 &   &   &   &   \\
PSS J0121+0347 & 4.130 & 2.546 & 4.041 &   &   &   &   \\
$\Diamond$ SDSS J0124+0044 & 3.840 & 2.077 & 3.758 & 2.2610 & 20.70 & 0.15 & Ly-$\alpha$ \\
$\Diamond$ PSS J0132+1341 & 4.160 & 2.833 & 4.067 & 3.9360 & 20.40 & 0.15 & Ly-$\alpha$, Ly-$\beta$, Ly-$\gamma$, Ly-$\delta$ \\
$\Diamond$ PSS J0133+0400 & 4.150 & 2.850 & 4.100 & 3.6920 & 20.70 & 0.10 & Ly-$\alpha$, Ly-$\beta$ \\
$\Diamond$  &   &   &   & 3.7725 & 20.70 & 0.10 & Ly-$\alpha$, Ly-$\beta$ \\
BRI J0137-4224 & 3.970 & 2.513 & 3.889 &   &   &   &   \\
SDSS J015339.60-001104.8 & 4.190 & 2.825 & 4.110 &   &   &   &   \\
PSS J0211+1107 & 3.980 & 2.438 & 3.891 &   &   &   &   \\
PMN J0214-0518 & 3.990 & 2.554 & 3.895 &   &   &   &   \\
$\Diamond$ BR J0234-1806 & 4.310 & 2.957 & 4.218 & 3.6930 & 20.40 & 0.15 & Ly-$\alpha$, Ly-$\beta$ \\
PSS J0248+1802 & 4.420 & 2.858 & 4.350 &   &   &   &   \\
$\Diamond$ SDSS J025518.57+004847.4 & 4.010 & 2.702 & 3.921 & 3.9145 & 21.50 & 0.10 & Ly-$\alpha$, Ly-$\beta$, Ly-$\gamma$, Ly-$\delta$, Ly-$\epsilon$ \\
$\Diamond$  &   &   &   & 3.2555 & 20.90 & 0.10 & Ly-$\alpha$ \\
$\Diamond$ BR J0307-4945 & 4.720 & 3.130 & 4.622 & 4.4665 & 20.60 & 0.10 & Ly-$\alpha$, Ly-$\beta$, Ly-$\gamma$, Ly-$\delta$, Ly-$\epsilon$ \\
 &   &   &   & 3.5910 & 20.50 & 0.15 & Ly-$\alpha$ \\
BR J0311-1722 & 4.040 & 2.562 & 3.951 &   &   &   &   \\
BR 0401-1711 & 4.230 & 2.858 & 4.141 &   &   &   &   \\
BR J0415-4357 & 4.070 & 2.800 & 3.990 & 3.8080 & 20.50 & 0.20 & Ly-$\alpha$, Ly-$\beta$, Ly-$\gamma$ \\
$\Diamond$ BR 0424-2209 & 4.320 & 2.751 & 4.242 & 2.9825 & 21.40 & 0.15 & Ly-$\alpha$ \\
BR 0523-3345 & 4.410 & 2.817 & 4.297 &   &   &   &   \\
BR J0529-3526 & 4.410 & 2.817 & 4.329 &   &   &   &   \\
BR J0529-3552 & 4.170 & 2.825 & 4.087 & 3.6840 & 20.40 & 0.15 & Ly-$\alpha$, Ly-$\beta$ \\
BR J0714-6455 & 4.460 & 2.776 & 4.374 &   &   &   &   \\
$\Diamond$ SDSS J074711.15+273903.3 & 4.170 & 2.710 & 4.049 & 3.4235 & 20.90 & 0.10 & Ly-$\alpha$, Ly-$\beta$ \\
$\Diamond$  &   &   &   & 3.9010 & 20.60 & 0.15 & Ly-$\alpha$, Ly-$\beta$, Ly-$\gamma$, Ly-$\delta$, Ly-$\epsilon$ \\
SDSS J075552.41+134551.1 & 3.670 & 2.085 & 3.587 &   &   &   &   \\
$\bigstar$ SDSS J080050.27+192058.9 & 3.96 & 2.7264 & 3.899 & 3.9465 & 20.40 & 0.10 & Ly-$\alpha$, Ly-$\beta$, Ly-$\gamma$, Ly-$\delta$, Ly-$\epsilon$ \\
SDSS J081855.78+095848.0 & 3.670 & 2.406 & 3.580 & 3.3060 & 21.00 & 0.10 & Ly-$\alpha$, Ly-$\beta$, Ly-$\gamma$ \\
SDSS J083322.50+095941.2 & 3.750 & 2.044 & 3.639 &   &   &   &   \\
SDSS J083510.92+065052.8 & 3.990 & 2.735 & 3.925 &   &   &   &   \\
SDSS J083941.45+031817.0 & 4.250 & 2.883 & 4.144 &   &   &   &   \\
SDSS J092041.76+072544.0 & 3.640 & 2.060 & 3.570 & 2.2380 & 20.90 & 0.15 & Ly-$\alpha$ \\
SDSS J093556.91+002255.6 & 3.750 & 2.249 & 3.669 &   &   &   &   \\
SDSS J093714.48+082858.6 & 3.700 & 2.118 & 3.626 &   &   &   &   \\
SDSS J095937.11+131215.4 & 4.060 & 2.702 & 4.008 &   &   &   &   \\
\hline
\label{tab:xq}
\end{tabular}
\end{table*}

\begin{table*}
\contcaption{XQ-100 DLA catalog.}
\begin{tabular}{@{}llllllll}
\hline
Name & \zem & \zmin & \zmax & \zabs & \logNHI & \elogNHI & Lines covered \\
\hline
SDSS J101347.29+065015.6 & 3.790 & 2.397 & 3.729 &   &   &   &   \\
$\spadesuit$ SDSS J101818.45+054822.8 & 3.520 & 2.299 & 3.441 &   &   &   &   \\
SDSS J102040.62+092254.2 & 3.640 & 2.093 & 3.564 & 2.5920 & 21.50 & 0.10 & Ly-$\alpha$ \\
$\spadesuit$ SDSS J102456.61+181908.7 & 3.530 & 2.159 & 3.450 & 2.2980 & 21.30 & 0.10 & Ly-$\alpha$ \\
SDSS J103221.11+092748.9 & 3.990 & 2.619 & 3.903 &   &   &   &   \\
SDSS J103446.54+110214.5 & 4.270 & 2.422 & 4.183 &   &   &   &   \\
SDSS J103730.33+213531.3 & 3.630 & 1.665 & 3.550 &   &   &   &   \\
SDSS J103732.38+070426.2 & 4.100 & 2.225 & 4.043 &   &   &   &   \\
SDSS J104234.01+195718.6 & 3.640 & 2.044 & 3.554 &   &   &   &   \\
SDSS J105340.75+010335.6 & 3.650 & 1.937 & 3.587 &   &   &   &   \\
SDSS J105434.17+021551.9 & 3.970 & 2.603 & 3.889 &   &   &   &   \\
SDSS J105705.37+191042.8 & 4.100 & 2.661 & 4.044 & 3.3735 & 20.30 & 0.10 & Ly-$\alpha$, Ly-$\beta$ \\
$\Diamond$ SDSS J105858.38+124554.9 & 4.330 & 2.570 & 4.253 & 3.4315 & 20.60 & 0.10 & Ly-$\alpha$, Ly-$\beta$ \\
SDSS J110352.73+100403.1 & 3.610 & 2.200 & 3.531 &   &   &   &   \\
SDSS J110855.47+120953.3 & 3.670 & 2.447 & 3.601 & 3.5460 & 20.80 & 0.15 & Ly-$\alpha$, Ly-$\beta$, Ly-$\gamma$, Ly-$\delta$, Ly-$\epsilon$ \\
$\Diamond$  &   &   &   & 3.3965 & 20.70 & 0.10 & Ly-$\alpha$, Ly-$\beta$, Ly-$\gamma$ \\
SDSS J111008.61+024458.0 & 4.120 & 2.364 & 4.062 &   &   &   &   \\
SDSS J111701.89+131115.4 & 3.620 & 2.208 & 3.546 &   &   &   &   \\
SDSS J112617.40-012632.6 & 3.610 & 2.225 & 3.558 &   &   &   &   \\
SDSS J112634.28-012436.9 & 3.740 & 2.430 & 3.687 &   &   &   &   \\
SDSS J113536.40+084218.9 & 3.830 & 1.780 & 3.755 &   &   &   &   \\
$\spadesuit$ SDSS J120210.08-005425.4 & 3.590 & 2.159 & 3.517 &   &   &   &   \\
SDSS J124837.31+130440.9 & 3.720 & 2.315 & 3.644 &   &   &   &   \\
SDSS J124957.23-015928.8 & 3.630 & 2.406 & 3.553 &   &   &   &   \\
SDSS J130452.57+023924.8 & 3.650 & 2.257 & 3.572 &   &   &   &   \\
SDSS J131242.87+084105.1 & 3.740 & 2.027 & 3.653 & 2.6600 & 20.50 & 0.10 & Ly-$\alpha$, Ly-$\beta$ \\
2MASSi J1320299-052335 & 3.700 & 1.904 & 3.640 &   &   &   &   \\
SDSS J132346.05+140517.6 & 4.040 & 2.241 & 3.971 &   &   &   &   \\
BR J1330-2522 & 3.950 & 2.282 & 3.867 &   &   &   &   \\
SDSS J133150.69+101529.4 & 3.850 & 2.323 & 3.772 &   &   &   &   \\
$\spadesuit$ SDSS J133254.51+005250.6 & 3.510 & 2.323 & 3.434 &   &   &   &   \\
SDSS J133653.44+024338.1 & 3.800 & 1.887 & 3.722 &   &   &   &   \\
SDSS J135247.98+130311.5 & 3.700 & 2.035 & 3.629 &   &   &   &   \\
SDSS J1401+0244 & 4.440 & 2.916 & 4.319 &   &   &   &   \\
$\spadesuit$ SDSS J141608.39+181144.0 & 3.590 & 2.266 & 3.518 &   &   &   &   \\
$\spadesuit$ SDSS J144250.12+092001.5 & 3.530 & 1.780 & 3.458 &   &   &   &   \\
SDSS J144516.46+095836.0 & 3.520 & 1.599 & 3.487 &   &   &   &   \\
SDSS J150328.88+041949.0 & 3.660 & 2.118 & 3.615 &   &   &   &   \\
$\spadesuit \Diamond$ SDSS J151756.18+051103.5 & 3.560 & 2.249 & 3.480 & 2.6885 & 21.40 & 0.10 & Ly-$\alpha$ \\
$\spadesuit$ SDSS J152436.08+212309.1 & 3.610 & 2.052 & 3.525 &   &   &   &   \\
SDSS J154237.71+095558.8 & 3.990 & 2.257 & 3.904 &   &   &   &   \\
SDSS J155255.03+100538.3 & 3.730 & 2.529 & 3.644 & 3.6010 & 21.10 & 0.10 & Ly-$\alpha$, Ly-$\beta$, Ly-$\gamma$, Ly-$\delta$, Ly-$\epsilon$ \\
SDSS J1621-0042 & 3.700 & 2.101 & 3.634 &   &   &   &   \\
$\Diamond$ SDSS J163319.63+141142.0 & 4.330 & 2.438 & 4.277 & 2.8820 & 20.30 & 0.15 & Ly-$\alpha$ \\
CGRaBS J1658-0739 & 3.740 & 2.537 & 3.671 &   &   &   &   \\
$\Diamond$ PSS J1723+2243 & 4.520 & 3.056 & 4.440 & 3.6980 & 20.50 & 0.10 & Ly-$\alpha$ \\
$\Diamond$ 2MASSi J2239536-055219 & 4.560 & 2.949 & 4.465 & 4.0805 & 20.60 & 0.10 & Ly-$\alpha$, Ly-$\beta$, Ly-$\gamma$, Ly-$\delta$ \\
$\Diamond$ PSS J2344+0342 & 4.240 & 2.693 & 4.162 & 3.2200 & 21.30 & 0.10 & Ly-$\alpha$ \\
BR J2349-3712 & 4.210 & 2.850 & 4.133 &   &   &   &   \\
\hline
\multicolumn{8}{l}{$\bigstar$ = PDLA, $\Diamond$ = DLA already identified in a previous survey, $\spadesuit$ = Color biased sight-line.}
\end{tabular}
\end{table*}

\section{Literature samples}
\label{sec:combined}

\begin{table}
\begin{minipage}{150mm}
\caption{Abbreviations used for literature samples.}
\begin{tabular}{@{}ll}
\hline
Id & Reference \\
\hline
P03 & \citet{2003MNRAS.346.1103P} \\
Z05 & \citet{2005MNRAS.364.1467Z} \\
R06 & \citet{2006ApJ...636..610R} \\
PW09 & \citet{2009ApJ...696.1543P} \\
G09 & \citet{2009AA...508..133G} \\
B12 & \citet{2012ApJ...749...87B} \\
N12 & \citet{2012AA...547L...1N} \\
D13 & \citet{2013MNRAS.433.1398D} \\
R13 & \citet{2013MNRAS.435.2693R} \\
C15 & \citet{2015MNRAS.452..217C} \\
N15 & \citet{2015ApJN} \\
\hline
\label{tab:references}
\end{tabular}
\end{minipage}
\end{table}

Since the early work by \citet{1986ApJS...61..249W}, numerous surveys have catalogued DLAs over a range of redshifts \citep[e.g.][]{1991ApJS...77....1L, 1995ApJ...454..698W, 1996ApJ...468..121S, 2000ApJ...543..552S, 2001AA...379..393E, 2001AJ....121.1799P, 2004PASP..116..622P, 2005ApJ...635..123P, 2006ApJ...636..610R, 2009AA...508..133G, 2009ApJ...696.1543P, 2009AA...505.1087N, 2012AA...547L...1N, 2015MNRAS.452..217C, 2015ApJN} in order to trace the cosmic evolution of neutral hydrogen gas in galaxies.  These surveys are extremely heterogenous and have been conducted with a variety of telescope apertures, both from space and on the ground, and at a range of spectral resolutions.  There are also considerable duplications between surveys, and both the naming conventions and the presentation of the data in the literature mean that assembling a combined sample is a considerable challenge.  Nonetheless, in this work we have attempted to assemble such a combined sample from the major DLA catalogs that are currently available, focusing on $z>2$, the redshift range where ground-based surveys have most effectively contributed.  We review these catalogs in turn below.

For comparison purposes, we also use \citet[][hereafter N12]{2012AA...547L...1N} data in the high redshift range, \citet[][hereafter R06]{2006ApJ...636..610R} and \citet[][hereafter N15]{2015ApJN} for intermediate redshifts and the 21cm samples from the local Universe by \citet[][hereafter R13]{2013MNRAS.435.2693R}, \citet[][hereafter D13]{2013MNRAS.433.1398D}, \citet[][hereafter Z05]{2005MNRAS.364.1467Z}, and \citet[][hereafter B12]{2012ApJ...749...87B}. For reference, all abbreviations for the literature catalogs used in this work are summarized in Table \ref{tab:references}.

\subsection{The Peroux et al. (2003) compilation}

A compilation of approximately the first decade and a half of DLA surveys is presented by  \citet{2003MNRAS.346.1103P}, hereafter the P03 sample. The P03 sample combines
the high redshift DLA survey of \citet{2001AJ....121.1799P} with DLAs from 25 separate papers, several of which are themselves compilations from other surveys. Based on their statistical sample of 713 quasars and 114 DLAs, this sample was used by \citet{2003MNRAS.346.1103P, 2005MNRAS.363..479P} to conclude that the total amount of neutral gas is conserved from \z=2 to \z$\sim$5.

In the process of duplication checking (described in more detail in Section \ref{sec:cs}) and checking the original references of the P03 compilation, we noted a number of inconsistencies with the original reference papers, such as the values of emission redshifts adopted for the computation of absorption statistics.  We checked each of these inconsistencies manually, and concluded that they are likely due to typographical errors, and we have corrected them, as summarized in Table \ref{tab:peroux}. We also identified 4 QSOs that are duplicates, but with different names, within the P03 compilation; the duplicated sightlines have been removed. In two cases, we could not find the original reference for a given QSO, and were therefore unable to verify the properties of the QSO/absorber; these two sightlines were also removed from the sample. We summarize all of the modifications made to the P03 compilation in Table \ref{tab:peroux}. We recomputed \zmax\ to be 5000 \kms\ bluewards of \zem, in order to be consistent with the threshold set for the XQ-100 sample.

\begin{table}
\begin{minipage}{150mm}
\caption{Modifications to the P03 catalog}
\begin{tabular}{@{}ll}
\hline
Id & Comment \\
\hline
BR B0331-1622 & $\equiv$ BR J0334-1612. Removed. \\
BR B0401-1711 & $\equiv$ BR J0403-1703. Removed. \\
Q 0007-000 & $\equiv$ Q 0007-0004. Removed. \\
Q 1600+0729 & $\equiv$ BR J1603+0721. Removed. \\
Q 0101-3025 & \zem\ changed from 4.073 to 3.164. \\
Q 0041-2607 & \zem\ changed from 2.79 to 2.46. \\
Q 0201+3634 & \zem\ changed from 2.49 to 2.912. \\
Q 2359-0216 & \zem\ changed from 2.31 to 2.81. \\
MG 1559+1405 & Reference not found. Removed. \\
MG 2254+0227 & Reference not found. Removed. \\
\hline
\label{tab:peroux}
\end{tabular}
\end{minipage}
\end{table}

\subsection{The Prochaska \& Wolfe (2009) SDSS DLA sample}

\citet[][hereafter PW09]{2009ApJ...696.1543P} used automated search algorithms to identify DLAs in the Sloan Digital Sky Survey Data Release 5 (SDSS DR5). A SNR requirement
of 4 was adopted, and DLAs are included in their statistical sample if they are at least 3000 \kms\ from the background QSO. DLAs have also been identified in more recent data releases, specifically the DR7 \citep{2009AA...505.1087N} and DR9 \citep{2012AA...547L...1N}, the former of which has a public catalog of identified DLAs. However, our computation of \Odla\ requires additional details of the minimum and maximum redshifts of the DLA search for every sightline, which is not available for the DR7 and DR9 samples. The DR5 sample of PW09 is therefore the largest of the individual literature samples considered in this work, containing 7472 QSOs with 738 DLAs.  We do, however, compare the results of our combined sample to that of \citet{2012AA...547L...1N} later in this paper.

PW09 find that the mass density of neutral gas has decreased by a factor of about two between redshifts of 3.5 and 2.5.  A similar rate of gas content decline was determined
from the later SDSS data releases by \citet{2009AA...505.1087N, 2012AA...547L...1N}.  However, \citet{2009AA...505.1087N} suggest that the analysis of PW09 may be biased against the detection of the lowest redshift DLAs, which could be rectified by implementing a velocity buffer to the minimum redshift used to compute DLA statistics.  We implement this buffer, which has a value of 10,000 \kms, at the low redshift end of each SDSS spectrum.  The new values of \zmin\ listed in Table \ref{tab:final_cs}, which describes our final combined DLA sample, include this velocity buffer. We also recomputed \zmax\ to be 5000 \kms\ bluer than \zem, for consistency with the XQ-100 sample.

\subsection{The Guimaraes et al. (2009) sample}

The broad \Lya\ wings of DLAs, and high absorber equivalent widths means that these galaxy scale absorbers can be easily detected in relatively low resolution spectra.
Indeed, most surveys have been performed at typical resolutions of $R \sim$ 1000 - 2000. However, there are advantages to pursuing absorption line surveys at higher resolution, such as the ability to push down to the sub-DLA regime \citep[e.g.,][]{2007ApJ...656..666O, 2013AA...556A.141Z} and to assess the increasing potential for contamination (blends) at higher redshifts.  The trade-off is, of course, the increased exposure times necessary to reach a fixed SNR.

\citet[][hereafter G09]{2009AA...508..133G} presented the first systematic DLA survey for \Odla measurements performed with an intermediate resolution spectrograph, namely the Echellette Spectrograph and Imager \citep[ESI, mounted at the 10m Keck telescope,][]{2002PASP..114..851S}. ESI's resolution ($R \sim$ 4500) is quite similar to that of X-shooter. However, the most notable difference between the two instruments is the wavelength coverage. Whereas X-shooter extends from the atmospheric cut-off to the K-band, ESI has no coverage below 4000 \AA\ and a greatly reduced efficiency from $\sim$ 4000 -- 4300 \AA.

A total of 99 QSOs (77 considered for their statistical analysis) with emission redshifts ranging from \z=4 to \z=6.3 were observed by G09, leading to the detection of 100 absorbers with \logNHI$>$19.5, of which 40 are DLAs.  DLAs at least 5000  \kms\ from the QSO redshift were included in the statistical sample.  Based on the DLA sample, G09 find that there is a decline in \Ohi\ at $z>3.5$. This decline is also present if the sub-DLAs (which increase the total gas mass density by about 30 per cent) are combined with the DLAs. The discrepancy with the results of \citet{2009ApJ...696.1543P} is suggested by G09 to come from the difficulties in establishing the damping nature of the systems with the high density \Lya\ 
forest in this very high redshift range. However, generations of previous surveys have demonstrated that low number statistics at the redshift boundary of the survey can also lead to an apparent turnover of \Ohi.  Of the 40 DLAs in the G09 sample, there are 6 duplicates with the XQ-100 sample.

\subsection{The Crighton et al. (2015) sample}

The Giant Gemini GMOS (GGG) survey observed 163 $z>4.4$ QSOs with GMOS-N and GMOS-S at the Gemini Observatory \citep{2014MNRAS.445.1745W}. Like the G09 sample, the main focus of the GGG DLA sample was the assessment of \Ohi\ at high redshifts. However, the GMOS spectra are of significantly lower resolution than the ESI spectra used by G09 and \citet{2015MNRAS.452..217C} carefully assess the contamination by both false positives (blends) and missed DLAs through a variety of blind tests and comparisons with repeated observations at higher spectral resolution.  Despite potential concerns of blending and low resolution, \citet{2015MNRAS.452..217C} conclude that the required correction factors are minimal. DLAs are included in the GGG statistical sample if they are at least 5000 \kms\ from the background QSO.

Although \citet{2015MNRAS.452..217C} are not able to accurately identify and fit sub-DLAs, in their estimate of \Ohi\ they make a uniform correction for the contribution of these lower column density systems. We do not make this correction \textit{a priori}, but rather use the DLA catalog of \citet{2015MNRAS.452..217C} directly in our combined sample.  At \zabs$\sim$5, the \Ohi\ derived from GGG is formally consistent with the SDSS measurements of \citet{2012AA...547L...1N} at \z $\sim$ 3. However, a power law of the form $(1+z)^{0.4}$, describing a slowly decreasing \Ohi\ towards lower redshifts is consistent with data spanning 0 $<$ \z\ $<$ 5.

\subsection{The combined sample}\label{sec:cs}

\begin{table}
\centering
\begin{minipage}{80mm}
\caption{Data-sets included to build up our combined sample.}
\begin{tabular}{llllll}
 \hline
 Id & No. QSOs & No. DLAs & No. PDLAs \footnote{PDLAs ($v$ $<$ 5000) not included in Combined Sample (CS)} & $\sum\Delta X_{i}$ \\
 \hline
 XQ100 & 100 & 38 & 3 & 536 \\
 G09 & 68 & 34 & 0 & 378 \\
 GGG & 154 & 43 & 0 & 553 \\ 
 PW09 & 4983 & 559 & 50 & 8529 \\ 
 P03 & 397 & 68 & 0 & 1494 \\
 \hline
 CS & 5702 & 742 & 53 & 10434 \\
 \hline
\label{tab:cs}
\end{tabular}
\end{minipage}
\end{table}

\begin{figure}
\includegraphics[width=0.49\textwidth]{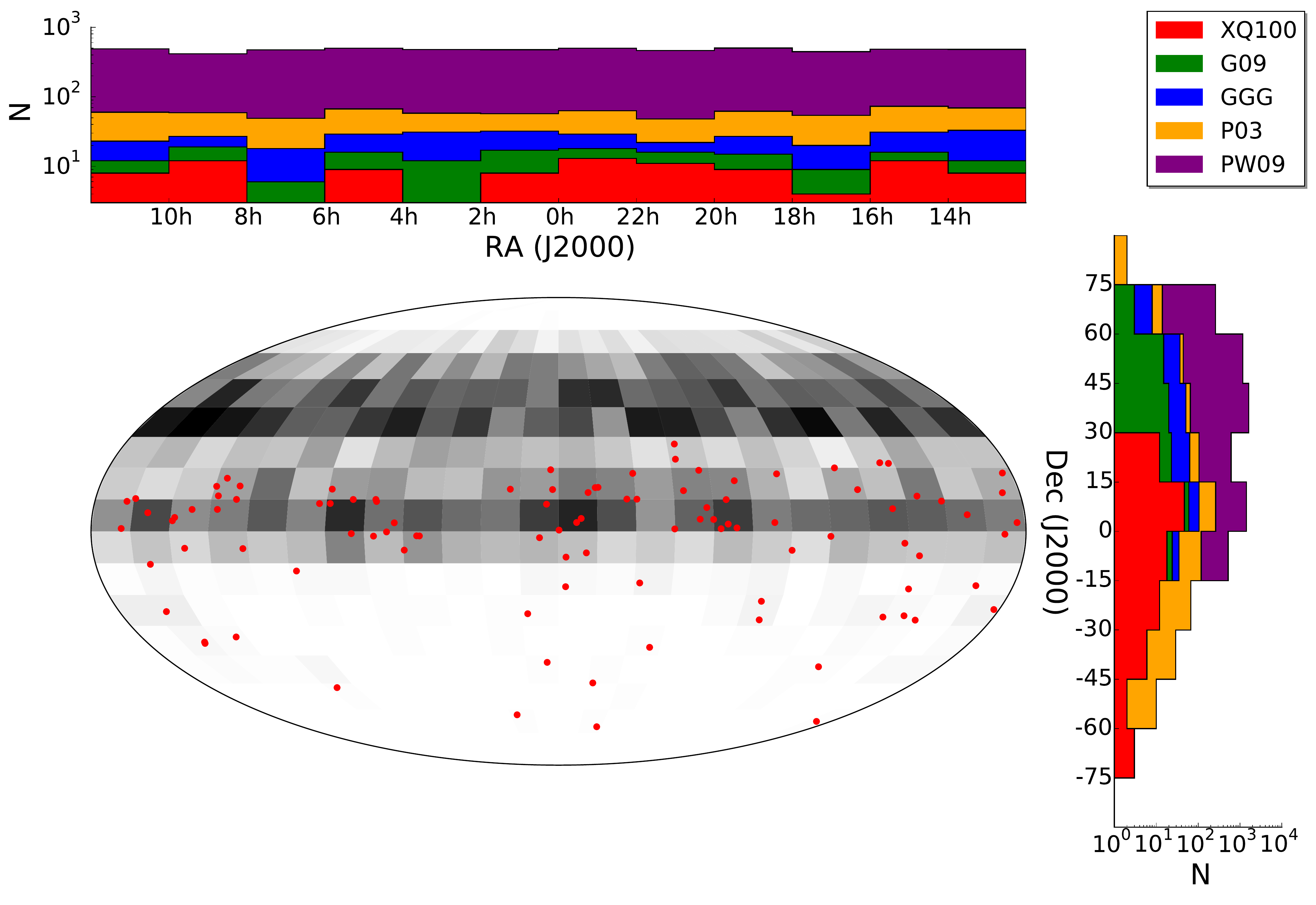} 
\caption{Sky position of the QSOs in the total combined sample. 1D histograms for each celestial coordinate and a 2D histogram density plot on the celestial globe are presented. The gray scale 2D histogram for the combined sample density plot represents the number of sources per surface unit (150 sq degrees), ranging from 0 (white) to 58 (black).  The mean number of quasars per surface unit is 13.2.  The red points over-plotted are the positions of the XQ-100 QSOs.  The propensity of northern sky coverage is driven by the SDSS.}
\label{fig:sky}
\end{figure}

The reviews provided above highlight the sensitivity of \Odla\ to a variety of possible systematics, including blending/contamination, SNR, robust definitions of the search path and, perhaps most importantly, sample size. Since the few, rare, high \NHI\ absorbers contribute to \Ohi\ (and \Odla) appreciably, large samples are required to statistically assess \Ohi\ at any given redshift.  We have therefore combined the major surveys and compilations described above, in order to minimize fluctuations in the determination of \Ohi\ due to inadequate sampling of the column density distribution function.

Prior to combining the literature samples and the XQ-100 sample, it is necessary to check for, and remove, duplicate DLAs. Duplication checking was achieved by obtaining the sky coordinates and redshift of all quasars in all samples from NED, SIMBAD or SDSS, since naming conventions between surveys are heterogeneous. For the join procedure, we made XQ-100 the initial base sample. The catalogs were assessed in the order listed in Table \ref{tab:cs}, which was adopted as an approximate ranking of spectral quality (SNR and resolution). The base sample was compared with each subsequent catalog in Table \ref{tab:cs} by performing a coordinate cross-matching with a positional tolerance of $10\arcsec$. All matches were removed from the last table and the resulting combined (duplicate free) catalog was used as the base table for the next iteration (i.e. with the next catalog in Table \ref{tab:cs}).

In Table \ref{tab:final_cs} we present the final combined sample (CS) with all duplicates removed. The final combined sample contains a total of 742 DLAs, spanning a redshift from 1.673 to 5.015. A comparison of the sky positions of the combined literature sample and the XQ-100 sample is shown in Figure \ref{fig:sky}.

\begin{table*}
\caption{Combined sample catalog. The full table is available in the online version.}
\begin{tabular}{@{}lllllllll}
\hline
Name & RA (J2000) & Dec (J2000) & \zem & \zmin & \zmax & \zabs & \logNHI & \elogNHI \\
\hline
FBQS J0000-1021 & 0.21084 & -10.3655 & 2.640 & 2.307 & 2.604 &   &   &   \\
$[$HB89$]$ 2359+068 & 0.41920 & 7.1650 & 3.238 & 1.632 & 3.203 &   &   &   \\
SDSS J000143.41+152021.4 & 0.43088 & 15.3393 & 2.638 & 2.307 & 2.602 &   &   &   \\
$[$HB89$]$ 2359+003 & 0.44475 & 0.6663 & 2.897 & 1.714 & 2.857 &   &   &   \\
LBQS 2359-0216B & 0.45833 & -1.9945 & 2.810 & 1.747 & 2.779 & 2.0951 & 20.70 & 0.20 \\
 &   &   &   &   &   & 2.1537 & 20.30 & 0.20 \\
FBQS J0002+0021 & 0.58799 & 0.3637 & 3.057 & 2.307 & 3.016 &   &   &   \\
SDSS J000300.34+160027.6 & 0.75146 & 16.0077 & 3.675 & 3.428 & 3.629 &   &   &   \\
... & ... & ... & ... & ... & ... & ... & ... & ... \\
\hline
\label{tab:final_cs}
\end{tabular}
\end{table*}

\section{Analysis}

\subsection{Redshift path coverage}

\begin{figure}
\includegraphics[width=0.49\textwidth]{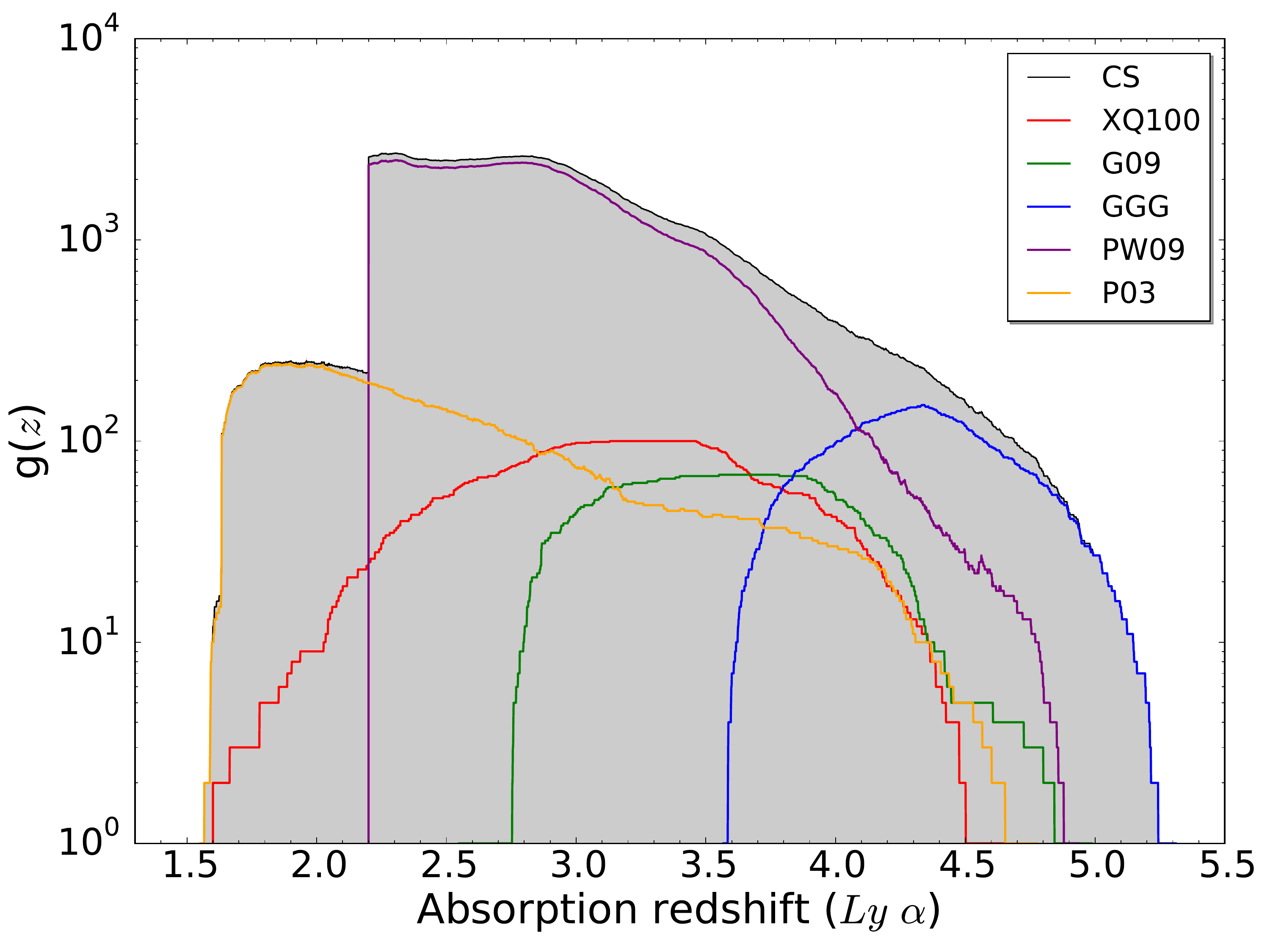} 
\caption{Redshift path for all samples used in this work and for the total combined sample (CS).}
\label{fig:g}
\end{figure}

We begin our analysis by computing the redshift path covered by both the individual samples described in the previous section, as well as for the combined sample. Although we will ultimately use only the combined sample for our determination of \Ohi, it is instructive to see how the various sub-samples contribute as a function of redshift.

The cumulative number of sightlines that could contain an absorber at a given redshift  is defined as \citep{1991ApJS...77....1L}
\begin{equation}
g(z) = \sum\limits_{i} H(z^{i}_{max}-z)H(z-z^{i}_{min})
\end{equation}
where $H$ is the Heaviside function. Then, the total redshift path surveyed is
\begin{equation}
\Delta z = \int \limits ^{\infty}_{0} g(z) dz = \sum \limits _{i} (z_{max}^{i} - z_{min}^{i}) 
\end{equation}

For the XQ-100 sample, the lower limit $z^{i}_{min}$ for the $i^{th}$ quasar was chosen as the minimum redshift where the SNR$>$ 7.5. This SNR threshold was determined from our data 
as a conservative value where we start to easily identify damping wings.  As described in Section \ref{sec:xq100}, the maximum DLA search redshift, $z^{i}_{max}$, was set conservatively to be 5000 \kms\ bluewards of $z^{i}_{em}$.

The resulting XQ-100 $g(z)$ curve is shown in Fig. \ref{fig:g} together with the data of the other samples used in this paper. For the rest of the samples, we kept published \zmin\ but we recomputed \zmax\ to be 5000 \kms\ bluewards \zem. This accounts also for possible uncertainties in the determination of emission redshifts, which are inhomogeneously computed \citep[see, e.g.,][and references therein]{2010MNRAS.405.2302H}. The 5000 \kms\ threshold therefore provides a safe buffer against the inclusion of PDLAs that are likely to affect our statistical sample.

Fig. \ref{fig:g} demonstrates the complementarity of the various previous surveys and the advantage of combining them all together. The XQ-100 survey probes from \z=1.6 up to \z=4.5, with the majority of the absorption path in the \z=[3.0-3.5] range.  The previous moderate resolution DLA survey by G09 extends to slightly higher redshifts than XQ-100.  However, due to ESI's lack of blue sensitivity and the emission redshift distribution of this sample, there is no absorption path coverage below $z\sim 2.7$. The PW09 sample dominates at moderate redshifts, but the P03 sample and GGG samples are major contributors at the lowest and highest redshifts, respectively.

\subsection{DLA distribution function}
\label{sec:DLAfxn}

\begin{figure}
\includegraphics[width=0.49\textwidth]{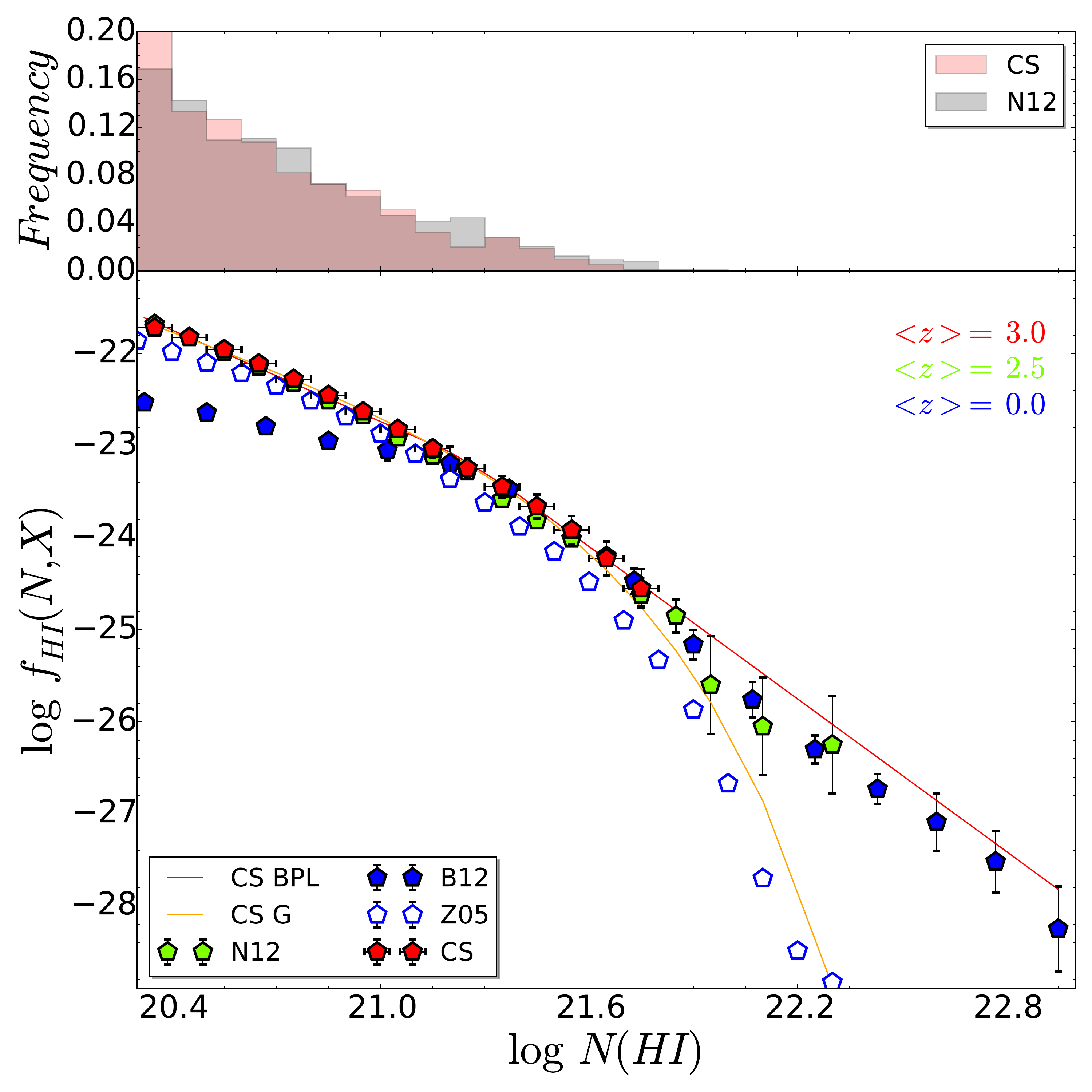}
\caption{Sample frequency and distribution function for the whole combined sample compared with N12 (at similar mean redshift) and Z05+B12 representing the local universe. Model fits to a broken powerlaw (BPL, red) and a gamma function (G, orange) are also plotted.}
\label{fig:fxn}
\end{figure}

\begin{table*}
\centering
\caption{Column density distribution (binned evaluations) for the complete CS ($<z>=2.99$), and split in 2 and 3 redshift bins}
\begin{tabular}{c|cc|cccc|cccccc}
\hline
\logNHI & \multicolumn{6}{c}{$\log$ \fnx} \\
 & \multicolumn{1}{c}{-------------------} & \multicolumn{2}{c}{--------------------------------------} & \multicolumn{3}{c}{---------------------------------------------------------} \\
 & \multicolumn{1}{c}{\z=2.99} & \multicolumn{1}{c}{\z=2.57} & \multicolumn{1}{c}{\z=3.47} & \multicolumn{1}{c}{\z=2.44} & \multicolumn{1}{c}{\z=2.95} & \multicolumn{1}{c}{\z=3.69} \\
\hline
20.3 & -21.72 $\pm$ 0.05 & -21.79 $\pm$ 0.08 & -21.66 $\pm$ 0.07 & -21.82 $\pm$ 0.1 & -21.69 $\pm$ 0.08 & -21.66 $\pm$ 0.08 \\
20.4 & -21.82 $\pm$ 0.05 & -21.89 $\pm$ 0.08 & -21.76 $\pm$ 0.07 & -21.91 $\pm$ 0.1 & -21.81 $\pm$ 0.09 & -21.76 $\pm$ 0.08 \\
20.5 & -21.95 $\pm$ 0.06 & -22.02 $\pm$ 0.08 & -21.89 $\pm$ 0.07 & -22.04 $\pm$ 0.1 & -21.96 $\pm$ 0.09 & -21.88 $\pm$ 0.08 \\
20.6 & -22.11 $\pm$ 0.06 & -22.17 $\pm$ 0.09 & -22.05 $\pm$ 0.08 & -22.18 $\pm$ 0.1 & -22.13 $\pm$ 0.1 & -22.02 $\pm$ 0.09 \\
20.7 & -22.28 $\pm$ 0.06 & -22.33 $\pm$ 0.09 & -22.23 $\pm$ 0.08 & -22.35 $\pm$ 0.11 & -22.31 $\pm$ 0.11 & -22.19 $\pm$ 0.1 \\
20.8 & -22.45 $\pm$ 0.07 & -22.51 $\pm$ 0.1 & -22.4 $\pm$ 0.09 & -22.53 $\pm$ 0.12 & -22.48 $\pm$ 0.12 & -22.36 $\pm$ 0.1 \\
20.9 & -22.63 $\pm$ 0.07 & -22.7 $\pm$ 0.11 & -22.56 $\pm$ 0.09 & -22.74 $\pm$ 0.14 & -22.65 $\pm$ 0.13 & -22.53 $\pm$ 0.11 \\
21.0 & -22.82 $\pm$ 0.08 & -22.92 $\pm$ 0.12 & -22.74 $\pm$ 0.1 & -22.94 $\pm$ 0.15 & -22.84 $\pm$ 0.14 & -22.71 $\pm$ 0.12 \\
21.1 & -23.03 $\pm$ 0.09 & -23.13 $\pm$ 0.14 & -22.95 $\pm$ 0.12 & -23.13 $\pm$ 0.16 & -23.07 $\pm$ 0.16 & -22.92 $\pm$ 0.14 \\
21.2 & -23.24 $\pm$ 0.11 & -23.32 $\pm$ 0.15 & -23.18 $\pm$ 0.13 & -23.28 $\pm$ 0.17 & -23.31 $\pm$ 0.18 & -23.14 $\pm$ 0.15 \\
21.3 & -23.45 $\pm$ 0.12 & -23.49 $\pm$ 0.16 & -23.41 $\pm$ 0.15 & -23.43 $\pm$ 0.18 & -23.56 $\pm$ 0.19 & -23.35 $\pm$ 0.17 \\
21.4 & -23.66 $\pm$ 0.13 & -23.68 $\pm$ 0.17 & -23.63 $\pm$ 0.17 & -23.61 $\pm$ 0.19 & -23.78 $\pm$ 0.21 & -23.55 $\pm$ 0.18 \\
21.5 & -23.92 $\pm$ 0.15 & -23.92 $\pm$ 0.19 & -23.88 $\pm$ 0.19 & -23.85 $\pm$ 0.2 & -23.99 $\pm$ 0.21 & -23.78 $\pm$ 0.2 \\
21.6 & -24.22 $\pm$ 0.18 & -24.2 $\pm$ 0.21 & -24.16 $\pm$ 0.21 & -24.11 $\pm$ 0.22 & -24.2 $\pm$ 0.22 & -24.03 $\pm$ 0.21 \\
21.7 & -24.56 $\pm$ 0.21 & --- & -24.41 $\pm$ 0.22 & --- & --- & -24.27 $\pm$ 0.22 \\
\hline
\label{tab:FNX}
\end{tabular}
\end{table*}


The differential column density distribution, that represents the number of absorbers between $N$ and $N$+$dN$ and \z\ and \z+$dz$, is defined \citep{1991ApJS...77....1L} as 
\begin{equation}
f_{HI}(N,X) dN dX = \frac{m}{\Delta N \sum_{i} \Delta X_{i}} dN dX
\label{eq:fxn}
\end{equation}
where $m$ is the number of systems in a column density bin and the absorption distance
\begin{equation}
\DX = \int\limits_{\zmin}^{\zmax} \frac{H_{0}}{H(z)} (1+z)^{2} dz
\label{eq:dx}
\end{equation}
is a quantity conveniently defined to give the distribution values in a comoving frame.

The results for the whole sample, as well as fits for the sample split in 2 or 3 different redshift bins are tabulated in Table \ref{tab:FNX}. In Figure \ref{fig:fxn} we present the sample frequency and the distribution function of the combined sample in bins of 0.1 dex. We also compare our points with the results from the SDSS/BOSS by N12, and with the samples from the local Universe by B12 and Z05. Our bias-corrected values for the combined sample\footnote{We explain the bias correction method in detail in section \ref{sec:XQ100curves}} are in very good agreement with N12 results, although the size of our sample is not large enough to have many absorbers with column densities higher than \logNHI$\sim$21.7.  The combined sample is also in good agreement with B12 at the high column density end (\NHI\ $>$ 21), but the B12 sample has a paucity of low column density systems compared to the combined sample and N12.  However, the B12 sample is limited to a small number of Local group galaxies and does not widely sample the low redshift universe.  At the low column density end of the distribution function, DLA surveys are in much better agreement with Z05.

We have fitted a broken power law \citep[e.g.][]{2009ApJ...696.1543P} to our binned data using the Nelder-Mead algorithm:

\begin{equation}
\fnx = \left\lbrace
  \begin{array}{ll}
     k_{d} \left( \frac{N}{N_{d}} \right) ^{\alpha_{1d}} & N < N_{d} \\
     k_{d} \left( \frac{N}{N_{d}} \right) ^{\alpha_{2d}} & N \geq N_{d} \\
  \end{array}
  \right.
\label{eq:bpl}
\end{equation}
and a gamma function \citep[e.g.][]{2003MNRAS.346.1103P}
\begin{equation}
\fnx = k_{g} \left( \frac{N}{N_{g}} \right) ^{\alpha_{g}} e^{N/N_{g}}
\label{eq:gam}
\end{equation}
The coefficients from the distribution function fits are given in Table \ref{tab:IntOmega}.
Both expressions reproduce well our observed points, but when extrapolating to the highest column densities the broken power law describes much better the N12 and B12 distributions.  The gamma function appears to significantly under-predict the frequency of the highest column density absorbers at high redshift. We note that these results are also consistent with the exponent $\alpha_{2}\sim-3$ expected by \cite{1995ApJ...454..698W} for self-similar `disks'.

\subsection{Hydrogen mass density (\Ohi) curves}
\label{sec:XQ100curves}

The main objective of the current work is to assess the redshift evolution of the gas mass density \Ohi, which is defined as the first moment of the distribution function

\begin{equation}
\Ohi = \frac{H_{0}}{c} \frac{m_{H}}{\rho_{c}} \int_{N_{min}}^{N_{max}} N f_{HI}(N, X) dN
\label{eq:ohi}
\end{equation}
where $\rho_{c} = 3 H_{0}^{2} / 8 \pi G$ is the critical density for which the spatial geometry of the universe is flat. Usually, the discrete limit
\begin{equation}
\int_{N_{min}}^{N_{max}} N f_{HI}(N,z) dN = \frac{\sum_{j} N_{j}}{\sum_{i} \Delta X_{i}}
\label{eq:approx}
\end{equation}
where $i$ refers to the QSO and $j$ to the DLA, is taken to compute this quantity. Here, we are assuming that the sample represents properly the real population characteristics and/or the errors coming from this discretization are negligible compared with the sampling errors. We will check these assumptions in section \ref{sec:binning}. For the computation of \Odla, log $N_{min}$ is taken to be 20.3.

\subsubsection{Error estimations}

\begin{figure}
\includegraphics[width=0.45\textwidth]{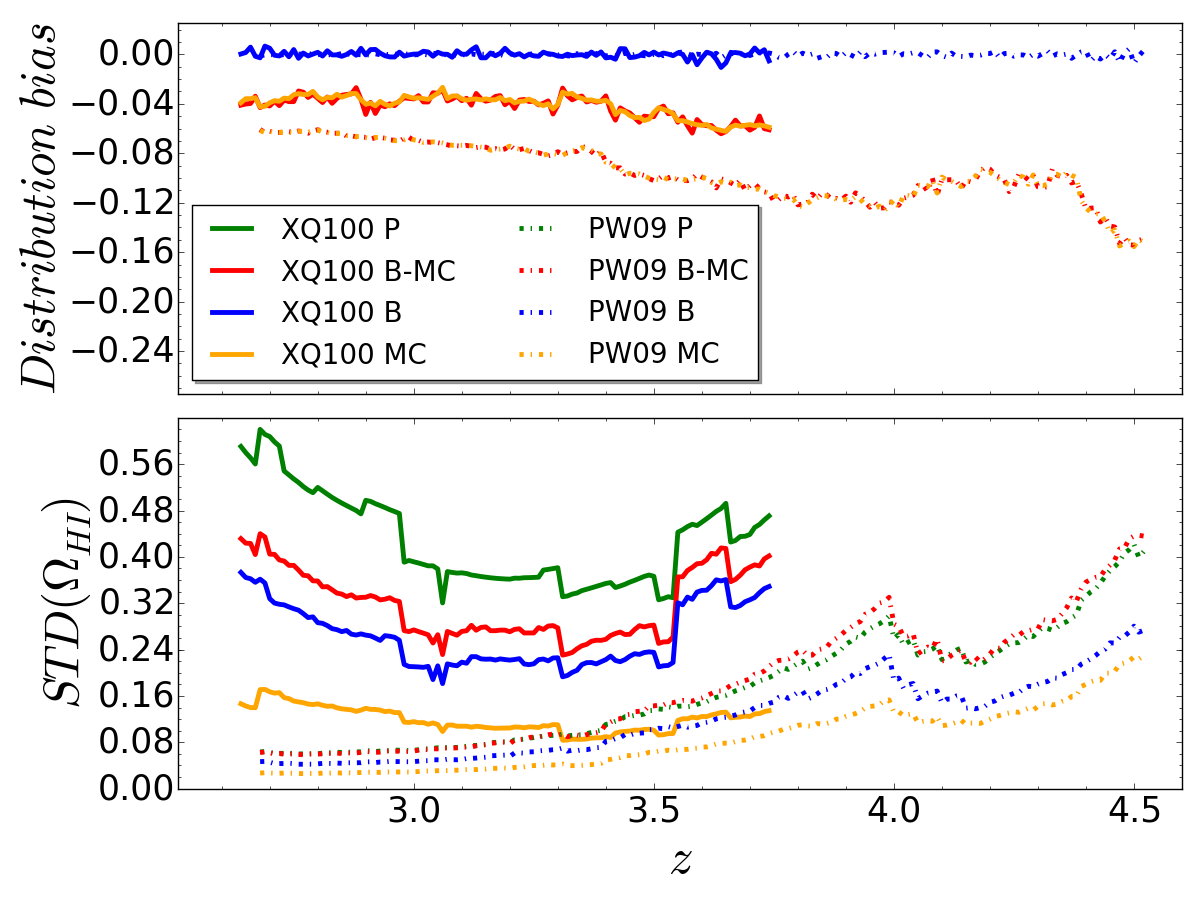}
\caption{Representation of the standard deviation (lower panel) and statistical distribution bias (upper panel) as a function of redshift for the XQ-100 and PW09 samples. The y-axis represents deviations in \Ohi$\times10^{3}$.}
\label{fig:errors}
\end{figure}

The error bars associated with \Odla\ may arise from several uncertainties.  Here, we consider two main sources of uncertainty: the error associated with the fitted \NHI\ of a given DLA (which includes uncertainties associated with the continuum fitting), and limited sampling of the complete column density distribution function.  The relative importance of these errors, and attempts to quantify them, have been varied in the literature. It is therefore instructive to compare various methods of error estimation, in order to quantify their relative magnitude.

In some early work, \citep[e.g.][]{1996MNRAS.283L..79S, 2001AA...379..393E} a common approach to determine the error on \Odla\ was to assume the sampling process followed a Poisson distribution (hereafter the `P method'). This means that we are assuming a probability function of the form $P(X)=e^{-\lambda}\frac{\lambda^{x}}{x!}$ for finding absorbers with column density \NHI\ within the total absorption path \DX. Under this assumption, the error on \Odla\ is propagated as
\begin{equation}
\Delta \Odla = \Odla\frac{\sqrt{\sum\limits_{i=1}^{p} N_{i}^{2}}}{\sum\limits_{i=1}^{p} N_{i}}.
\label{eqn:p}
\end{equation}

An alternative approach is to use a bootstrap re-sampling (hereafter `B method') of the observations in a given redshift bin \citep[e.g.][]{2015MNRAS.452..217C}. The bootstrap method consists of building up a large number of \Odla\ values ($N_{\rm re-samples}=10,000$ in our computations) that resamples, with replacement, the observed DLA sample in a given redshift bin. If the observed DLA `pool' is large enough, and is a good representation of the population properties, then each bootstrap realization is analogous to performing another independent DLA survey with the same total absorption path and redshift interval. Then, this large set of simulated observations provides an easy way of computing the mean \Odla, its standard deviation and the distribution bias (the difference between the computed \Odla\ from the originally measured DLA distribution and the resampled \Odla\ distribution mean).  Although they provide a useful estimator of the uncertainties associated with survey size and the sampling of the column density distribution function, neither the P method nor the B method account for the fit uncertainties on an individual DLA.

In order to assess the uncertainty due to errors on the \NHI\ fitting, a Monte Carlo approach may be adopted (hereafter, the `MC method'). In this approach, the \Odla\ for the original sample of DLAs is recomputed 10,000 times, but the value of \NHI\ is perturbed on each iteration, by drawing values from within the Gaussian error distribution defined by the 1$\sigma$ errors of the fit. We note the possibility of a `boundary bias' in the application of the MC method. This bias can occur because DLAs that are barely above the limiting \NHI\ threshold of 20.3 may be lost from the re-sampled distribution after they are perturbed within their error distribution. In order that this sample migration occurs equally in the other direction (absorbers just below the log \NHI\ = 20.3 being boosted into the re-sampled distribution) we include in our error analysis tests all absorbers that have been identified in our sample, down to a limiting column density of log \NHI\ = 19.5.

 It is also possible to combine the uncertainties associated with population sampling (epitomized with the B method) and the fit uncertainties (MC method).  In the `B-MC method', the bootstrap re-sampling of the original population additionally includes the perturbation of each \NHI\ within its Gaussian error distribution.

In Figure \ref{fig:errors} we compare the standard deviations (lower panel) and bias distributions ( i.e. the difference between the actual \Odla\ computed for the original observed sample and the value derived from sampling either the population or its errors, top panel) of the various error estimation techniques.  Note that the P method does not entail re-sampling, so there is no measure of bias for this technique.  The error curves are computed using a fixed $\Delta X$=2.5 (we will discuss in the following subsection different approaches to binning). Due to the different characteristics of the surveys that comprise our combined sample, such as sample size and spectral resolution, it is useful to compare the errors from the different estimation techniques for some of the separate sub-samples. In Fig. \ref{fig:errors} we therefore show the error estimate comparisons for PW09 and XQ-100. The former is the largest component of our combined sample and will hence best sample the underlying \NHI\ distribution. However, this sample has a relatively relaxed SNR threshold (SNR $>$ 4) and was constructed from moderate resolution spectra. Conversely, the XQ-100 sample is of modest size, but has both relatively high spectral resolution and
SNR. Moreover, the wide wavelength coverage of X-shooter permits the simultaneous fitting of higher order Lyman lines. These factors are all advantageous for the reduction of \NHI\ fitting uncertainties.

Fig. \ref{fig:errors} reveals several interesting features of the different error estimators.  The P method generally exhibits the largest standard deviations of the four techniques investigated, particularly for smaller samples. Moreover, without any re-sampling, it is difficult to explicitly test the effect of sample size, and the P method does not account at all for errors in \NHI. The technique that best captures the uncertainties in the \NHI\ fit is the MC method. From the lower panel of Fig. \ref{fig:errors} it can be seen that the MC errors of PW09 are smaller than XQ-100, even though the fit uncertainties are generally smaller in the latter sample. The much greater size of the PW09 sample can apparently largely compensate for the slightly poorer fit accuracy. Comparing a subset of the PW09 survey with a size matched to XQ-100 would lead to larger uncertainties in the former. An important and non negligible effect that can be seen in the upper panel of Fig. \ref{fig:errors} is that any variant of the MC technique (i.e. either with or without bootstrapping) always yields a negative bias, i.e. the simulated \Odla\ is usually greater than the original. This is due to the error distribution which is asymmetric in linear space.

We now consider the contribution of population sampling to the error estimate, as encapsulated by the bootstrap B method. For an infinitely large survey that fully samples the underlying population, the error due to incomplete sampling becomes negligible and the error is dominated by the uncertainties in the individual \NHI\ measurements. In smaller surveys, the sampling error becomes dominant. We can see this effect by comparing the large PW09 sample to the more modest sized XQ-100 sample in the lower panel of Fig. \ref{fig:errors}. Here, we see that the error from the sampling alone (B method) is approximately twice that of the fitting errors alone (MC method) for the XQ-100 sample. However, the fitting errors approach the sampling errors for the much larger PW09 sample. In general, the larger the sample, the more important is the relative contribution of the fitting uncertainties, compared to population sampling alone. Therefore, where XQ-100 is dominated by the sampling uncertainties, sampling and fitting uncertainties contribute at a similar level in PW09. In contrast to the MC techniques, the top panel of Fig. \ref{fig:errors} shows that the B method does not lead to a bias in the re-sampled distributions, as it considers perfect determinations of the column densities.

A combination of sampling and fitting errors seems to be required by our data, so we conclude that the best resampling technique to compute statistics is the B-MC method. However, since we have shown that the B-MC resampled distribution is biased (and could have non-negligible skewness), the standard deviation alone is an inadequate representation of the uncertainties.  We have therefore computed the 68 and 95 per cent confidence intervals, using the so-called bootstrap bias corrected accelerated method \citep[BCa, ][]{Efron:1987jw} on the B-MC resampled \Odla. In this non-parametric approach, the results are less affected by skewed distributions, and with this technique the bias mentioned before is also taken into account.

\subsubsection{Binning techniques}
\label{sec:binning}

\begin{table*}
\caption{Combined sample \Odla\ and \lx\ curves (\DX=2.5). The full table is available in the online version.}
\begin{tabular}{@{}llllllllllllll}
\hline
\z & \Dz & $\sum$\DX & n$_{abs}$ & \Odla\ ($\times 10^3$) & \multicolumn{2}{c}{{68\% \Odla\ C.I.}} & \multicolumn{2}{c}{{95\% \Odla\ C.I.}} & \lx & \multicolumn{2}{c}{{68\% \lx\ C.I.}} & \multicolumn{2}{c}{{95\% \lx\ C.I.}} \\
\hline
2.150 & 0.831 & 965 & 62 & 0.064 & 0.056 & 0.072 & 0.049 & 0.081 & 0.543 & 0.506 & 0.753 & 0.432 & 0.983 \\
2.160 & 0.831 & 997 & 64 & 0.064 & 0.056 & 0.072 & 0.049 & 0.081 & 0.530 & 0.494 & 0.729 & 0.422 & 0.948 \\
2.170 & 0.830 & 1023 & 66 & 0.064 & 0.057 & 0.072 & 0.050 & 0.081 & 0.527 & 0.489 & 0.727 & 0.422 & 0.940 \\
2.180 & 0.827 & 1094 & 68 & 0.062 & 0.055 & 0.070 & 0.048 & 0.078 & 0.532 & 0.495 & 0.721 & 0.425 & 0.917 \\
2.190 & 0.826 & 1140 & 68 & 0.060 & 0.052 & 0.067 & 0.046 & 0.075 & 0.511 & 0.476 & 0.695 & 0.407 & 0.885 \\
2.200 & 0.824 & 1219 & 72 & 0.059 & 0.052 & 0.066 & 0.046 & 0.073 & 0.507 & 0.479 & 0.689 & 0.413 & 0.857 \\
2.210 & 0.823 & 1271 & 78 & 0.061 & 0.054 & 0.068 & 0.048 & 0.076 & 0.504 & 0.477 & 0.682 & 0.414 & 0.848 \\
2.220 & 0.821 & 1317 & 81 & 0.061 & 0.055 & 0.068 & 0.049 & 0.076 & 0.504 & 0.481 & 0.682 & 0.418 & 0.849 \\
... & ... & ... & ... & ... & ... & ...  & ... & ... & ... & ... & ... & ... & ... \\
\hline
\label{tab:cs_curve}
\end{tabular}
\end{table*}

\begin{figure*}
\includegraphics[width=0.45\textwidth]{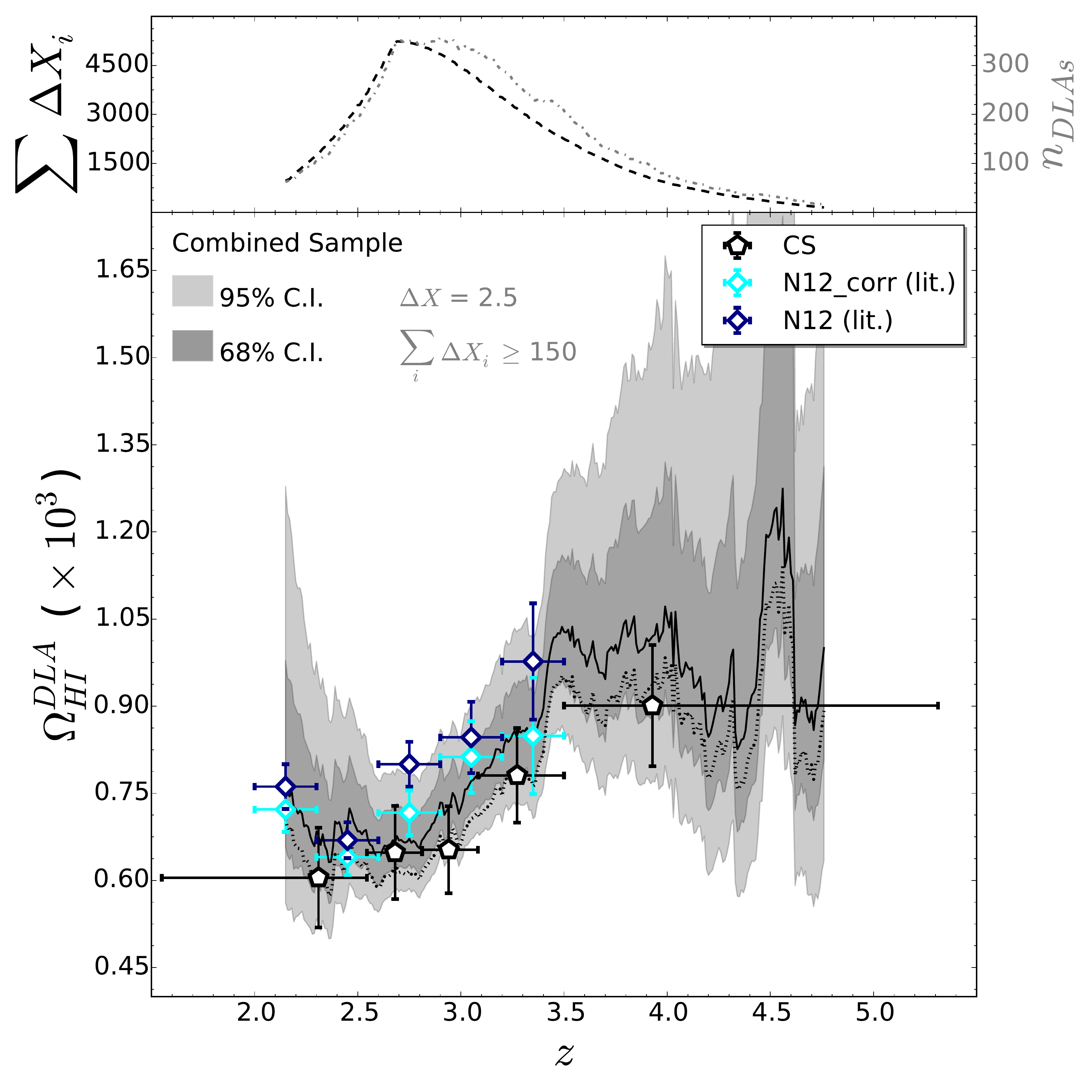}
\includegraphics[width=0.45\textwidth]{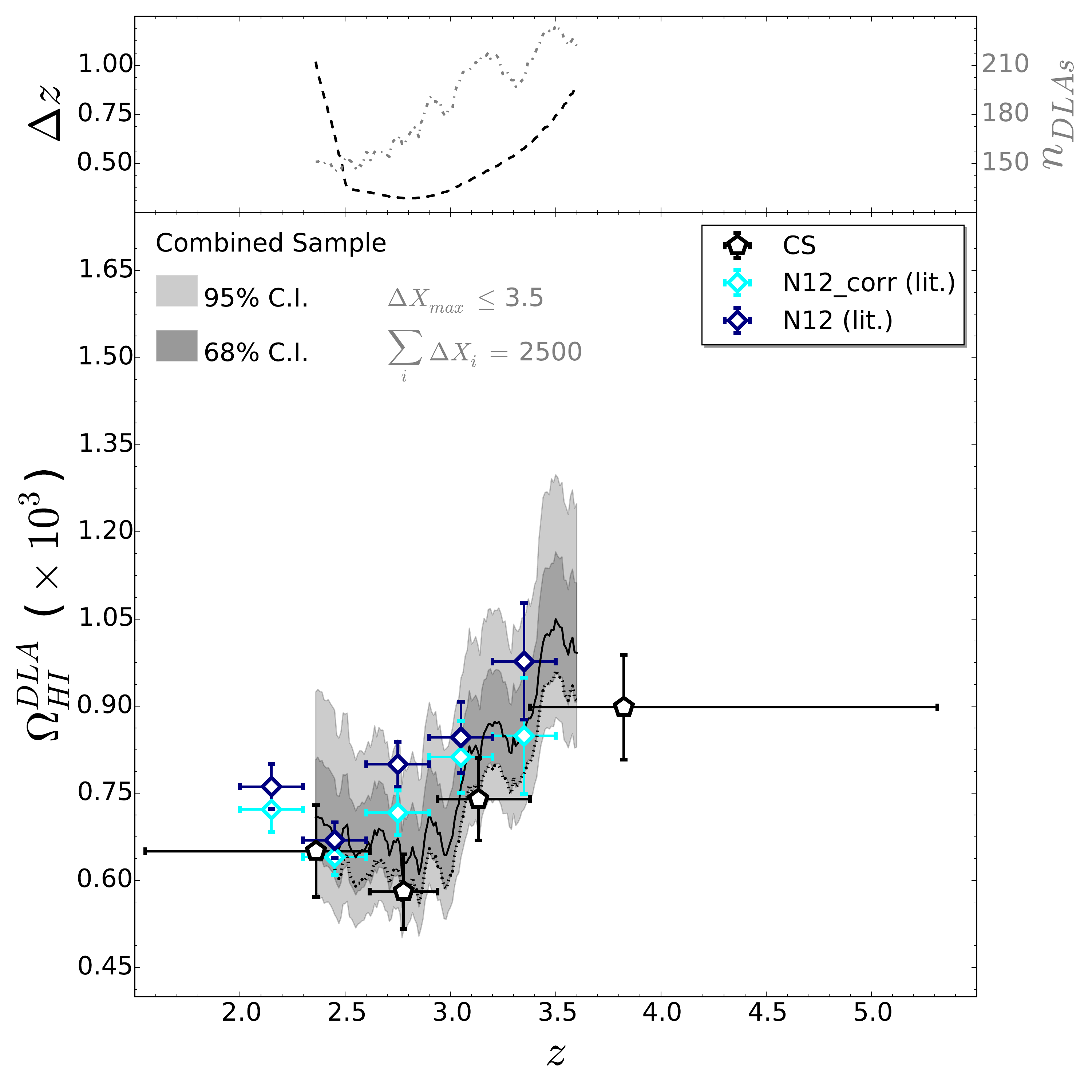}
\caption{\Odla curves for the combined sample using fixed comoving redshift intervals (left) and fixed total absorption path (right). 68 and 95 per cent confidence intervals derived from the BCa technique are shown in the shaded regions. Black line is the uncorrected curve. For comparison, we show the high column density correction applied to N12 results in order to have in both samples a maximum column density contribution of \logNHI=21.75, and the conventional representation of discrete bias-uncorrected \Odla\ points for the combined sample. Upper subpanel for each plot represents the redshift range or the total absorption path probed to build up the curve (black) and the number of DLAs used for the computations (grey).}
\label{fig:Otechs}
\end{figure*}

In order to study the redshift evolution of \Odla, there are two common procedures used in the literature to bin the observed data as a function of redshift. The first, and most common, is to select contiguous, and often equally spaced, \z-intervals based on the characteristics of the sample \citep[e.g., ][]{2009ApJ...696.1543P, 2012AA...547L...1N}. This aproach has the advantage of being intuitive and computationally straight-forward.  However, if the redshift path is very uneven, it can lead to large variations in the statistical uncertainties in \Ohi\ as a function of redshift.  The alternative method is to set the redshift intervals of each \Ohi\ bin such that it uniformly samples the total absorption path \citep[e.g., ][]{2009AA...505.1087N, 2009AA...508..133G}. The redshift bin sizes may therefore be heterogeneous, but each bin is statistically similar. However, both of these methods adopt contiguous, non-overlapping bins to discretize the data.

In this work, we take a different approach. Instead of computing \Odla\ in contiguous, non-overlapping intervals, we sample the redshift distribution finely using a sliding redshift 
window to compute an \Odla(\z) `curve'.  The computation of \Odla\ is still based on the summation of \NHI\ between a given \zmin\ and \zmax. However, we consider all \zmin\ and \zmax\ pairs over the full redshift range of our sample, beginning at the minimum redshift where \gz$>$0, and incrementing \zmin\ by 0.001. The \zmax\ at each point in the \Odla curve can be computed analogously to the two traditional methods described above: In the first technique, we simply define the \zmax\ of each \Odla\ curve point to be \zmax=\zmin+\Dz, where \Dz\ is determined by imposing a constant \DX\ (in the example below, we use \DX\ = 2.5) and using eq. \ref{eq:dx}. The resulting mean redshift of the interval is weighted by the \gz\ curve.  As a result of the weighting procedure, several (\zmin, \zmax) pairs can lead to a computation of \Odla\ at the same mean \z; when this occurs, we take the median \Odla\ of the duplications. Although simple in its approach, the uneven distribution of redshift coverage in our sample leads to some redshift intervals being better sampled than others. The second approach mitigates this effect by setting the \zmax\ dynamically for each \zmin, so that each (\z, \Odla) point has the same total absorption path contributing to it (in the example below, we use $\Sigma$ \DX\ = 2500). In this approach, we compute the \z\ for each redshift interval by again weighting for the \gz\ curve in that interval. As with the first method, when multiple (\zmin, \zmax) pairs lead more than one bin to have the same mean redshift, we take the median \Odla\ of the duplications.

In Figure \ref{fig:Otechs} we compare both binning methods for computing \Odla\ for the final combined sample, where the grey shaded regions show the 68 and 95 per cent confidence intervals determined from the BCa method described in the previous subsection.  The \Odla\ curves are shown in the main (lower) panels and the upper panels show absorption paths/intervals (black dashed line) and number of DLAs (grey dotted line) as a function of $z$. The first conclusion we can draw from a comparison of the curves in  Figure \ref{fig:Otechs} is that both binning methods yield curves of \Odla\ that are statistically equivalent in the range 2.2 $< z<$ 3.6 where they can both be computed.  In both cases, there is a scatter whose magnitude is inversely proportional to the number of absorbers used to compute statistics. This scatter originates from the limited sampling that draws different absorber distributions for each computed point. This effect is especially evident when high column density absorbers are included or not in a given bin. Therefore, although using a fixed total absorption path (right panel) theoretically smooths any uneven sampling, the contribution of the SDSS dominates in the redshift range where both techniques can be applied and the statistics are relatively uniform anyway.  One of the most notable differences between the binning methods shown in  Figure \ref{fig:Otechs} is that a fixed total absorption path (right panel) greatly restricts the redshift range over which the curve can be computed.  This is due to `running out' of redshift path as the curve is built up towards higher $z$.   This truncation can be mitigated by reducing the choice of $\Sigma$ \DX, although in turn this is a compromise in the uncertainty.  Using a fixed \DX\ interval is therefore our preferred binning method.

Having constructed the \Odla\ curves, that we tabulate for the CS in Table \ref{tab:cs_curve} for ease of reproduction, it is interesting to compare the results to the more classical binning approach.  For the combined sample, the binned-uncorrected values are shown with black symbols in Fig.  \ref{fig:Otechs}, as well as the bias-uncorrected curve plotted in black. Different redshift bins select different DLAs, which cause the small variations between both uncorrected data points and curve. Larger effects are corrected when taking into account fitting uncertainties, as we can observe when comparing the uncorrected curve with BCa confidence intervals. However, when we plot together these intervals and the N12 points (dark blue), the latter are systematically above (although still consistent within the confidence intervals) the values for the CS.  

We investigate whether the values measured by N12 are higher due to a more complete sampling of the high column density end of the distribution function by those authors, i.e. whether the combined sample's incompleteness at high \NHI\ may cause an under-estimate of \Odla.  For this purpose, we also computed \Odla\ by integrating \fnx. In Table \ref{tab:IntOmega} we show the results of integrating up to \logNHI=21.7 (the upper limit of our DLA sample, denoted in the table by $^{max}$) and to \logNHI=$\infty$. We observe, as expected, a very good agreement between our summed \Odla\ (e.g. 0.97 $\times 10^{-3}$ for the redshift unbinned CS) and the partially integrated one (0.94 $\times10^{-3}$), being  inside the derived confidence intervals for all bins. However, the difference between the partial and fully integrated values (1.16$\times10^{-3}$ for the full combined sample) indicates that our values may still be under-estimated due to our limited sample size. Our confidence intervals are actually rigorous as long as the bootstrap hypothesis is fulfilled, but we remark the fact that this technique cannot infer the information on the real population that is not contained in the combined sample. This hypothesis could be true if the contribution of the absorbers with higher column densities than \logNHI$\sim$21.7 (in our case) is negligible. However, the extrapolated broken power law fit results, supported by N12 (and B12) observations, suggest that these absorbers play an important role  ($\approx 20\%$) in the value of the hydrogen gas mass density. If we impose the same maximum \logNHI\ = 21.7 threshold for the N12 sample as applies to the combined sample, integrating the \fnx\ for the SDSS/BOSS sample yields a slightly lower value of \Odla\ that is now nicely consistent with the combined sample (cyan points in Fig. \ref{fig:Otechs}), confirming the equivalency of both samples over their common redshift range.

\subsection{DLA incidence rate}

\begin{figure*}
\includegraphics[width=0.45\textwidth]{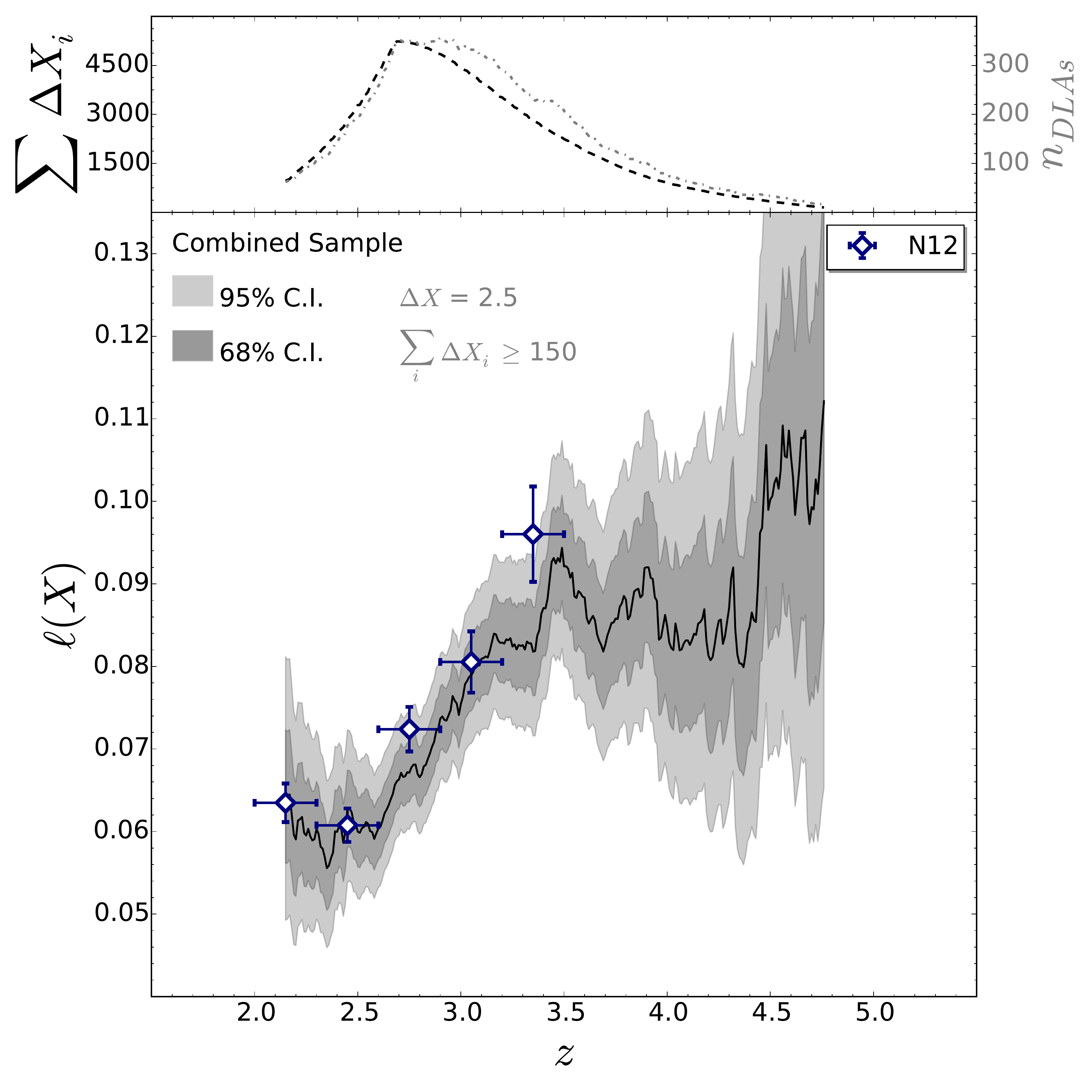}
\includegraphics[width=0.45\textwidth]{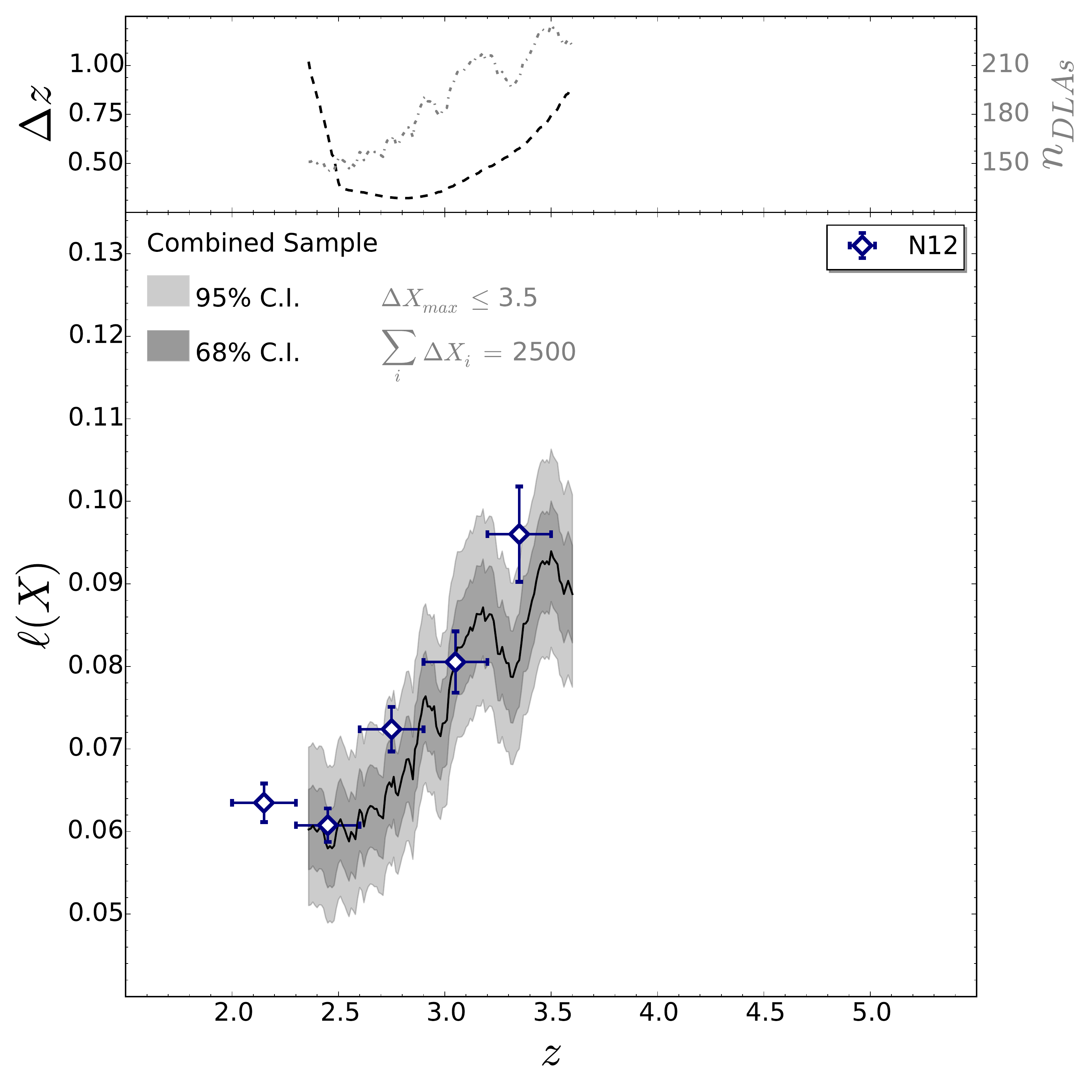}
\caption{\lx\ curves for the combined sample using fixed comoving redshift intervals (left) and fixed total absorption path (right). 68 and 95 per cent confidence intervals are derived assiming a Poisson distribution. Solid line represents the value computed directly from eq. \ref{eq:lx}}
\label{fig:lx}
\end{figure*}

The DLA incidence rate \lx, or line density, is defined as the zeroth moment of \fnx
\begin{equation}
\lx dX = \int_{N_{min}}^{N_{max}} f_{HI}(N,z) dN dX
\label{eq:incidence}
\end{equation}
Its discrete limit, commonly used to compute this quantity, is given by
\begin{equation}
\lx = \frac{m}{\sum_{i} \Delta X_{i}}
\label{eq:lx}
\end{equation}
i.e., it is the number of DLAs found per unit comoving redshift path.

In this analysis, for the error estimation we assumed a Poisson distribution. This is because the main variation we can expect using the B-MC technique is from the effect of the low \NHI\ absorbers crossing the DLA limit from one side to the other.  However, the major source of uncertainty in the column density distribution function is from the poor sampling of the highest column density absorbers. We show line density curves in Figure \ref{fig:lx}, built up using the same binning techniques as for \Odla. Here, we can observe again an excellent agreement with the N12 results: from $z \sim 2$ to 3.5 where the samples overlap  a significant increase in the line density of DLAs is seen.  At higher redshifts, the limited statistics prevent us from distinguishing whether \lx\ continues to rise, or flattens, although a steep evolution is not supported by the current dataset \citep{2015MNRAS.452..217C}.

\section{Discussion}

\begin{figure*}
\includegraphics[width=0.45\textwidth]{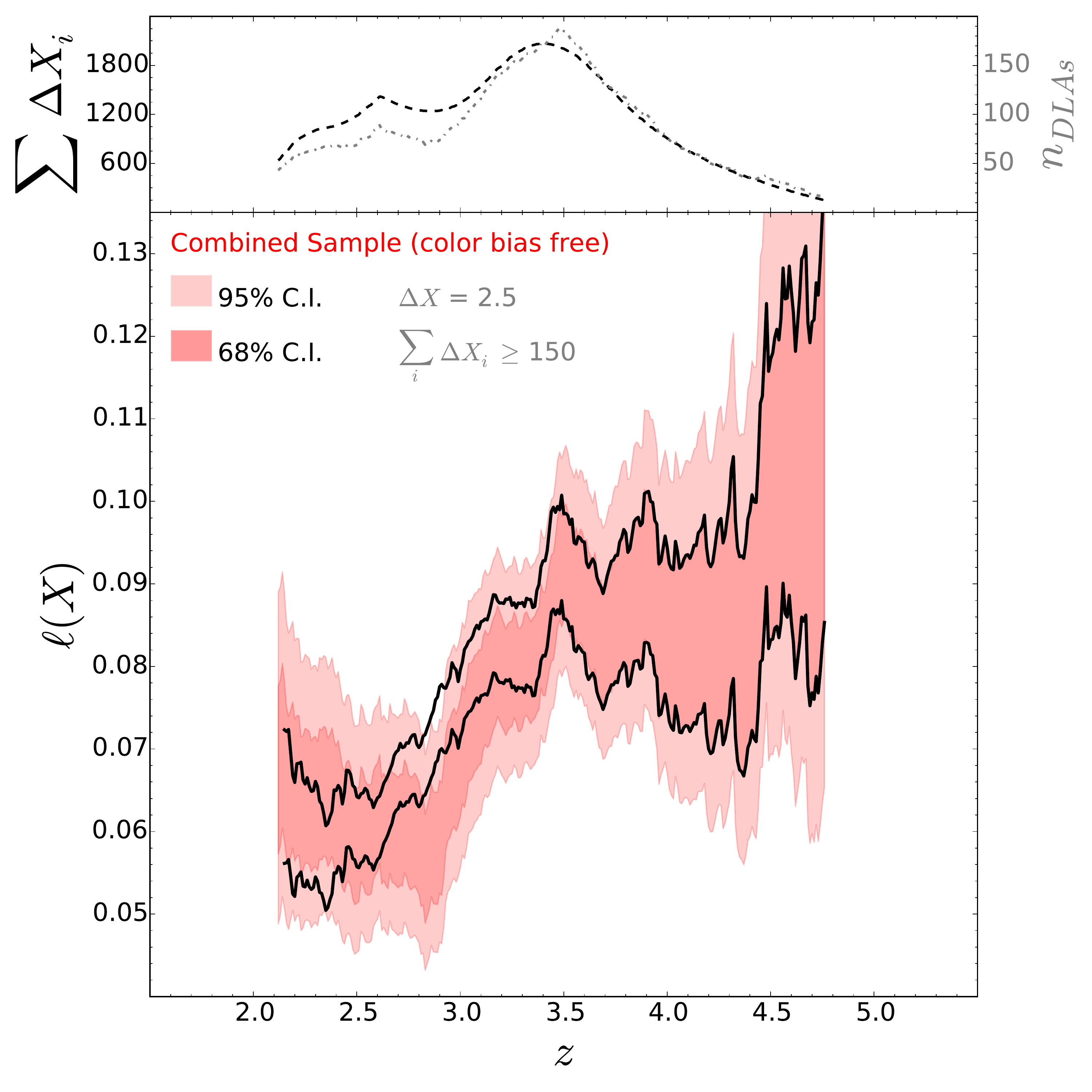}
\includegraphics[width=0.45\textwidth]{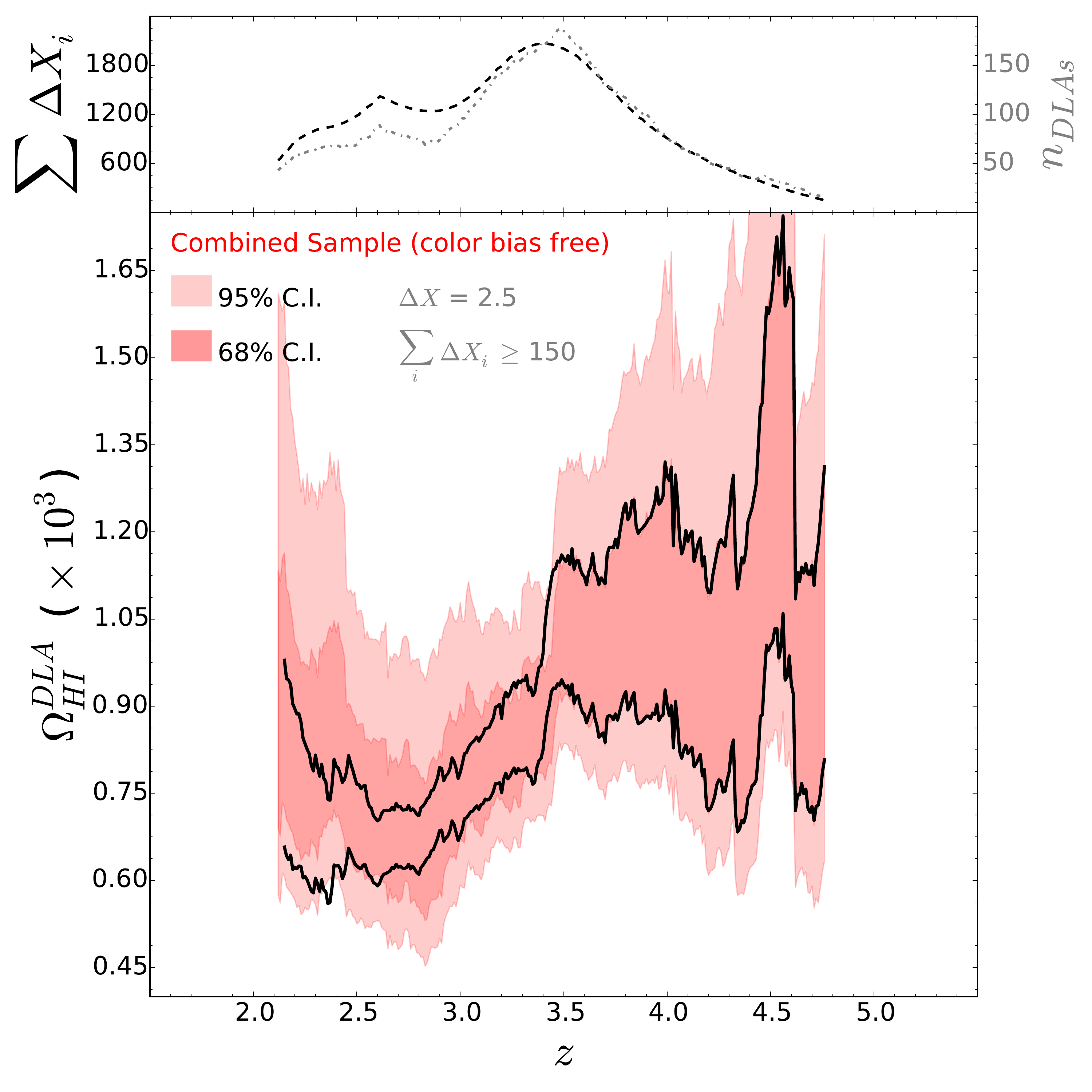}
\caption{\lx\ (left) and \Odla (right) curves for our color bias free sub-sample. The solid black lines represent the full CS 68\% confidence limits.}
\label{fig:cb}
\end{figure*}

We have presented a new survey for DLAs based on spectra obtained for the XQ-100 survey, an X-shooter Large Program to observe 100 $z>3.5$ QSOs (Lopez et al. in prep).  The 38 DLAs identified in the XQ-100 sample are combined with major literature samples compiled over the last $\sim$ 20 years, with close attention paid to duplications, yielding a total of 742 DLAs over a redshift range of approximately $1.6 < z< 5.0$.  The total redshift path of the combined sample is \DX\ = 10,434.  We have described a thorough assessment of error estimators, and the relative contributions of fitting and sampling errors appropriate for our combined sample.  A novel technique for binning the DLA statistics is presented, which yields continuous \Odla\ `curves' rather than discretized binned values.  A comparison with the limited redshift range covered by the \citet{2012AA...547L...1N} sample of DLAs in the SDSS/BOSS indicates that, despite the large size of our total sample, \Odla\ may still be under-estimated due to the absence of very high column density (\logNHI\ $>$ 21.7) DLAs. Looking at Table \ref{tab:IntOmega}, extrapolation of the whole combined sample indicates that the missing contribution of these absorbers is $\sim$20\% for our sample. However, this correction could be redshift-dependent. Statistics at large column densities are too poor when splitting the sample, and this could also lead to incorrect extrapolations, as inferred from Table \ref{tab:IntOmega}. Consequently, yet larger samples are needed in order to precisely constrain the impact of high column density DLAs on \Ohi\ evolution.

On the other hand, low column density DLAs are sufficiently well sampled to offer narrow \Odla\ confidence intervals. This means that relatively small effects in sample selection that have been previously neglected could now become important. One such example is the false positives/negatives issue, that might represent an important systematic uncertainty. Results from \citet{2015MNRAS.452..217C} and the excellent agreement within our CS and N12 distribution functions, that were obtained using different methods, suggest that still we are not able to distinguish this effect from the sampling uncertainties.

Another source of error could be due to the SDSS color-selection of QSOs \citep[see][]{2009ApJ...705L.113P, 2011ApJ...728...23W}. In order to investigate this effect, we built up a sub-sample of our CS composed only of `safe' quasars, i.e., we excluded 10 biased XQ-100 QSOs (flagged in Table \ref{tab:xq} with spades), SDSS quasars not flagged as selected by its FIRST counterpart with 2.7$<$\zem$<$3.6, and the P03 sample. Results are shown in Figure \ref{fig:cb}. We observe that, at this sample size (3152 QSOs and 402 DLAs), there is no strong evidence of the effects of this bias in \Odla. However, from the \lx\ plot we can see a small decrease in the incidence rate at \z$\sim$2.8, consistent with what we might expect, and suggesting that this color bias could be important in larger samples.

\begin{figure}
\includegraphics[width=0.49\textwidth]{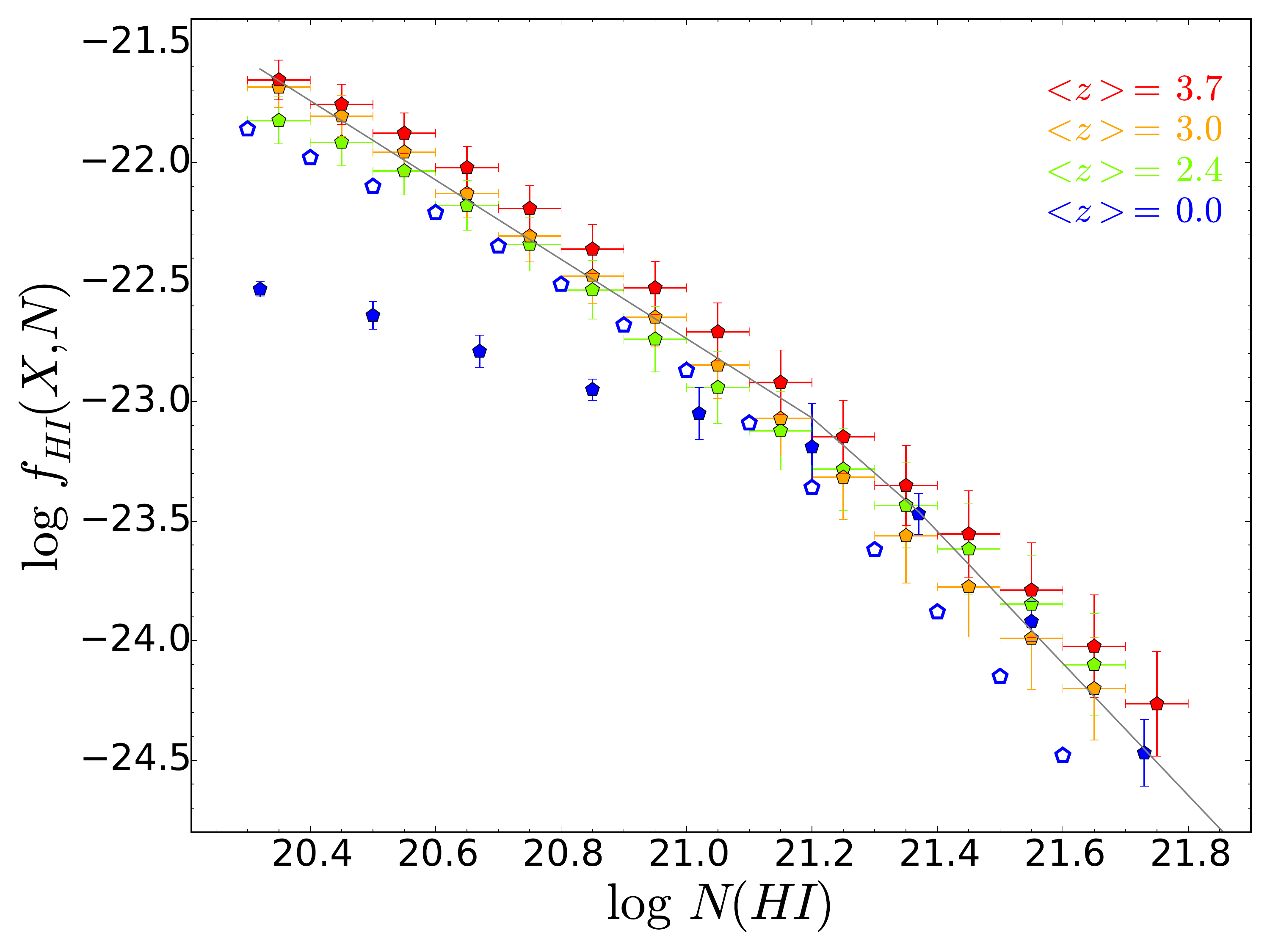}
\caption{Column density distribution function of the CS 3 redshift bins compared with N12 (green points), B12 (filled blue points) and Z05 (unfilled blue points). The black line is the broken power law (BPL) fit of the whole CS.}
\label{fig:Evolfxn}
\end{figure}

\begin{table*}
\centering
\caption{Double power law fitting parameters and first moment of the \fnx. All values of \Odla\ are in units of $10^{-3}$}
\begin{tabular}{ccccccccccccccccc}
\hline
\z & \zmin & \zmax & $n_{abs}$ & \DX & log $N_{d}$ & log $k_{d}$ & $\alpha_{1d}$ & $\alpha_{2d}$ & $\Odla^{max}_{d}$ & $\Odla^{\infty}_{d}$ & \Odla & \Odla 68\% C.I. \\
\hline
2.99 & 1.550 & 5.313 & 742 & 10434 & 21.27 & -23.18 & -1.66 & -2.77 & 0.94 & 1.16  & 0.97 & 0.93 - 1.03 \\
\hline
2.57 & 1.550 & 2.947 & 318 & 5219 & 21.27  & -23.26 & -1.67 & -2.41 & 0.60 & 1.02 & 0.65 & 0.60 - 0.71 \\
3.47 & 2.947 & 5.313 & 424 & 5219 & 21.27 & -23.12 & -1.65 & -2.74 & 0.83 & 1.05 & 0.85 & 0.80 - 0.92 \\
\hline
2.44 & 1.550 & 2.732 & 208 & 3482 & 21.27  & -23.26 & -1.63 & -2.16 & 0.62 & 1.65 & 0.67 & 0.61 - 0.74 \\
2.95 & 2.732 & 3.207 & 235 & 3482 & 21.27  & -23.26 & -1.77 & -2.64 & 0.63 & 0.86  & 0.64 & 0.58 - 0.70 \\
3.69 & 3.207 & 5.313 & 300 & 3482 & 21.27 & -23.08 & -1.62 & -2.53 & 0.93 & 1.34 & 0.95 & 0.88 - 1.05 \\
\hline
\label{tab:IntOmega}
\end{tabular}
\end{table*}

In Fig. \ref{fig:Evolfxn} we investigate whether the column density distribution function varies with redshift by comparing our \fnx\ results split in 3 redshift bins with Z05 and B12 representing the local universe.   Although the error bars for the 3 redshift bins of the CS overlap, there is a systematic trend for lower redshift intervals to have a lower normalization of \fnx\ for moderate and low column density absorbers (\logNHI\ $<$ 20.9).  At high \NHI\ there is more scatter amongst the points, due to poorer statistics, such that it is not possible to conclude if the evolution extends over the full column density range.

\begin{figure*}
\includegraphics[width=0.9\textwidth]{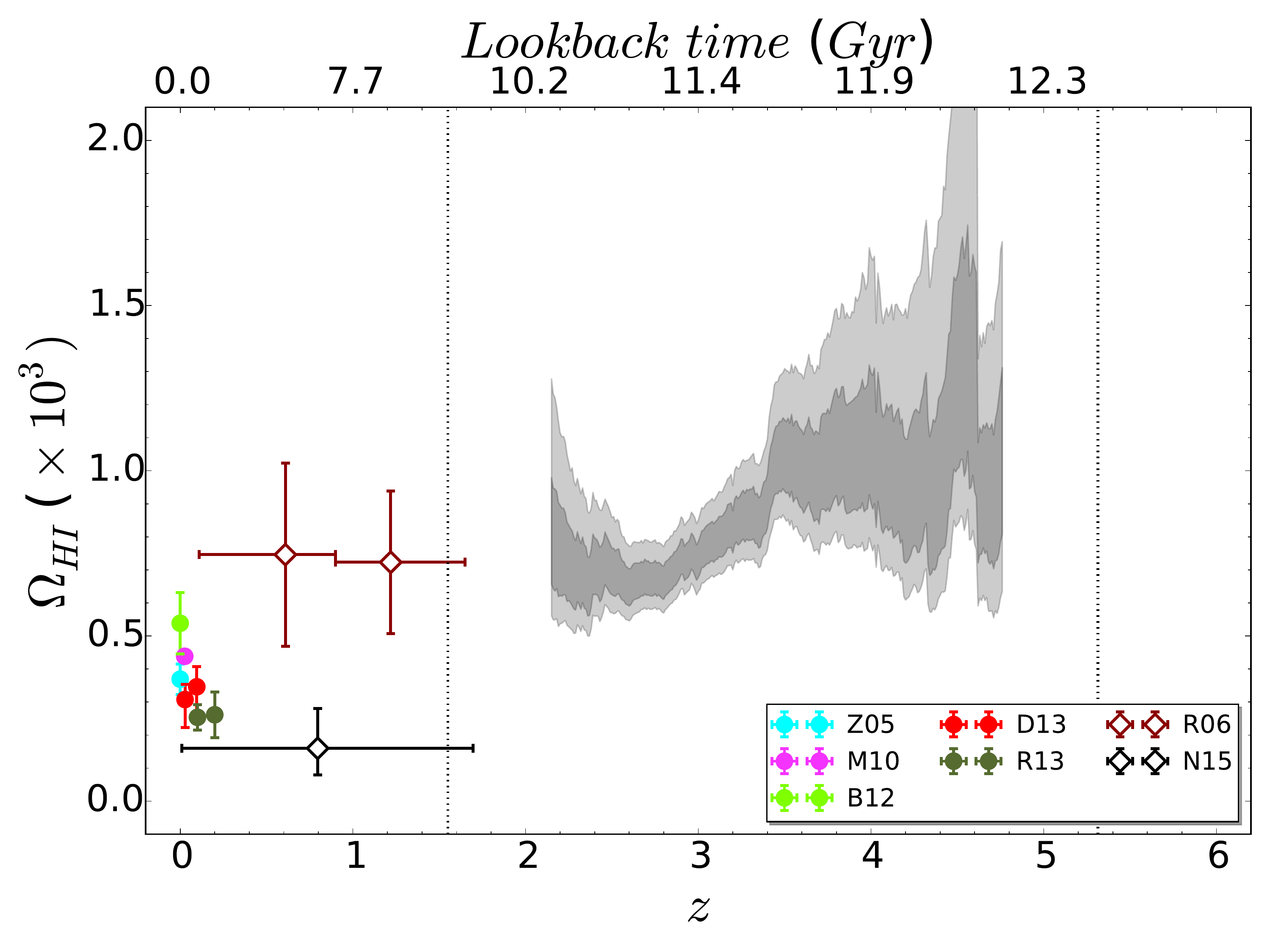} 
\caption{Evolution of the atomic HI gas in galaxies over the cosmic time. Data from 21cm emission line surveys (circles) and from QSO absorption systems (squares) are plotted. Empty points represents \Odla\ and filled \Ohi. Vertical lines are the redshift limits of our combined sample. The data for the CS curve is tabulated in Table \ref{tab:cs_curve}.}
\label{fig:evolution}
\end{figure*}

In Fig. \ref{fig:evolution} we plot the \Odla\ curve we derived down to $z \sim 2$, together with DLA surveys conducted at lower redshift \citep[see][and references therein]{2006ApJ...636..610R, 2015ApJN} and \Ohi\ from 21~cm emission surveys (the different selection techniques, sparse sampling at intermediate z, and contrast in measurement techniques mean that it is not appropriate to combine these measurements into our computation of the \Odla\ curve).

\begin{figure}
\includegraphics[width=0.45\textwidth]{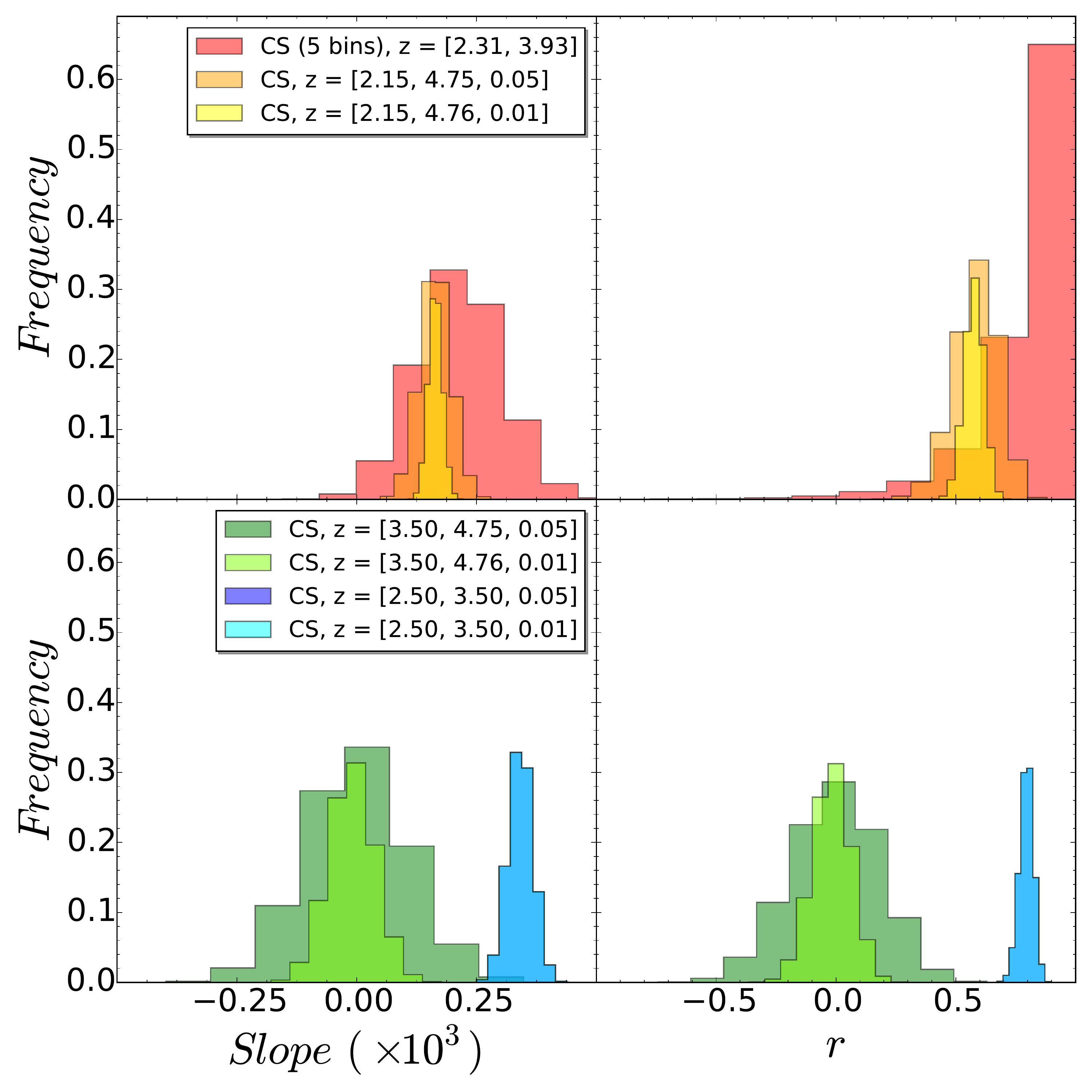}
\caption{Histograms of the results of the evolution test for different redshift ranges of the CS split in 5 non overlapping bins (top panels), and the CS curve (bottom panels). Redshifts intervals in the legend have the format [\zmin, \zmax] (bins) or [\zmin, \zmax, \Dz] (curves). The left hand panels show the distribution of slopes from a linear regression of 100,000 re-sampled $\Omega$ curves; the right hand panels show the distributions of the Pearson correlation coefficients, $r$.  Different colored histograms indicate different samples and redshift ranges, as given in the Figure legends.}
\label{fig:CSEvoltest}
\end{figure}

\begin{figure}
\includegraphics[width=0.45\textwidth]{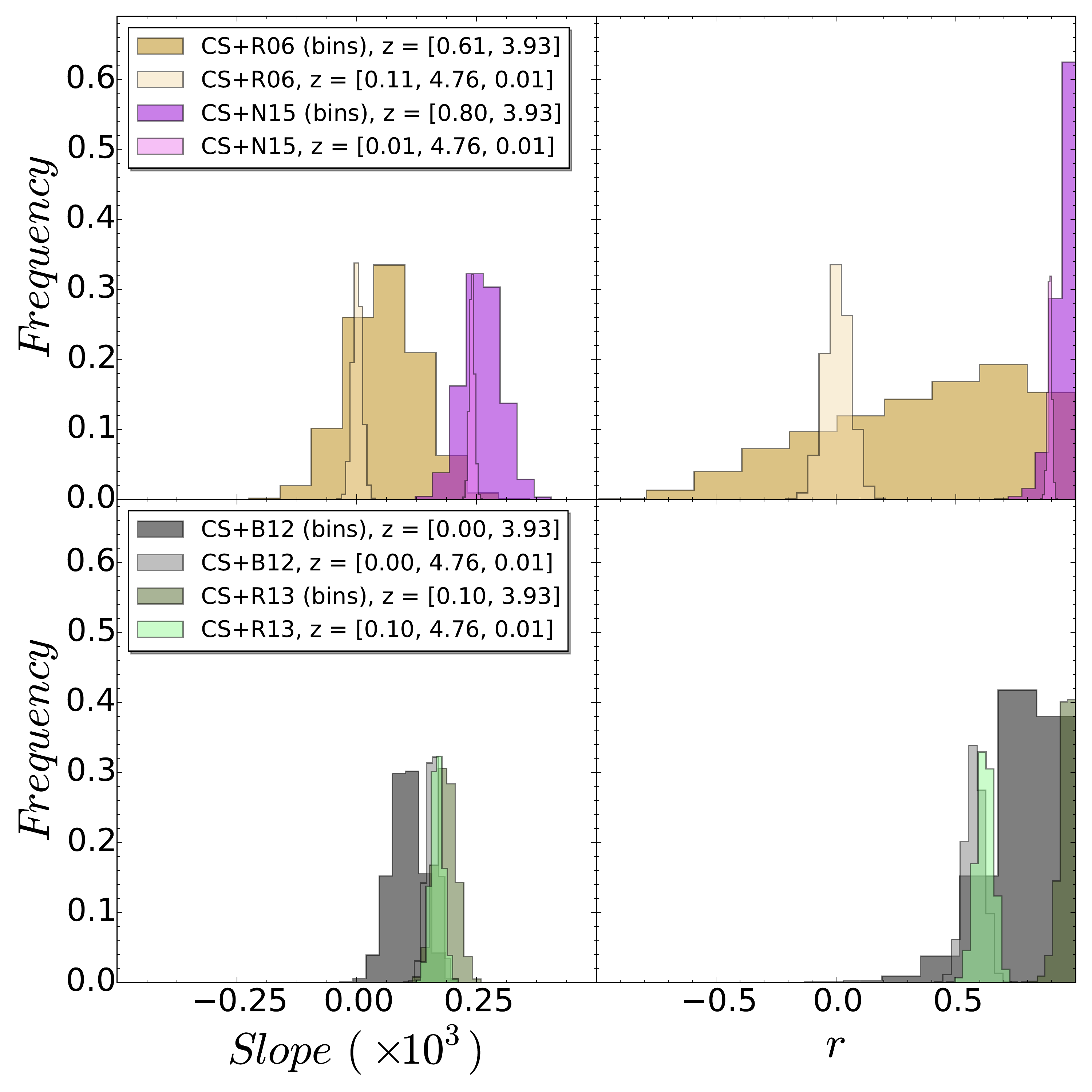}
\caption{Version of Figure \ref{fig:CSEvoltest} with histograms of the results of the evolution test for the CS plus intermediate redshift samples (top panels), and local 21cm samples (bottom panels).}
\label{fig:Evoltest}
\end{figure}

A critical debate in the literature, that has been ongoing since the first measurements of \Odla, is whether or not this quantity evolves with redshift. In order to \textit{statistically} investigate whether the data favors (or not) an evolution of the \HI\ content in the universe over cosmic time, we performed the following test: For each non overlapping redshift point of the CS split in 5 bins with the same \DX, we randomly selected one value of \Odla\ from its probability distribution. This was repeated using the \Odla\ curve, in order to compare the impact of using either our new `curve' methodology with traditional binning.  In order to include intermediate redshift DLA surveys or local 21cm emission surveys in our re-samples, we draw \Odla\ (or \Ohi) points assuming a Gaussian distribution within the quoted 1 $\sigma$ error bars of those works. For surveys with extended redshift coverage (i.e. R06 and N15) we also re-sampled evenly across the quoted range in $z$, using the same $\Delta z$ sampling as for the CS curve.  After each re-sampling of the \Odla\ curve, we performed a linear regression and computed the slope and Pearson's correlation coefficient, $r$, which tests for the significance of a correlation between two quantities (in this case, $z$ and \Odla). We tabulate in Table \ref{tab:Evol} the distribution of the slopes and $r$ for 100,000 iterations. Figure \ref{fig:CSEvoltest} considers only the CS, whereas Figure \ref{fig:Evoltest} include either intermediate or low redshift data.

Considering first only the CS, we observe a significant correlation for the whole sample, as shown by the red-to-yellow histograms, spanning approximately $2 < z< 5$. In Table \ref{tab:Evol} we can observe that the significance of the results depends on the redshift sampling of the \Odla\ results, since the larger the $z$ range, the greater the degrees of freedom we have. We therefore also establish the significance of this correlation, finding $\geq 3 \sigma$ to be independent of binning/sampling method. We note that in the \z-range from 2.5 to 3.5, where our statistics are best, the slopes and $r$ distributions are independent of the $\Delta z$ chosen.  From the curve analysis, the correlation is driven mostly by the data in the \z-range from 2.5 to 3.5, where the median correlation coefficient is $r=0.8$ (blue and cyan histograms). Taken together, the positive slopes and $r$ values indicate significant redshift evolution in \Odla\ within these redshift ranges. However, when considering only redshifts greater than 3.5 (orange histograms), the data do not show a significant correlation, although the statistics in this high redshift regime are considerably poorer.  The large confidence intervals associated with the \Odla\ curve at $z>3.5$ could be masking some mild evolution.

In upper panels of Figure \ref{fig:Evoltest} we now include R06 ($0.11 <z< 1.6$, purple histograms) and N15  ($0.01 <z< 1.7$, green histograms) data. If we use the R06 statistics to represent intermediate redshift DLAs, in combination with the CS, the correlation coefficient (median $r=0.0$) and slope (median value = 0.0) favors no evolution across the entire redshift range from $0.1 < z < 5$. Conversely, if the low value of \Odla\ reported by N15 is adopted, the evolution is very significant ($\geq 6.5 \sigma$, with median $r=0.9$ and slope = 0.24$\times 10^{-3}$).  The N15 and R06 samples both have important limitations. On the one hand, the results from R06 are based on the incidence of \MgII, a pre-selection technique which may bias the high column density end of \fnx\ towards higher values \citep[see][and references therein]{2015ApJN}. On the other hand, the N15 statistics are too poor to fully sample the column density distribution function and they had to correct their measurements assuming N12's \fnx\ to include contribution by missing DLAs with \logNHI$>$21.0. 

Given the relatively small sample sizes of DLAs at intermediate redshifts, and the potential biases and uncertainties described above, we next test for evoution by combining the CS with $z \sim 0$ measurements of \Ohi.   However, even for the local universe, \HI\ measurements are not absent of concerns on possible systematics, and  Fig. \ref{fig:evolution} shows a factor of two disagreement between surveys.  For example, \citet{2012ApJ...749...87B} claims that current low-\z\ measurements are strongly biased due to resolution effects, as the densest \HI\ clouds are much smaller than the typical spatial resolution and no self-absorption correction, which is important in this case, can be computed from the data-cubes. This effect was pointed out previously by \citet{2005MNRAS.364.1467Z}, but with their data-set they were not able to find significant deviations. However, \citet{2012ApJ...749...87B} builds up the local distribution function using high-resolution images ($\sim 100 pc$) of only three galaxies, assuming that the cloud distribution is representative of the whole local Universe, which may not be accurate.  We therefore repeat the evolution test, combining the CS with either R13 or B12, which show the lowest and and highest measurements of \Ohi\ respectively in the local universe.  In the lower panels of Fig.  \ref{fig:Evoltest} we show the distribution of slopes (left panel) and correlation coefficients (right panel) for the 100,000 resamplings when the CS is combined with either R13 (olive green histogram) or B12 (grey histogram).  Although the \Ohi\ values reported by these two works differ by a factor of two, when combined with the CS both exhibit a statistically significant correlation between \Ohi\ and $z$; in both cases the median correlation coefficient is $r=0.6$ with a significance $\geq 3 \sigma$.  The median slopes are also very similar, $\sim 0.17 \times 10^{-3}$, which correspond to a factor of $\sim$ 4 decrease in \Ohi\ between $z=5$ and $z=0$.

Despite decades of effort in compiling ever larger samples, our assessment of the galactic gas reservoir in DLAs is still limited by the missing high density absorbers and the traditional \NHI\ definition at \logNHI\ = 20.3.  It has been argued \citep[e.g.][]{2005MNRAS.363..479P} that, particularly at high redshifts, sub-DLAs (\logNHI\ $>$ 19.0) could contribute significantly to the atomic hydrogen budget. \citet{2015MNRAS.452..217C} made a statistical 20 per cent correction for sub-DLAs in their study of $z>3.5$ DLAs, based on results from \citet{2007ApJ...656..666O, 2009AA...505.1087N, 2010ApJ...718..392P, 2013AA...556A.141Z}.  The suggested redshift dependence of a sub-DLA correction means that a uniform correction to account for lower column density absorbers is not appropriate for our combined sample.  Moreover, the majority of DLA surveys that comprise our combined sample were conducted at insufficient resolution to robustly identify sub-DLAs, such that a `manual' assessment of the sub-DLA contribution to each is not possible.  Although a redshift dependent sub-DLA correction could produce some changes in the \Ohi\ curve shape, existing results indicate that this factor is unlikely to dramatically change the picture that we presented in this work.  In a future paper we will identify and investigate the nature of the sub-DLAs in the XQ-100 sample.

In closing, we note that the quality of the XQ-100 spectra are sufficient, both in terms of SNR and resolution, to directly determine elemental abundances for the DLAs presented here.  There are numerous DLAs of interest amongst the sample, including a candidate very metal poor DLA, several PDLAs, and some cases of multiple DLAs that lie very close in velocity space along a single line of sight.  All of these categories have been proposed to be chemically interesting \citep[e.g.][]{2001AA...380..117E, 2003AA...403..573L, 2010MNRAS.406.1435E, 2011MNRAS.417.1534C, 2011MNRAS.412..448E}.  In Berg et al. (in prep) we study the chemical abundances of all the XQ-100 DLAs, with a particular focus on these special cases, in the context of a large literature sample.

\section{Conclusions}

\begin{table*}
\centering
\caption{Results of the evolution tests for 3 cases: non overlapping curves, \Odla\ curve with $\Delta z$=0.01 and its resampled version with $\Delta z$=0.05. Slopes are in units of $10^{-3}$}
\begin{tabular}{cccccccccc}
\hline
 & \multicolumn{3}{c}{Non overlapping points} & \multicolumn{3}{c}{$\Delta z=0.05$} & \multicolumn{3}{c}{$\Delta z=0.01$} \\
 & \multicolumn{3}{c}{--------------------------------------} & \multicolumn{3}{c}{--------------------------------------} & \multicolumn{3}{c}{--------------------------------------} \\
Sample & Slope & $r$ & $\sigma$ & Slope & $r$ & $\sigma$ & Slope & $r$ & $\sigma$ \\
\hline
CS & 0.21 $\pm$ 0.09 & 0.86 $\pm$ 0.19 & 3.0 & 0.16 $\pm$ 0.03 & 0.59 $\pm$ 0.09 & 5.0 & 0.16 $\pm$ 0.01 & 0.58 $\pm$ 0.04 & 11.0 \\
CS ($\z \geq 3.5$) & --- & --- & --- & 0.00 $\pm$ 0.11 & 0.00 $\pm$ 0.18 & 0.0 & -0.01 $\pm$ 0.05 & -0.01 $\pm$ 0.08 & 0.0 \\
CS ($2.5 \leq \z \leq 3.5$) & --- & --- & --- & 0.34 $\pm$ 0.03 & 0.80 $\pm$ 0.03 & 9.0 & 0.34 $\pm$ 0.03 & 0.80 $\pm$ 0.03 & 21.0 \\
CS + R06 & 0.06 $\pm$ 0.07 & 0.42 $\pm$ 0.42 & 1.0 & --- & --- & --- & 0.00 $\pm$ 0.01 & 0.0 $\pm$ 0.05 & 0.0 \\
CS + N15 & 0.26 $\pm$ 0.04 & 0.95 $\pm$ 0.04 & 6.5 & --- & --- & --- & 0.24 $\pm$ 0.01 & 0.89 $\pm$ 0.01 & 41.0 \\
CS + B12 & 0.10 $\pm$ 0.03 & 0.80 $\pm$ 0.14 & 3.0 & --- & --- & --- & 0.16 $\pm$ 0.01 & 0.58 $\pm$ 0.04 & 11.0 \\
CS + R13 & 0.19 $\pm$ 0.02 & 0.96 $\pm$ 0.03 & 8.0 & --- & --- & --- & 0.17 $\pm$ 0.01 & 0.62 $\pm$ 0.04 & 13.0 \\
\hline
\label{tab:Evol}
\end{tabular}
\end{table*}

Based on the results of the XQ-100 survey, we report the detection of 38 intervening DLAs identified towards 100 $z>3.5$ QSOs.  This sample has been combined, after exhaustive checking for duplications and errors, with a literature sample of DLA surveys spanning the last $\sim$ 20 years.  The final combined sample contains 742 DLAs spanning the redshift range from $z \sim 1.6$ to 5.  We present statistical measures of the column density distribution function (\fnx, Fig \ref{fig:fxn}), DLA number density (\lx, Fig. \ref{fig:lx}), and the DLA HI gas content (\Odla, Fig. \ref{fig:evolution}), and present a thorough estimation of errors and potential biases (such as colour selection and incomplete sampling of the high column density end) on these quantities.  The main focus of this paper is the evolution of \Odla, and we present a novel technique for computing this quantity as a continuous function of redshift, with confidence intervals computed at every redshift point.  In order to statistically assess whether there is evolution in \Odla\ over cosmic time, we perform a bootstrap re-sampling of the \Odla\ curve and compute the slope and correlation coefficients ($r$) of 100,000 iterations.  For the combined sample, the most significant \Odla\ redshift evolution (median $r=0.8$) is found for the interval $2.5<z<3.5$.  However, at higher redshifts, the median slope of the bootstrap iterations is zero and median $r=0.0$, indicating no significant evolution, but improved statistics above $z \sim 3.5$ are still required to confirm this. Assessing evolution in \Odla\ down to lower redshifts is found to be highly dependent on the choice of sample used for the evolution test.  Combining the CS with the intermediate redshift sample of DLAs from \citet{2006ApJ...636..610R} yields a median correlation coefficient $r=0$, indicating that the cosmic gas density is not strongly evolving from $z \sim 0.1$ to 5.  However, this picture is challenged by an alternative survey for $0 < z < 1.6$ absorbers by \citet{2015ApJN} who find an \Odla\ value a factor of five lower than R06.  Adopting the N15 value of \Odla\ results in a highly significant evolution of galactic gas content (median correlation coefficient $r=0.9$). A more consistent picture is obtained when the CS is combined with $z \sim 0$ surveys.  Although these surveys exhibit a factor of $\sim$ two variation in their quoted \Ohi, both the highest \citep{2012ApJ...749...87B} and lowest \citep{2013MNRAS.435.2693R} yield a statistically significant ($r=0.6$) redshift evolution when combined with the high redshift data. The median slope obtained from out bootstrap re-sampling is $\sim 0.17 \times 10^{-3}$, corresponding to a factor of $\sim$ 4 decrease in \Ohi\ from $z=5$ to $z=0$.  Therefore, the greatest uncertainty in \Ohi\ measurements is in the intermediate redshift regime, which is currently beyond the reach of most 21cm surveys, but still poorly sampled by DLA studies. An accurate measure of \Odla\ at redshifts between 0.1 and 2 is therefore clearly of the utmost importance. Upcoming surveys with the Square Kilometre Array \citep[SKA,][]{2015aska.confE.167S} and its pathfinders present an exciting prospect for resolving the current uncertainty in the gas content of galaxies since $z \sim 1.5$.

\section*{Acknowledgments}

In memory of Javier Gorosabel: an exceptional supervisor, a brilliant scientist, and an even better human being. The XQ-100 team is indebted to the ESO staff for support through the Large Program execution process, and to the DARK cosmology centre for financial support of the XQ-100 team meeting.  We are grateful to Neil Crighton for sharing the DLA catalog of the GGG survey \citep{2015MNRAS.452..217C} in advance of publication, as well as to Marcel Neeleman \citep{2015ApJN} for also sharing their results before they become public.  RSR acknowledges financial support of the Spanish Government under his FPI grant and projects AYA 2009-14000-C03-01 and AYA 2012-39727-C03-01, as well as to Ikerbasque Fundation the contract that allowed him to finish this paper. RSR also thanks to UVIC, IAA-CSIC and the Dpto. de F\'{i}sica Aplicada I of the UPV/EHU for hosting him during the development of this work, and to Antx\'{o}n Alberdi, Carlos Barcel\'{o}, Sam Oates, Sabina Ustamujic, and especially to Antonio de Ugarte Postigo and Alberto Castro-Tirado for interesting and useful conversations on topics discussed in this manuscript.  SLE acknowledges the receipt of an NSERC Discovery Grant which supported this research.  JXP is supported by NSF grant AST-1109447. SL has been supported by FONDECYT grant number 1140838 and partially by PFB-06 CATA. VD and IP acknowledge support from the PRIN INAF 2012 "The X-shooter sample of the quasar spectra at z-3.5: Digging into cosmology and galaxy evolution with quasar absorption lines". The Dark Cosmology Centre is funded by the DNRF.

\bibliographystyle{mnras}
\bibliography{omega}

\appendix

\clearpage

\section{DLA fits}
\label{sec:DLAfits}

\includegraphics[width=0.5\textwidth]{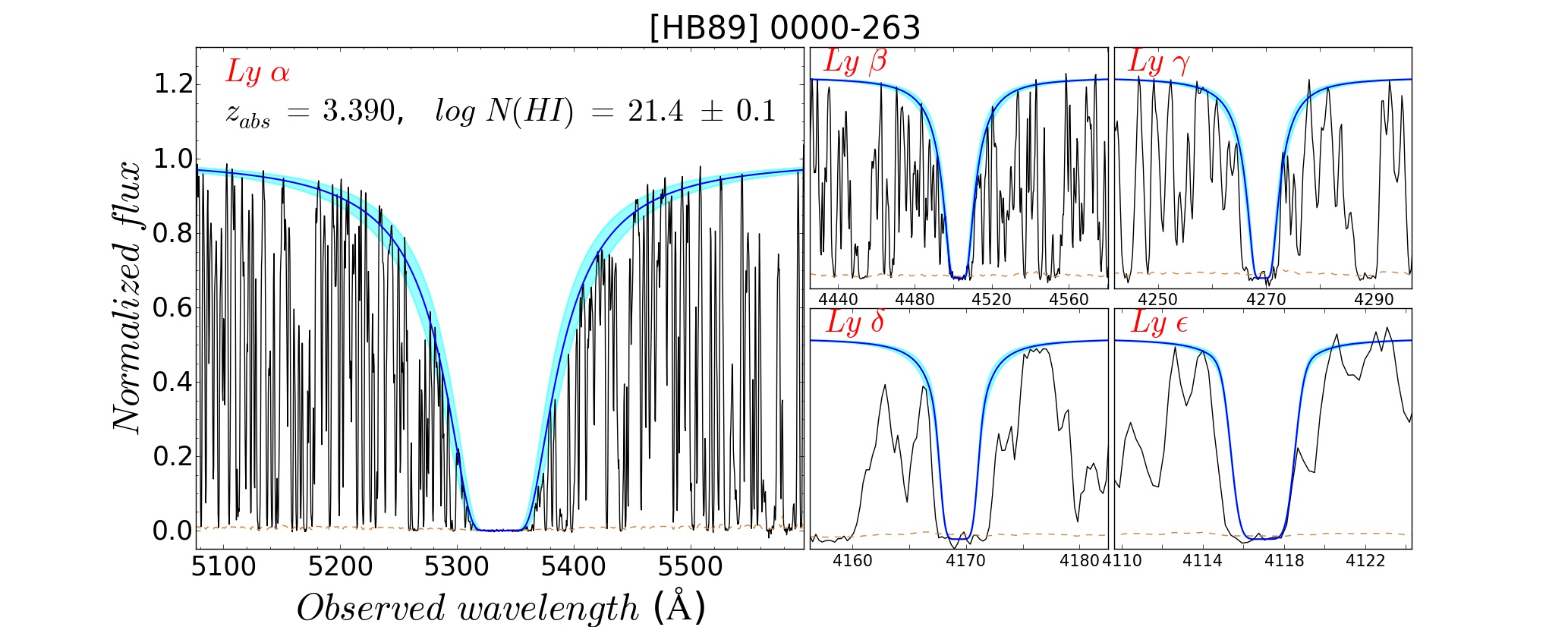} \\ 
\includegraphics[width=0.5\textwidth]{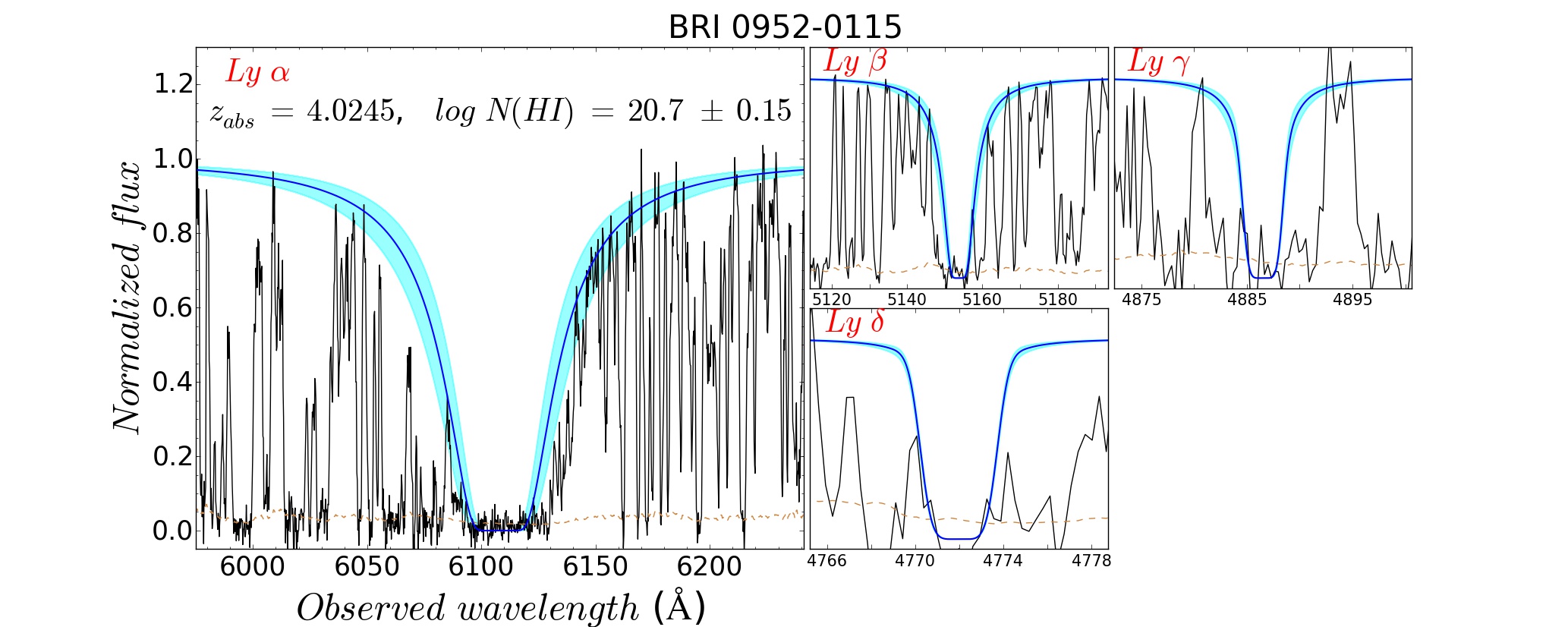} \\ 
\includegraphics[width=0.5\textwidth]{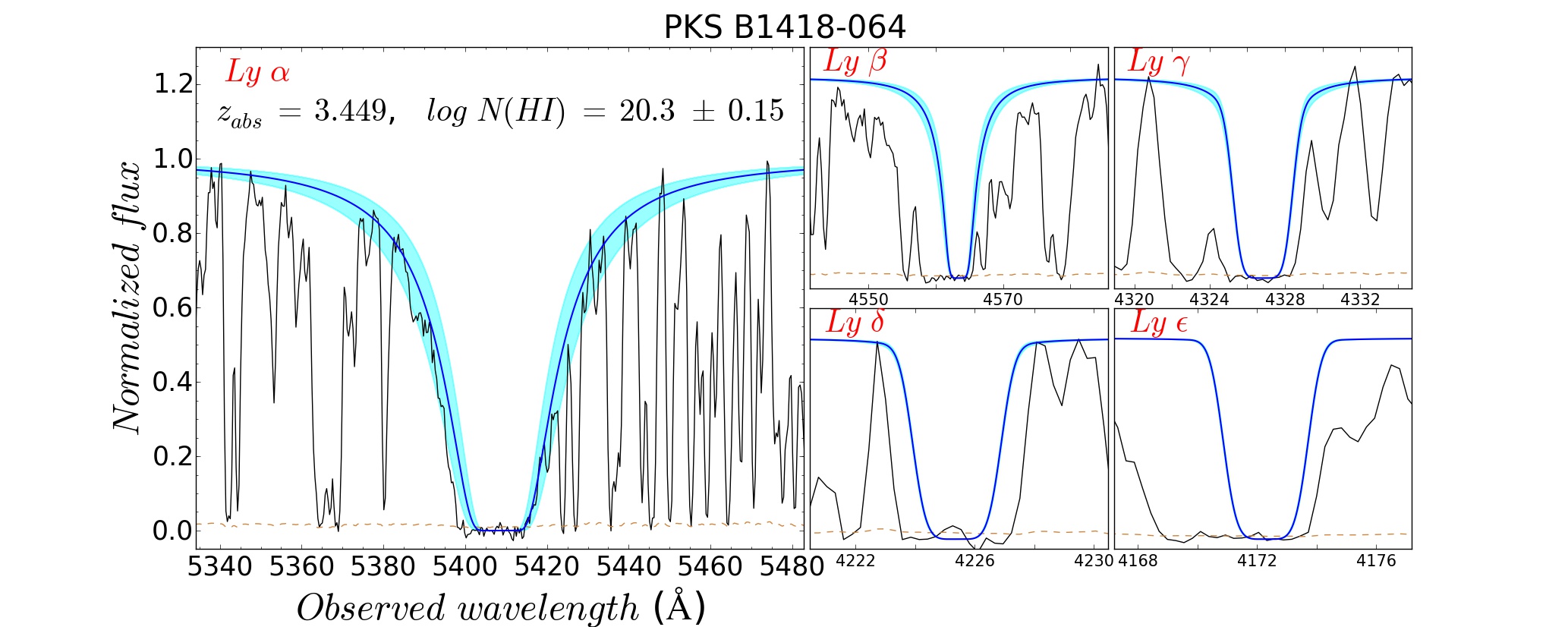} \\ 
\includegraphics[width=0.5\textwidth]{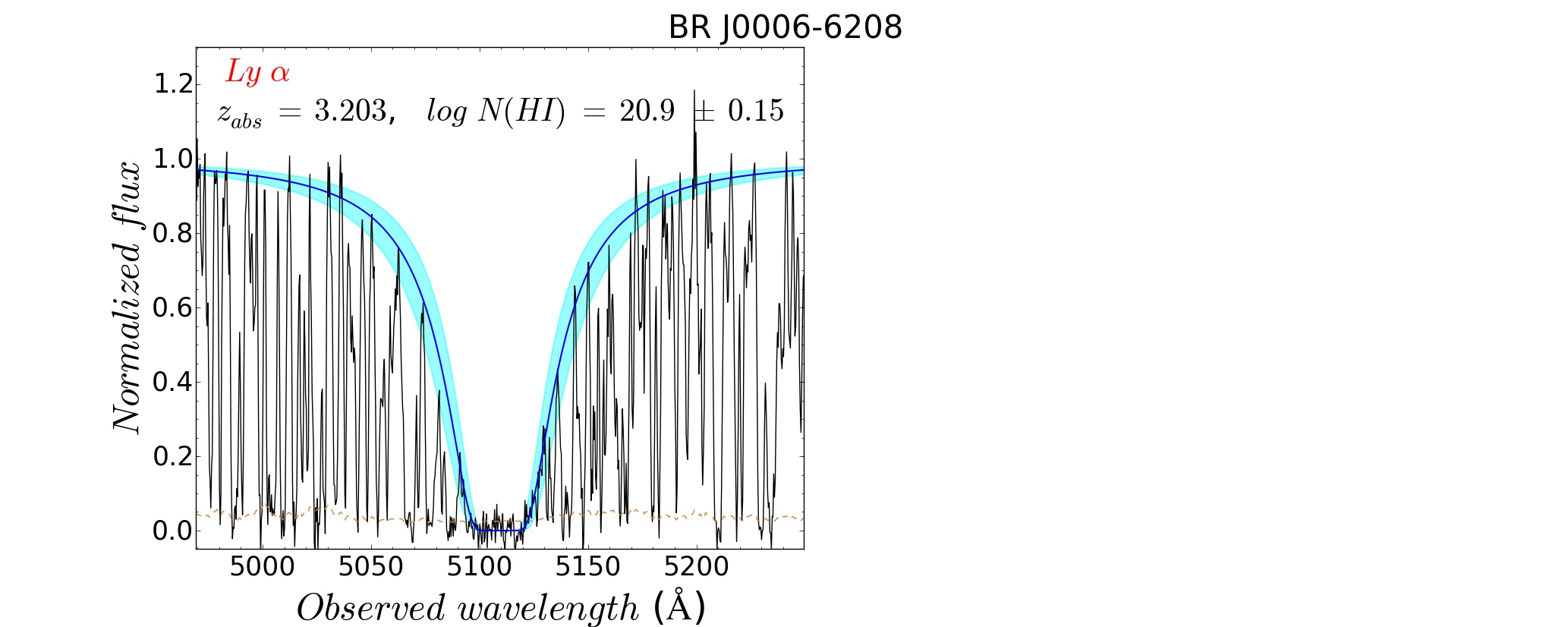} \\ 
\includegraphics[width=0.5\textwidth]{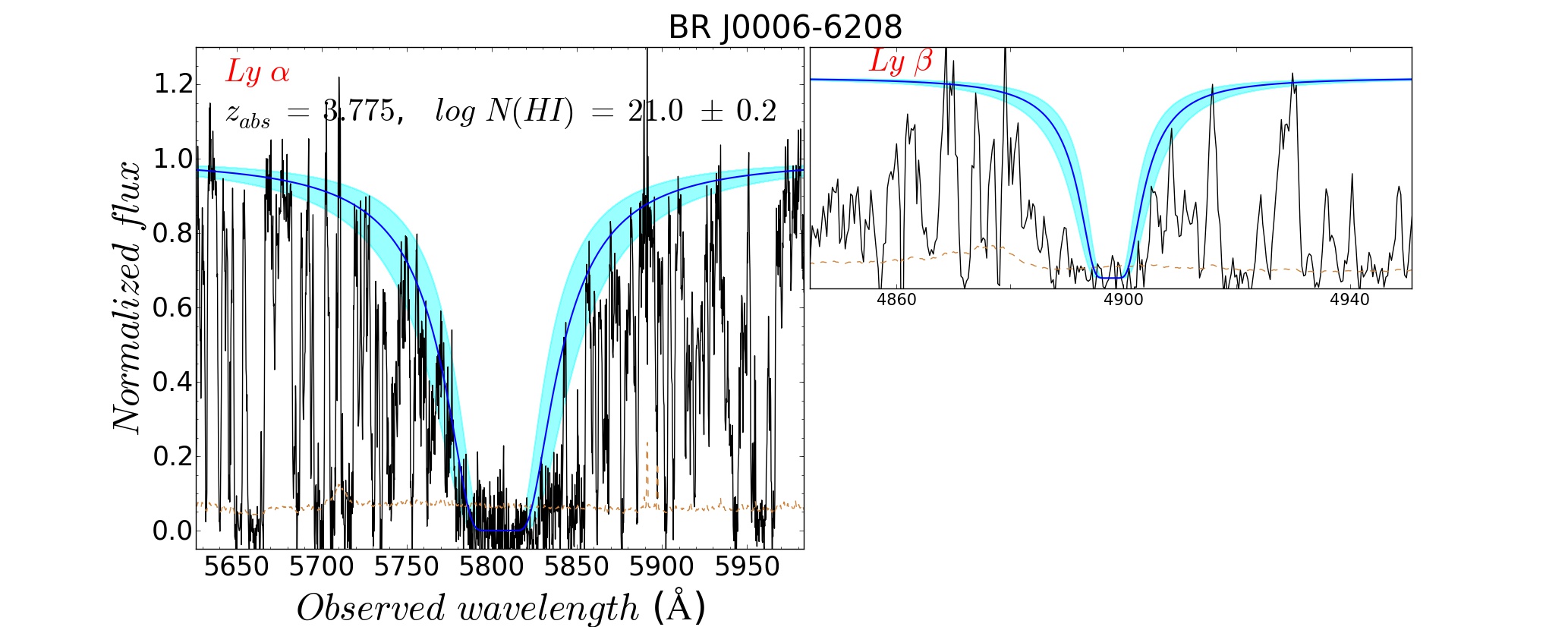} \\ 
\includegraphics[width=0.5\textwidth]{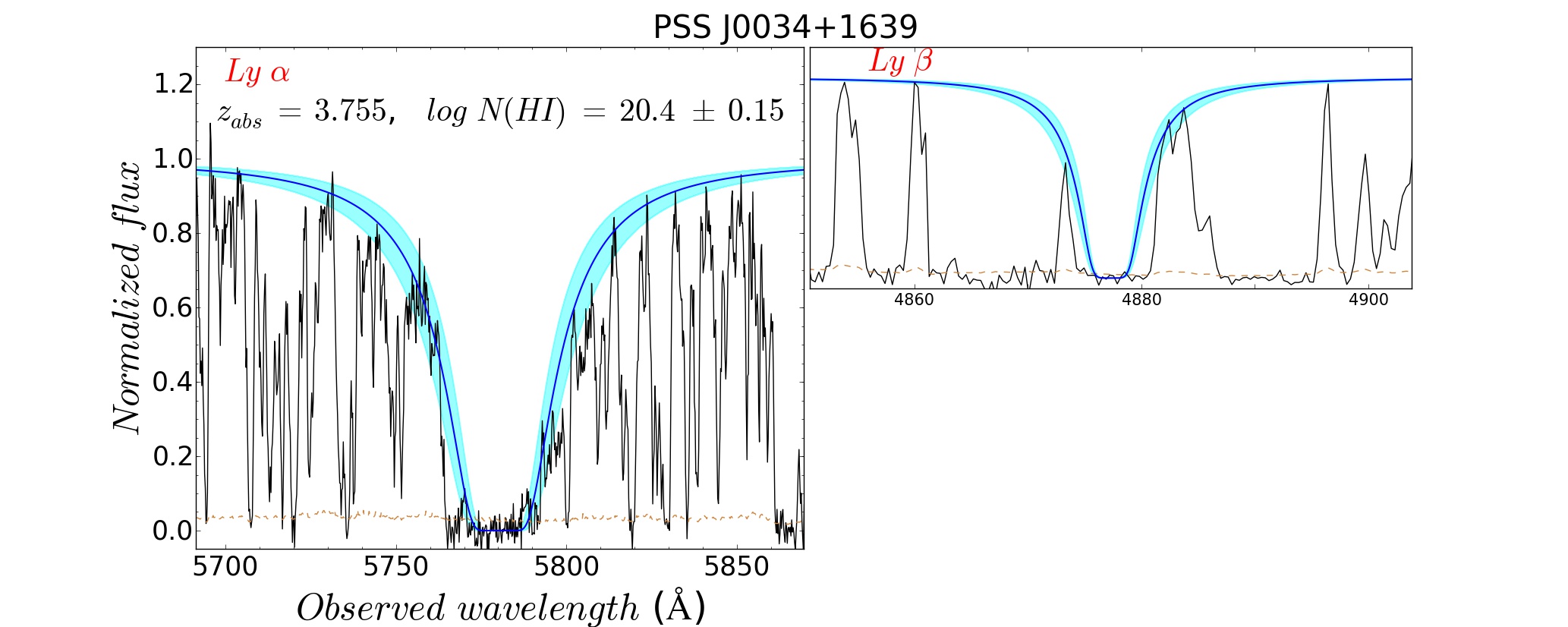} \\ 
\includegraphics[width=0.5\textwidth]{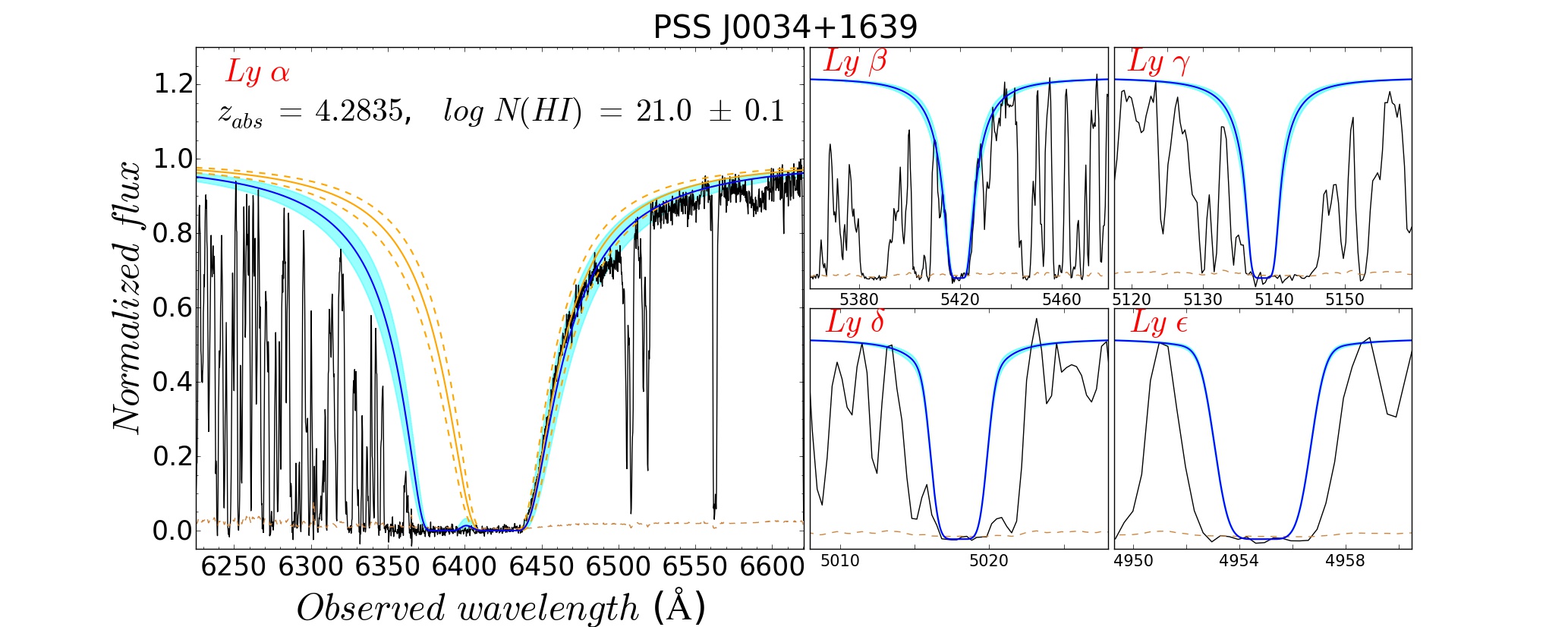} \\ 
\includegraphics[width=0.5\textwidth]{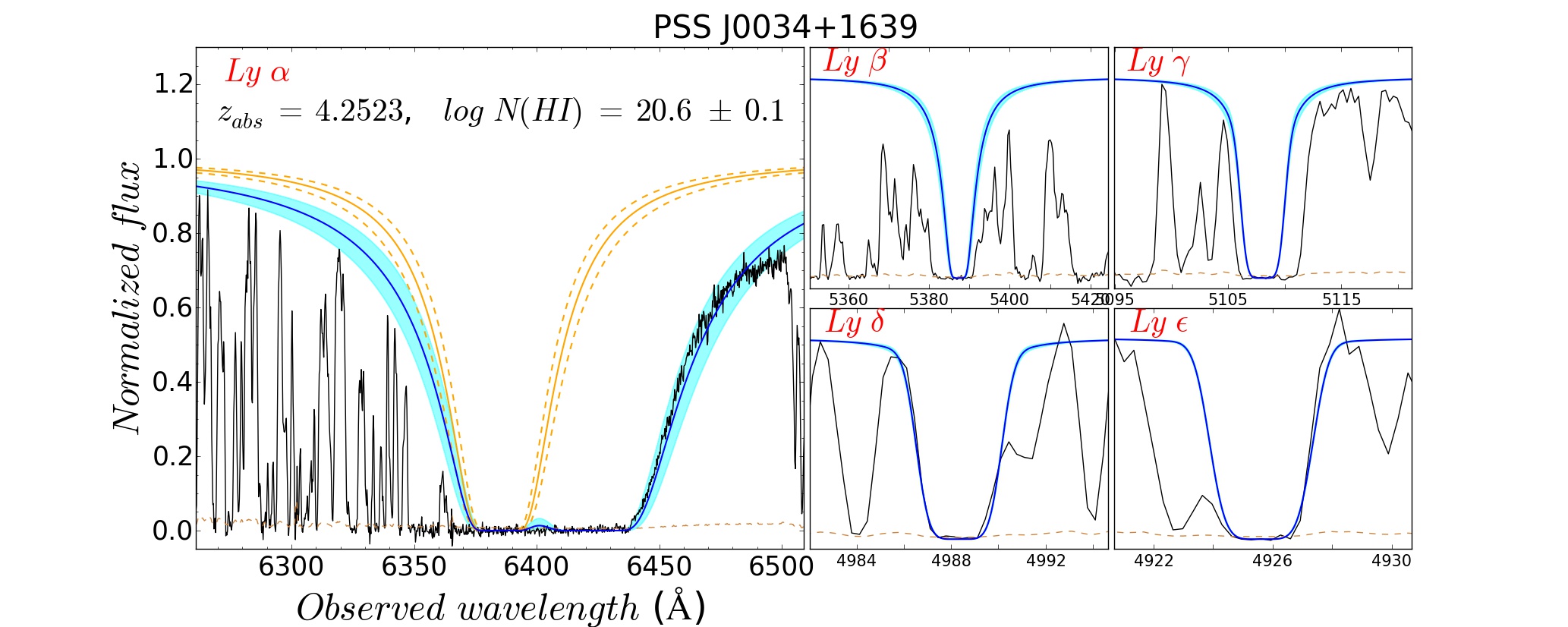} \\ 
\includegraphics[width=0.5\textwidth]{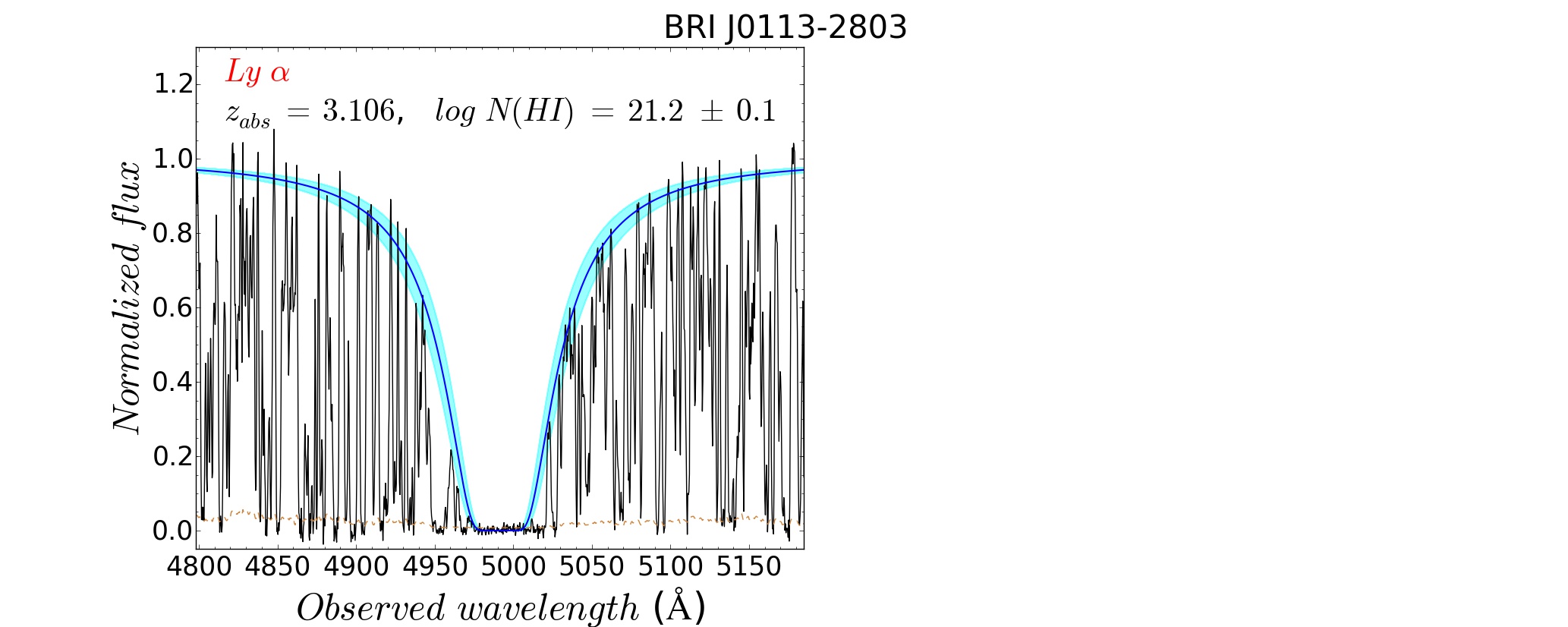} \\ 
\includegraphics[width=0.5\textwidth]{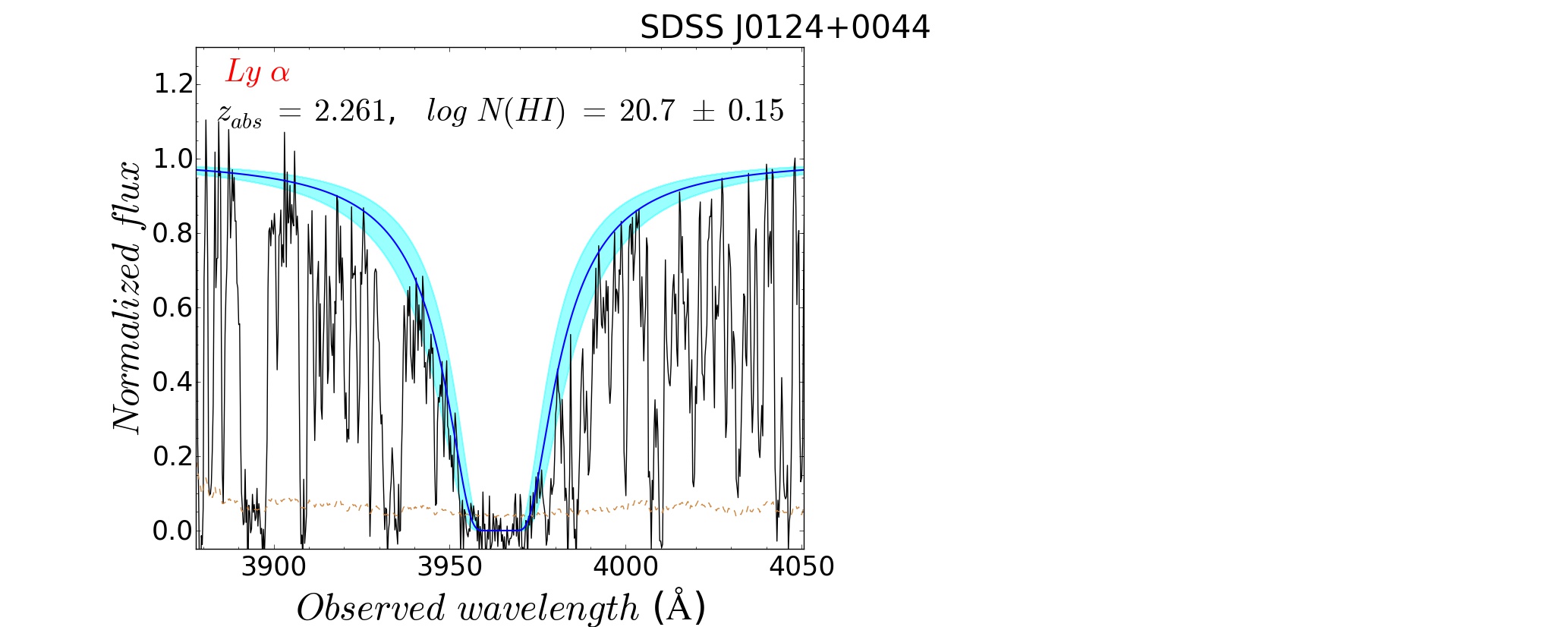} \\ 
\includegraphics[width=0.5\textwidth]{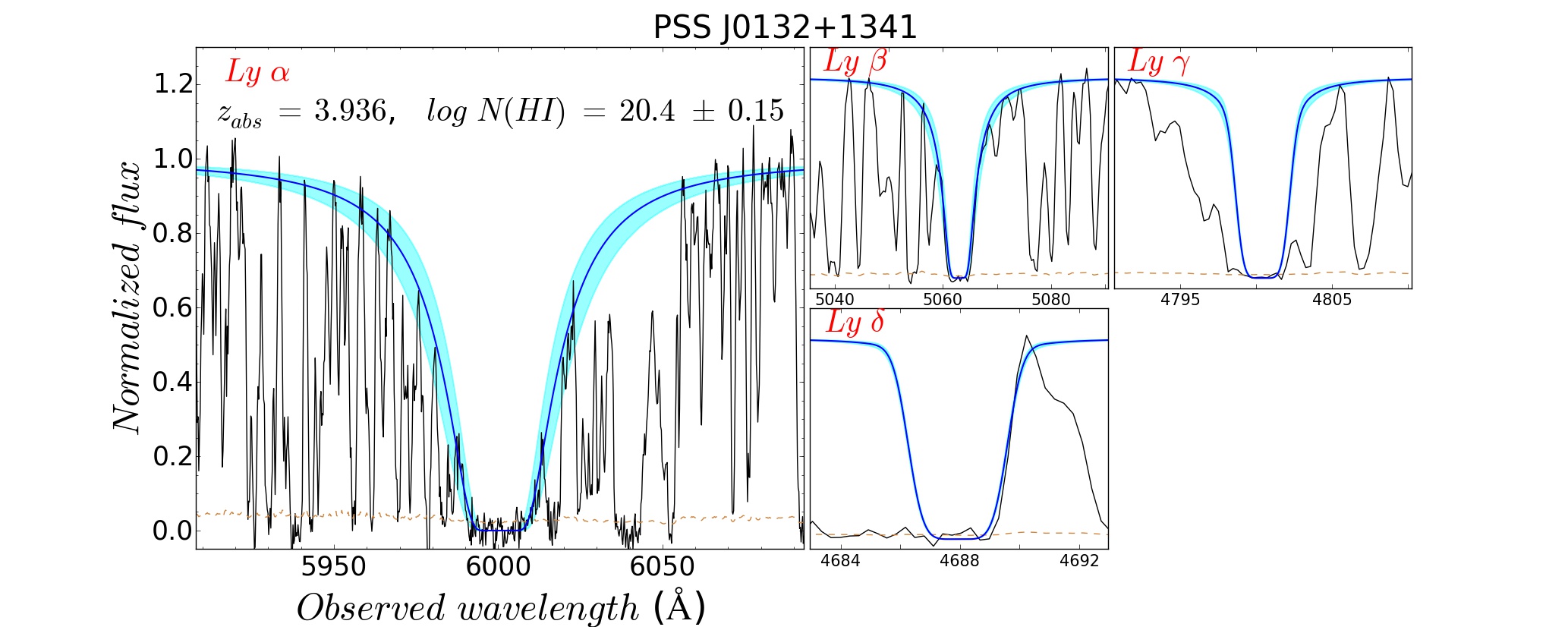} \\ 
\includegraphics[width=0.5\textwidth]{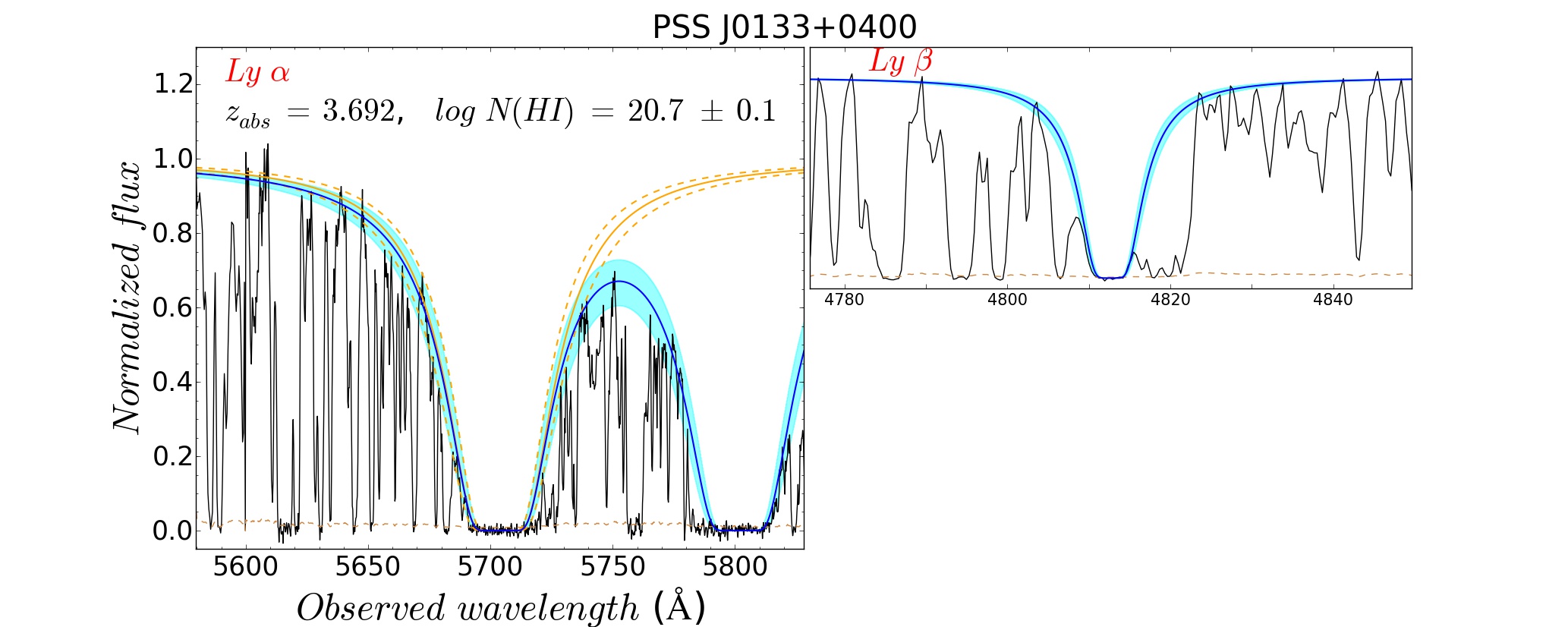} \\ 
\includegraphics[width=0.5\textwidth]{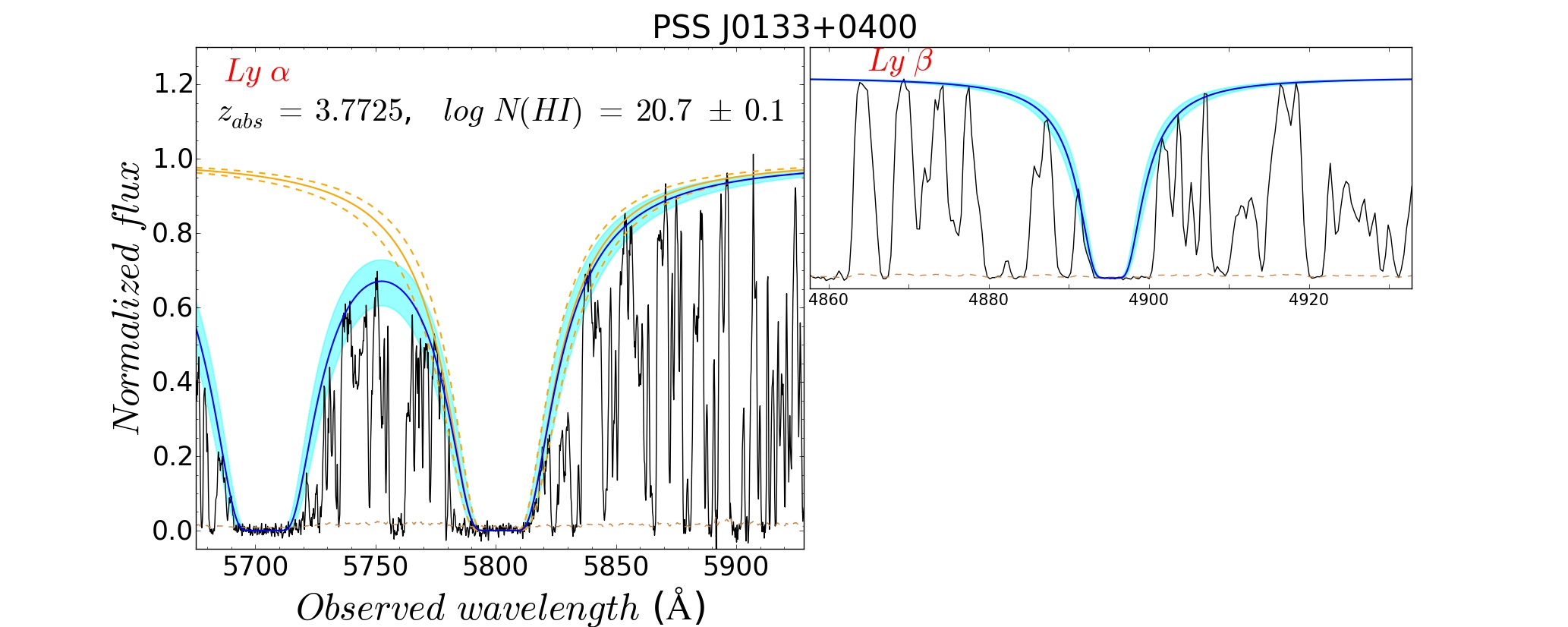} \\ 
\includegraphics[width=0.5\textwidth]{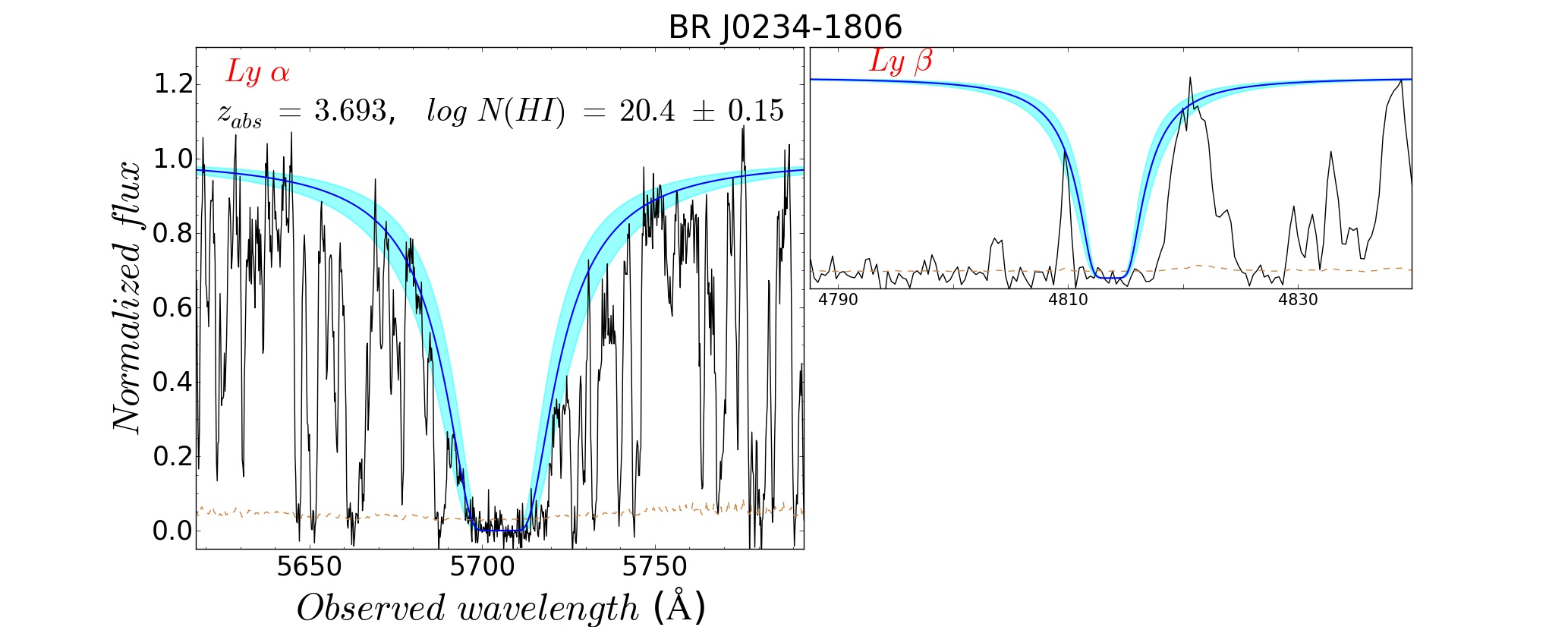} \\ 
\includegraphics[width=0.5\textwidth]{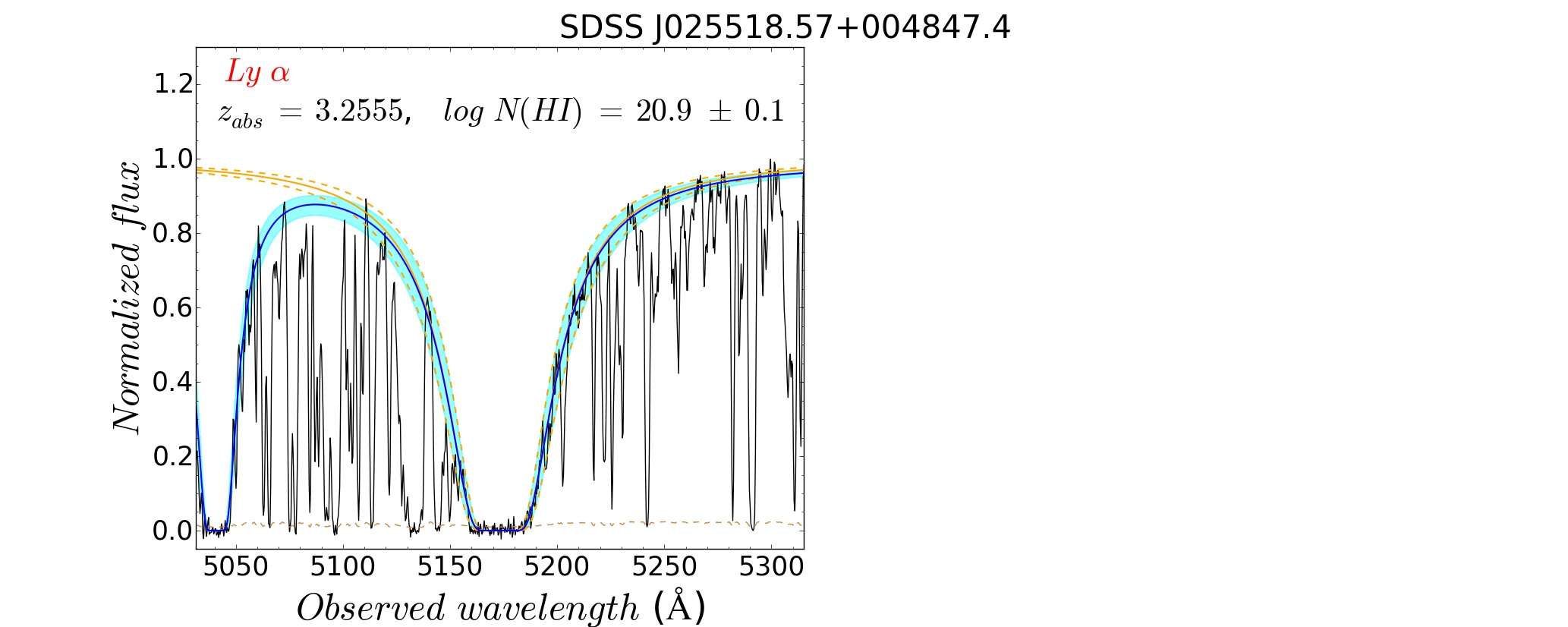} \\ 
\includegraphics[width=0.5\textwidth]{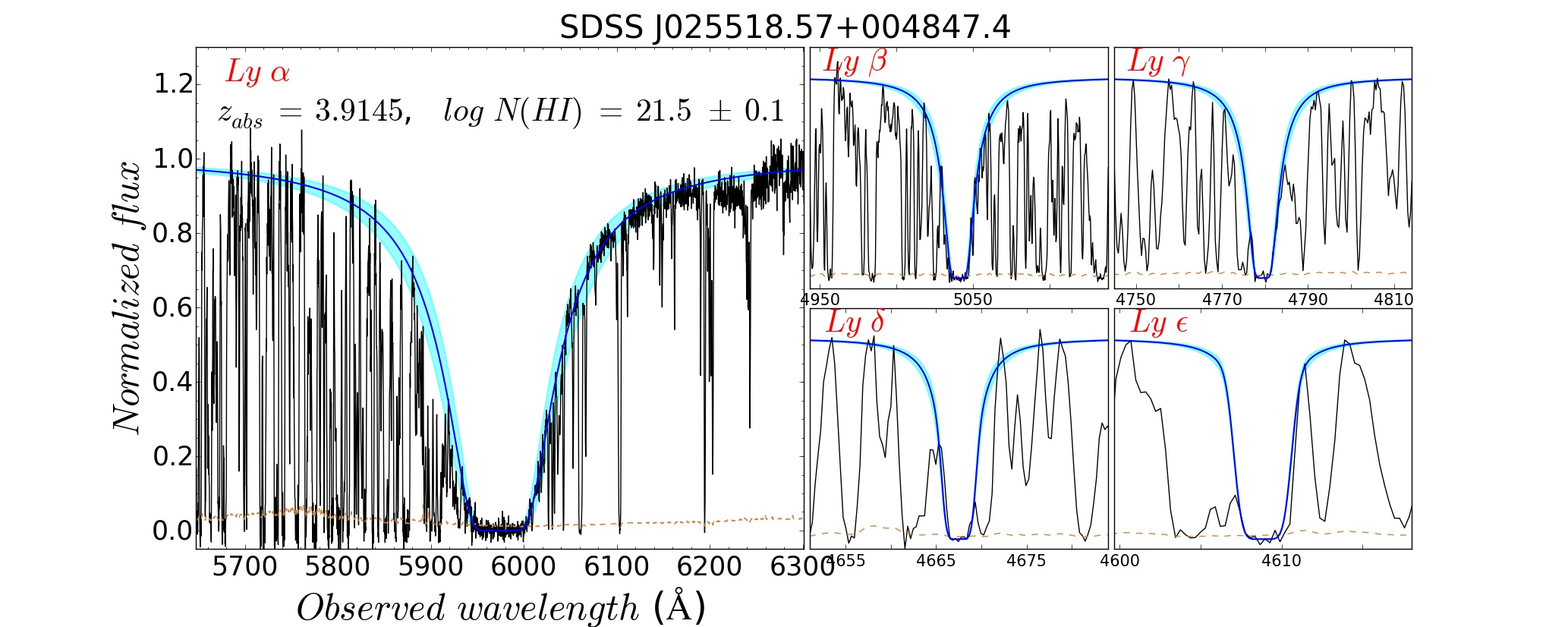} \\ 
\includegraphics[width=0.5\textwidth]{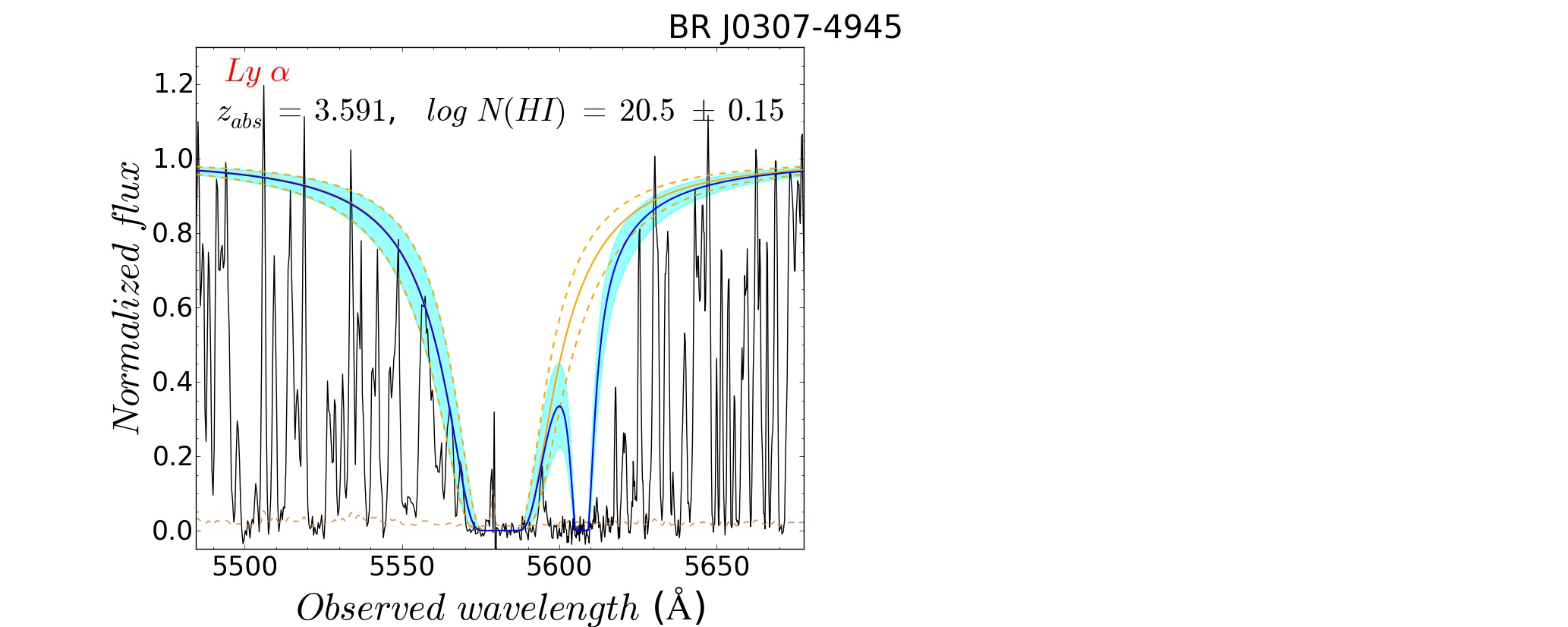} \\ 
\includegraphics[width=0.5\textwidth]{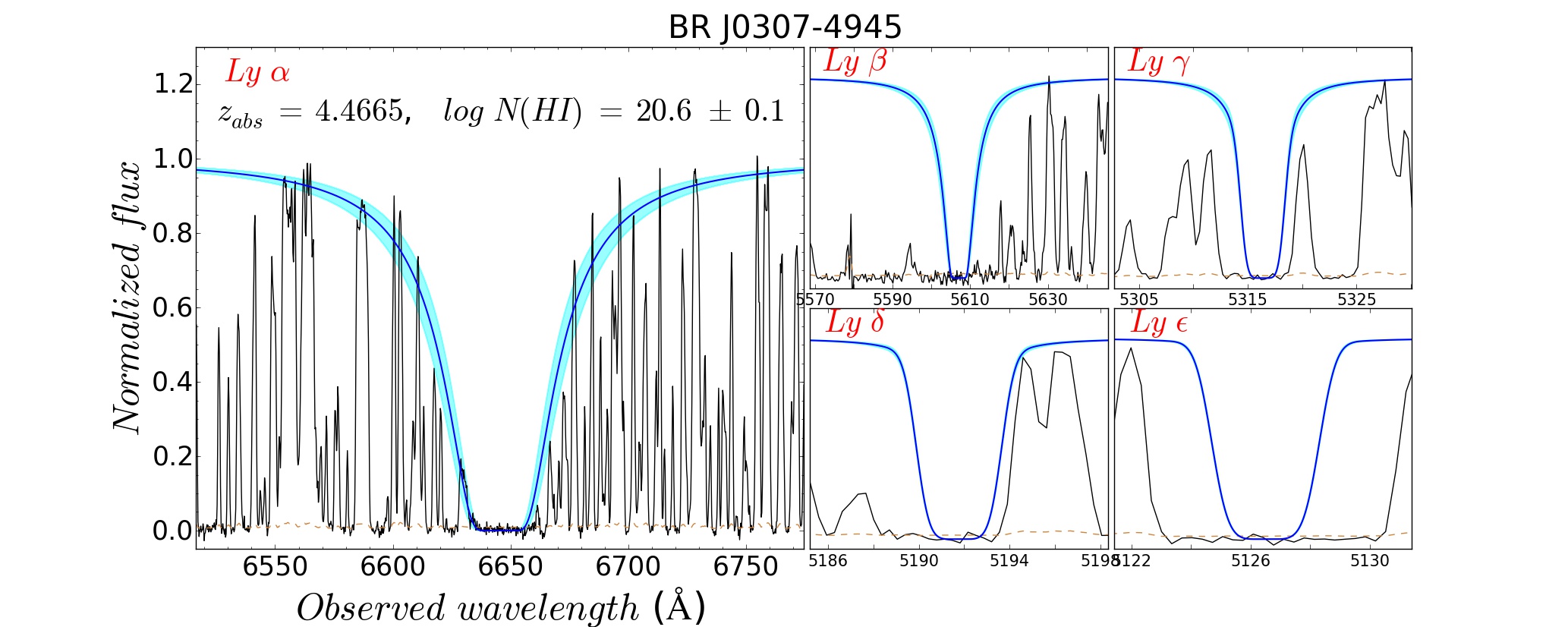} \\ 
\includegraphics[width=0.5\textwidth]{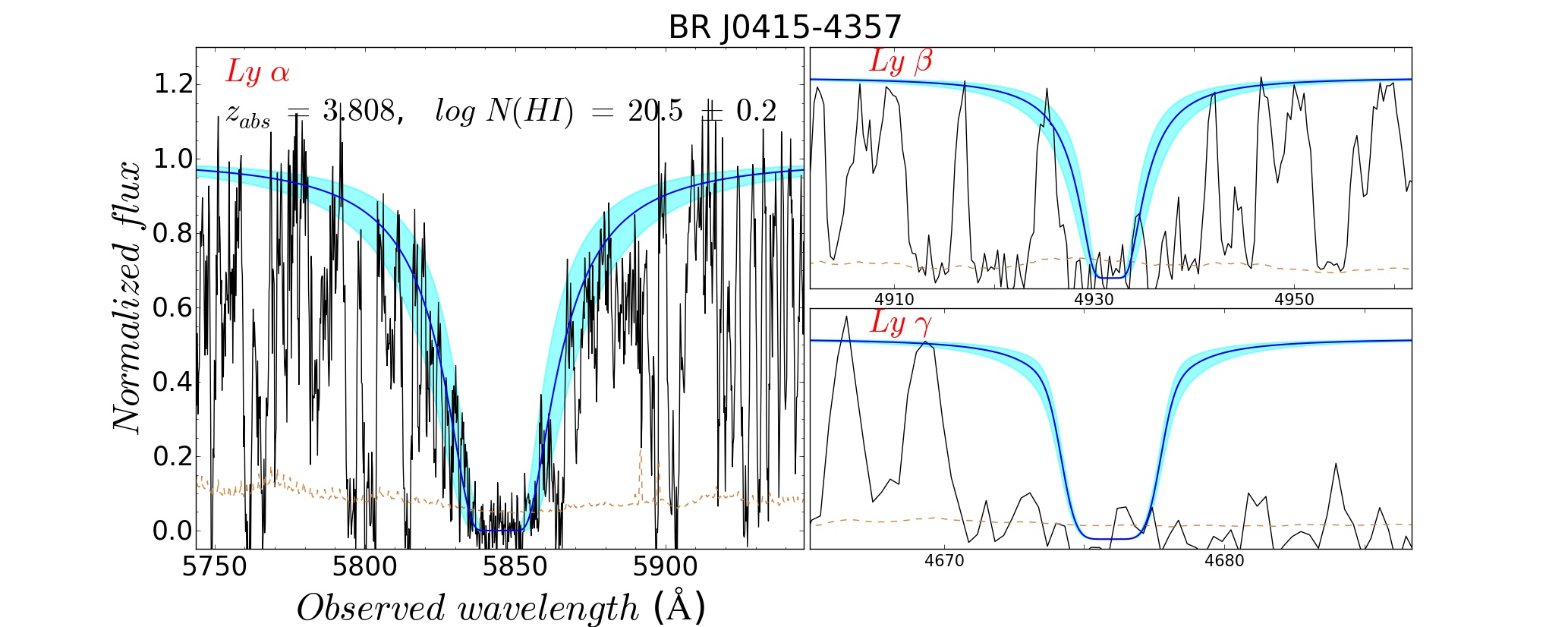} \\ 
\includegraphics[width=0.5\textwidth]{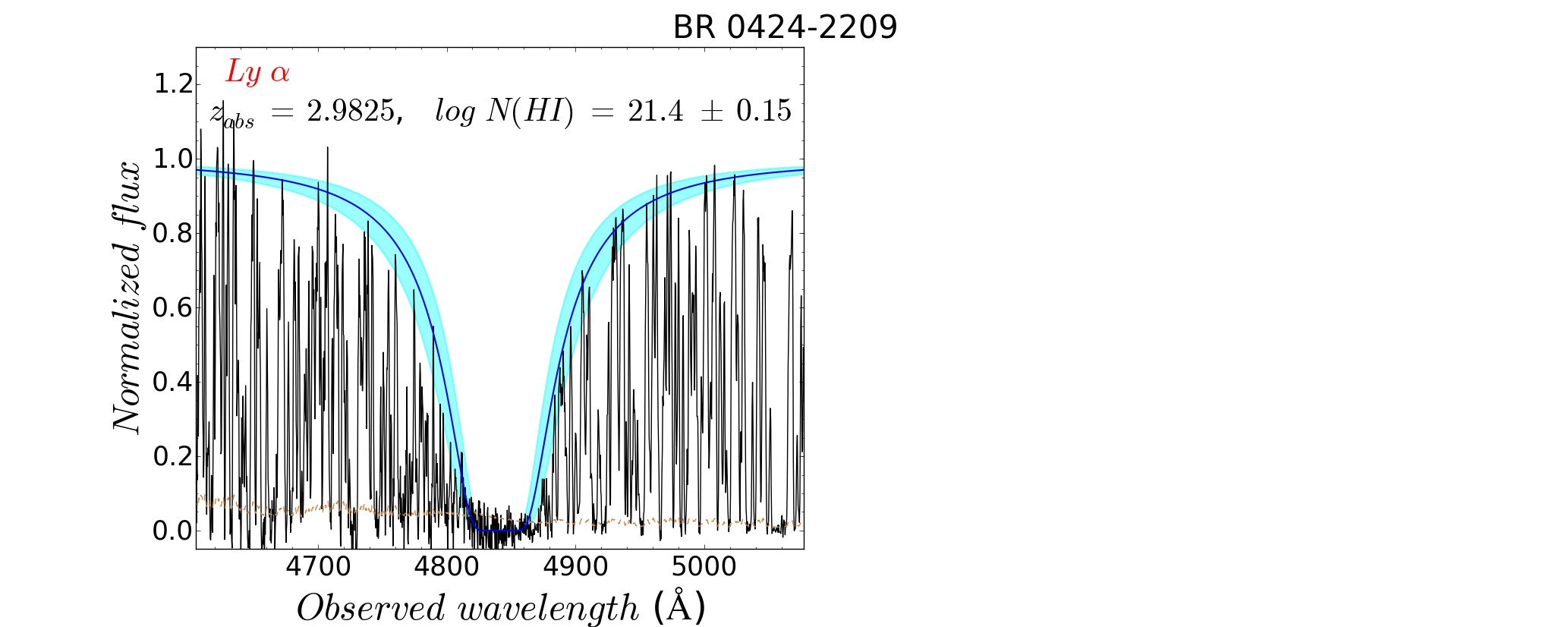} \\ 
\includegraphics[width=0.5\textwidth]{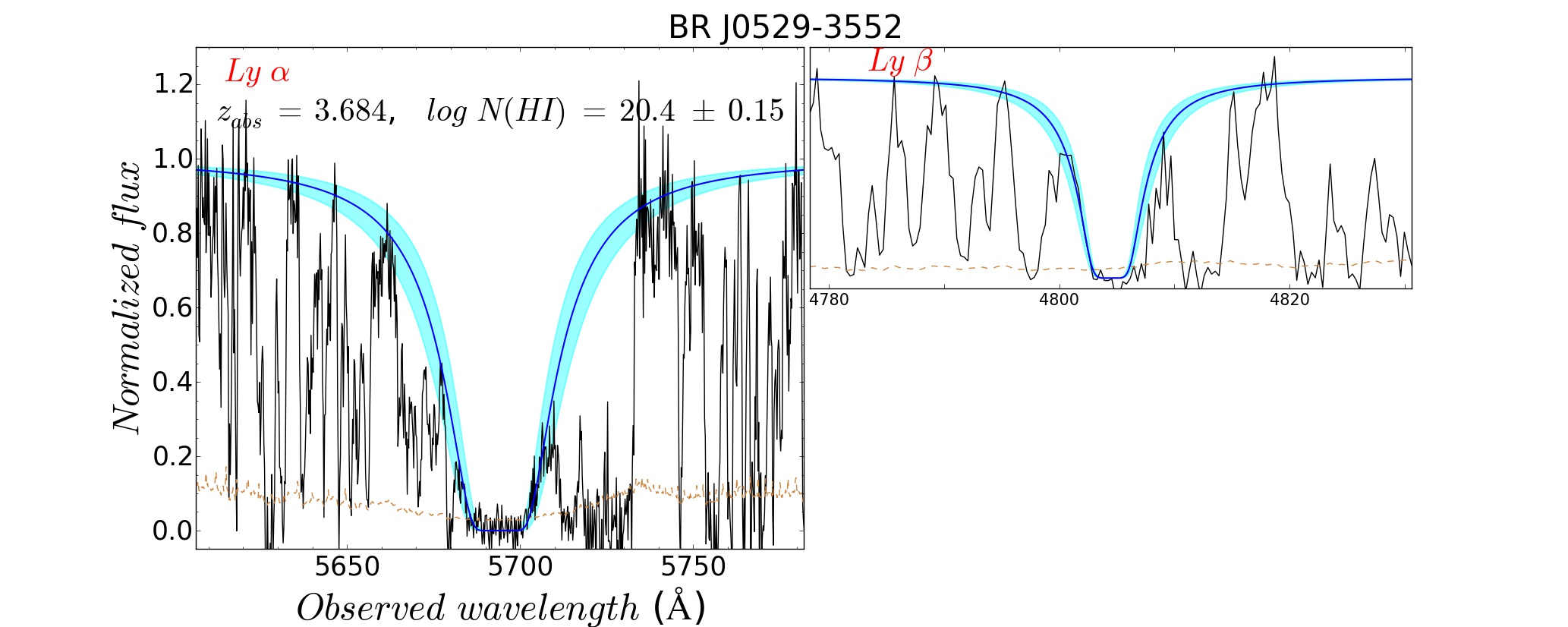} \\ 
\includegraphics[width=0.5\textwidth]{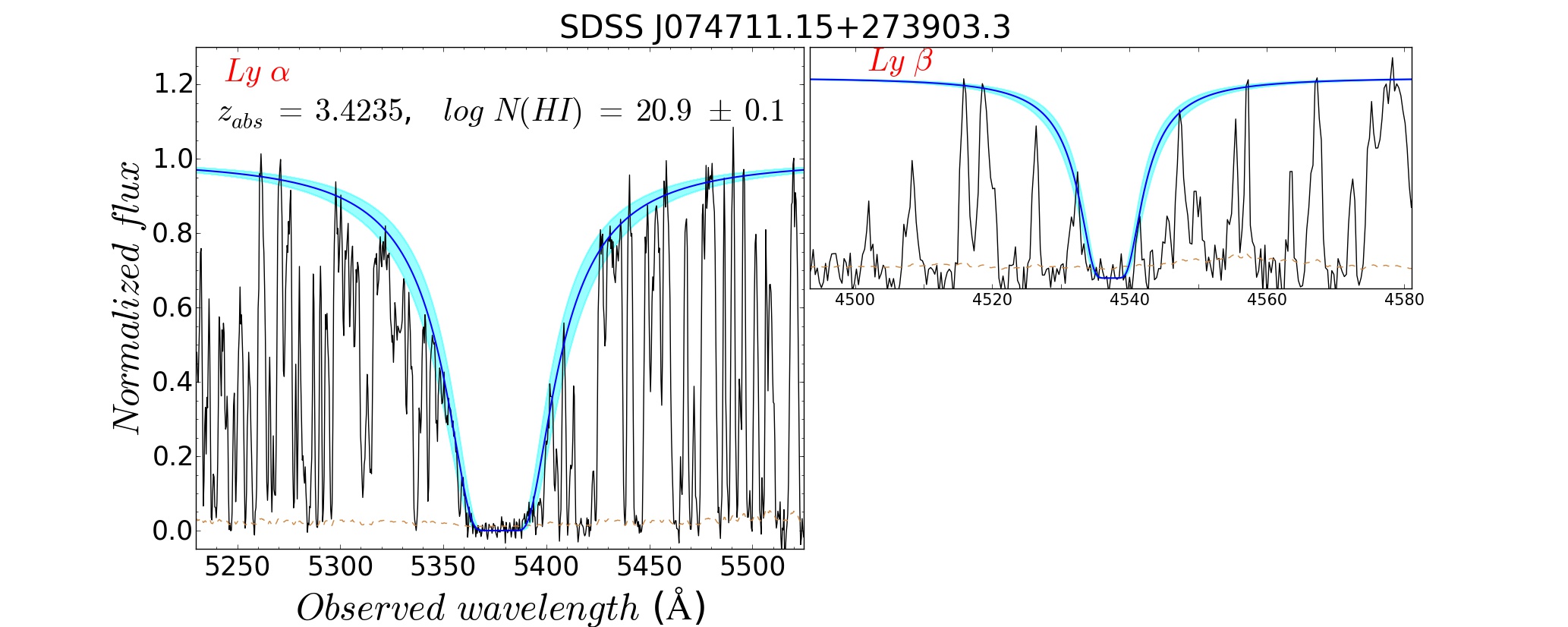} \\ 
\includegraphics[width=0.5\textwidth]{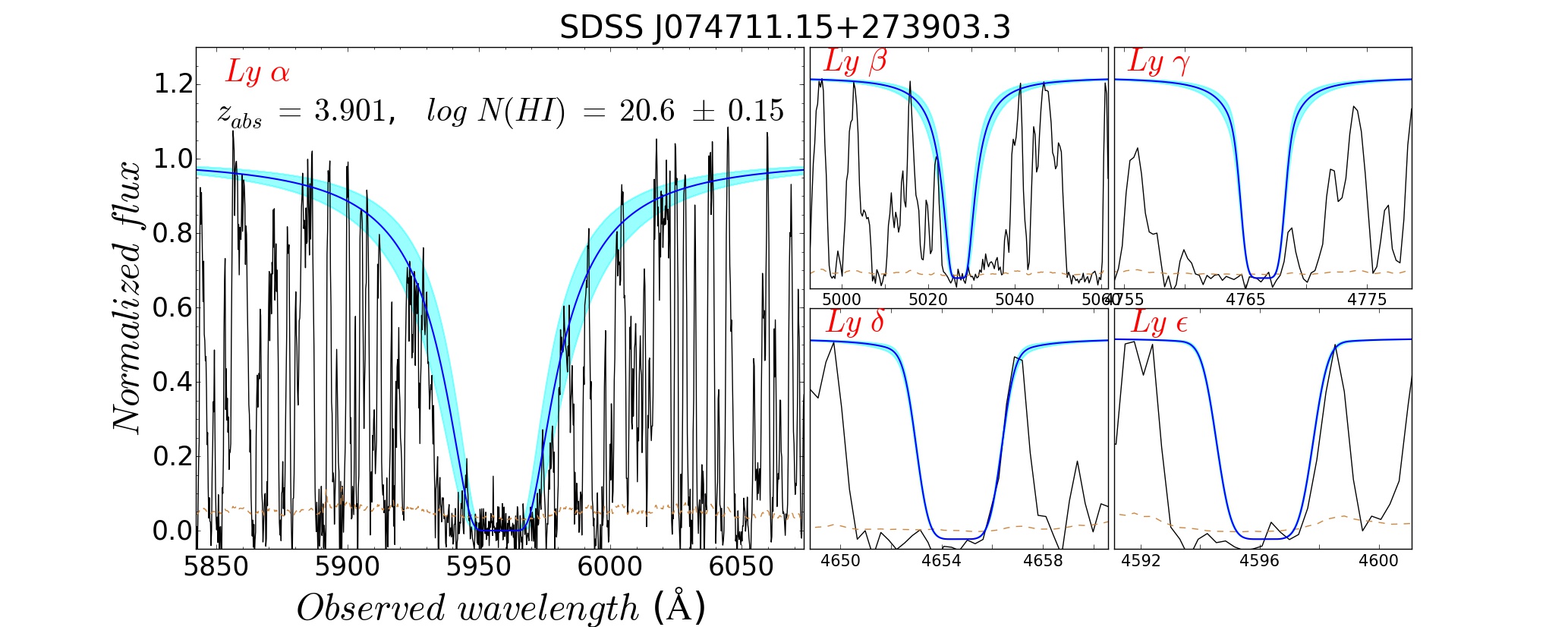} \\ 
\includegraphics[width=0.5\textwidth]{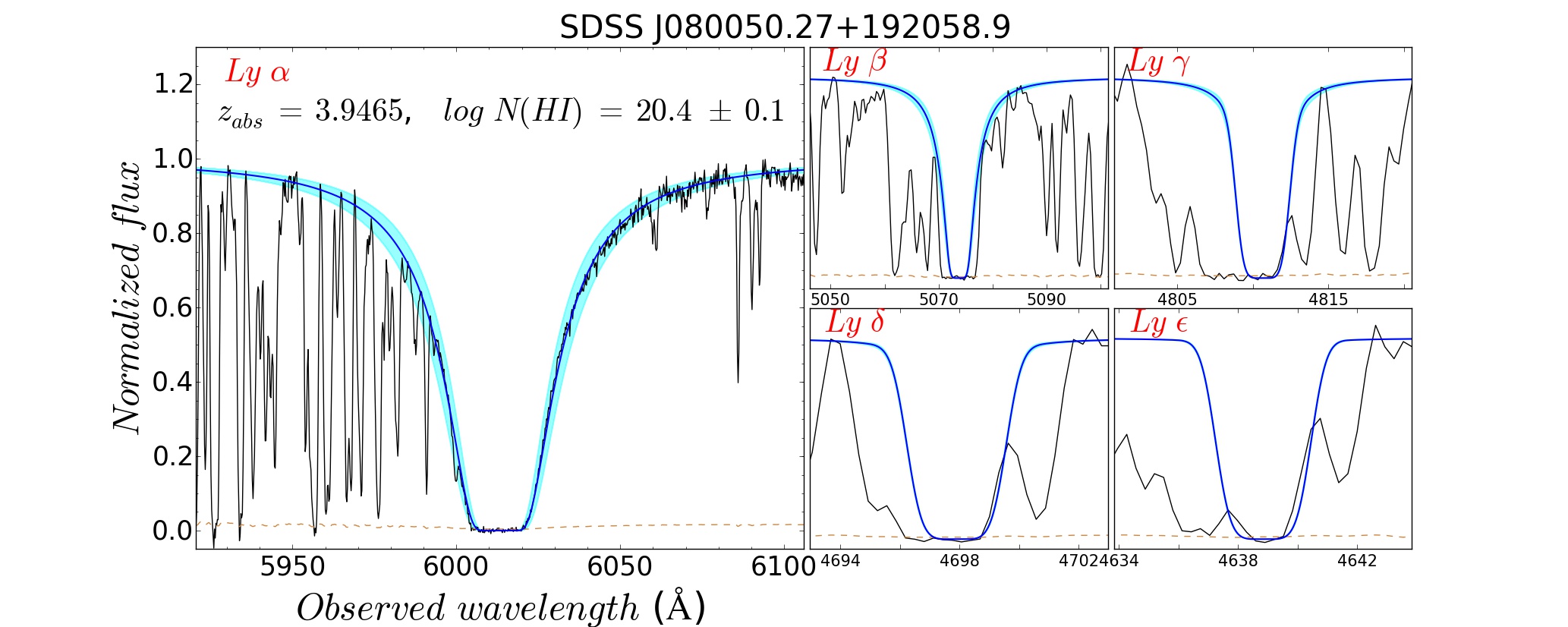} \\ 
\includegraphics[width=0.5\textwidth]{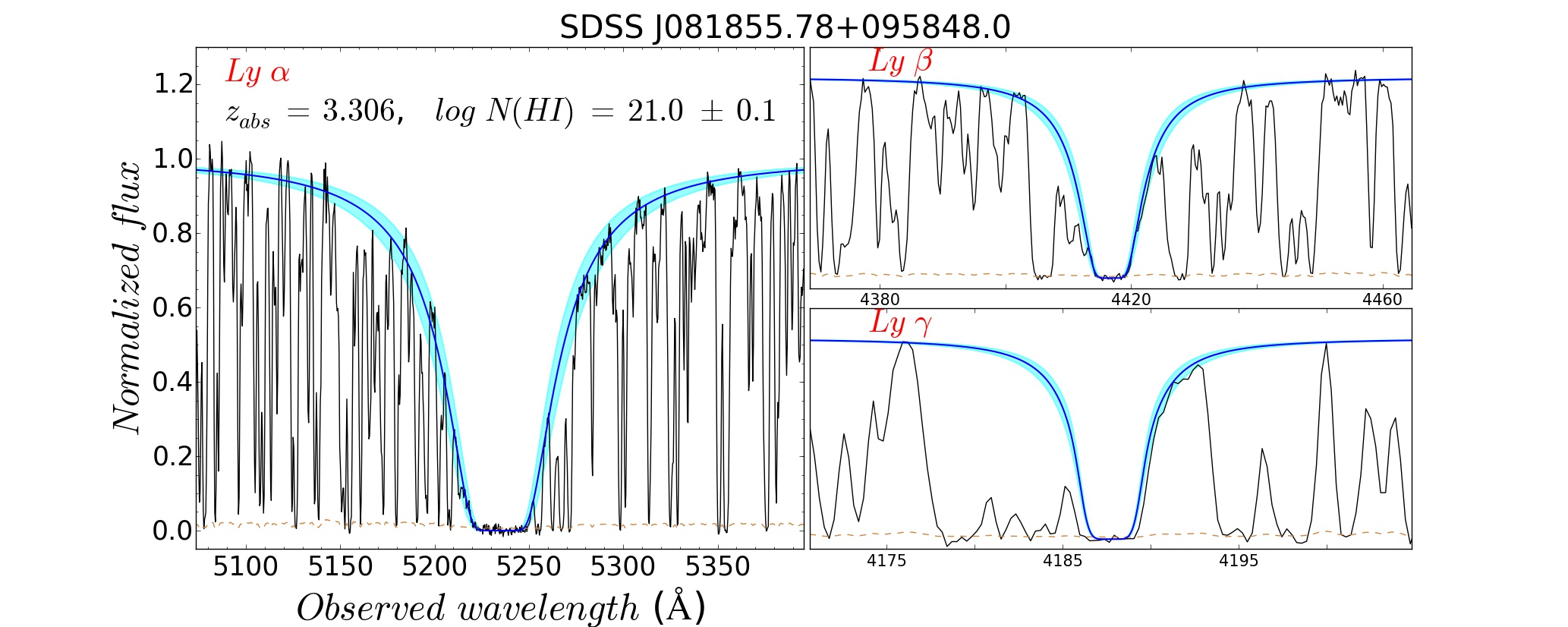} \\ 
\includegraphics[width=0.5\textwidth]{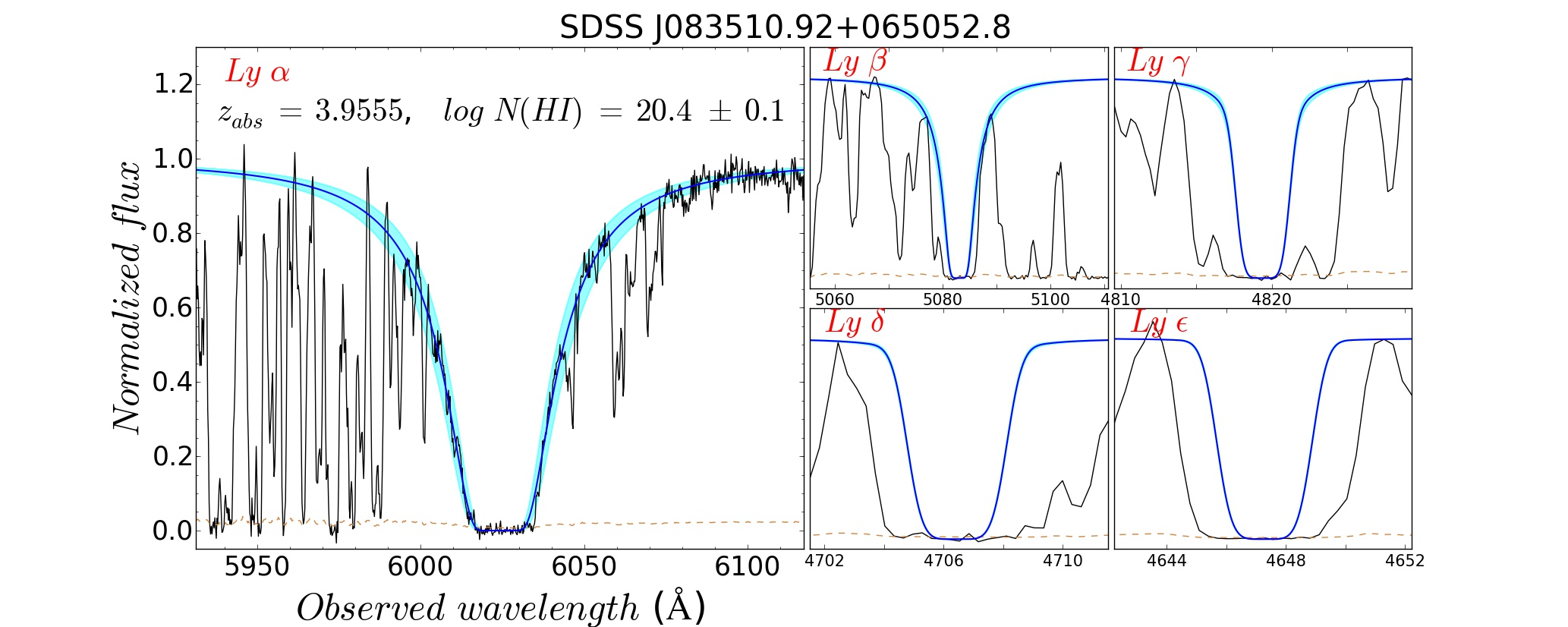} \\ 
\includegraphics[width=0.5\textwidth]{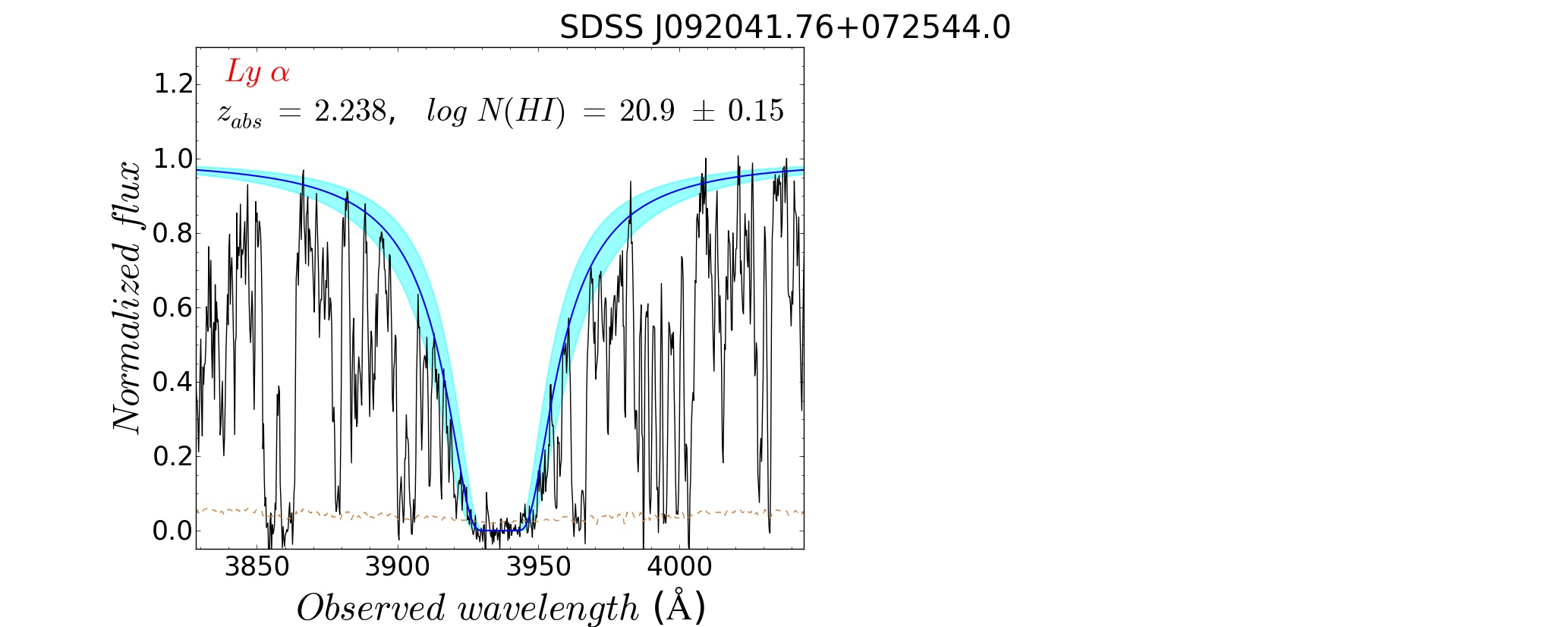} \\ 
\includegraphics[width=0.5\textwidth]{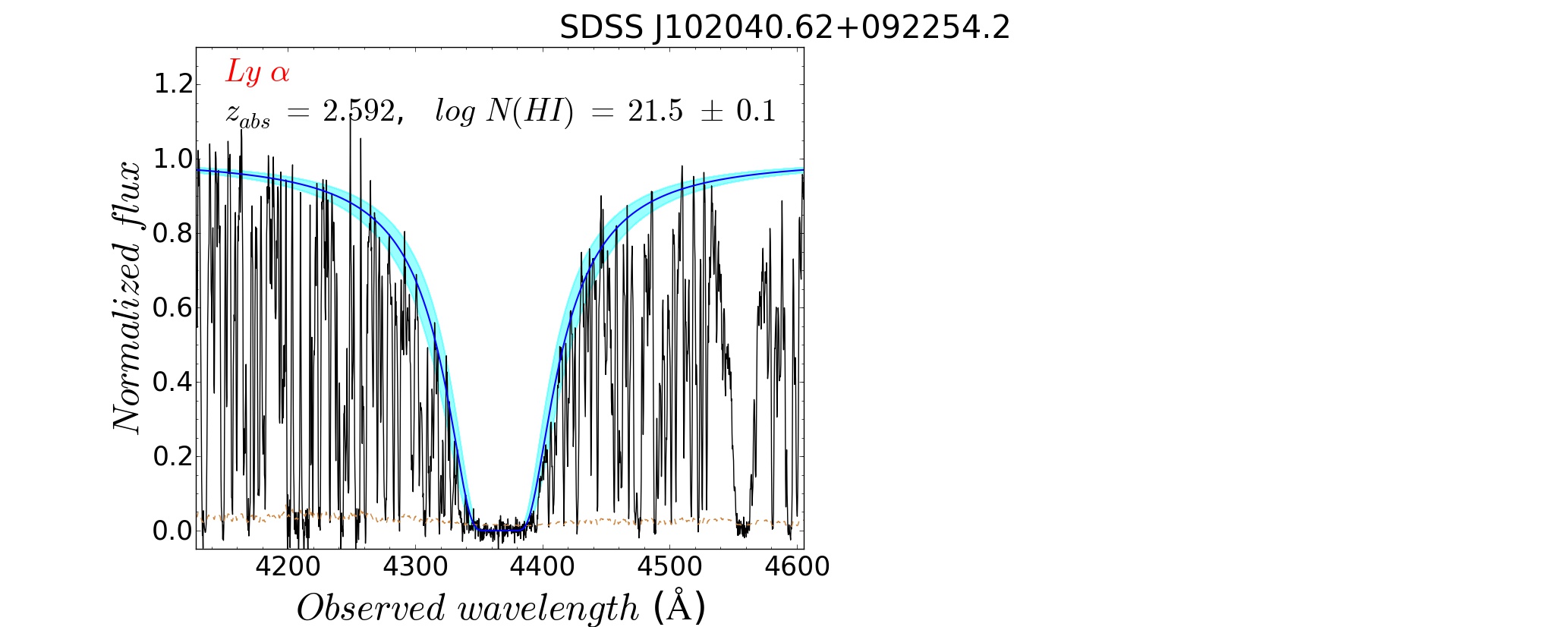} \\ 
\includegraphics[width=0.5\textwidth]{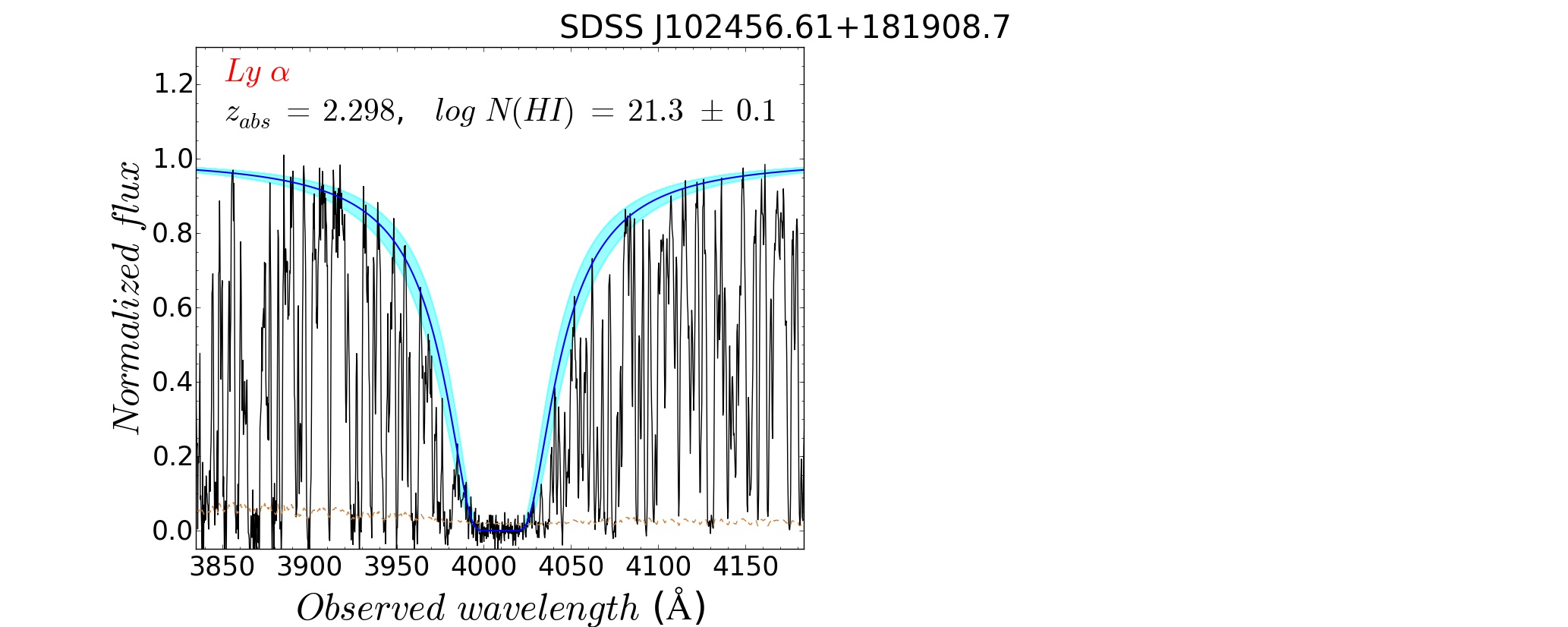} \\ 
\includegraphics[width=0.5\textwidth]{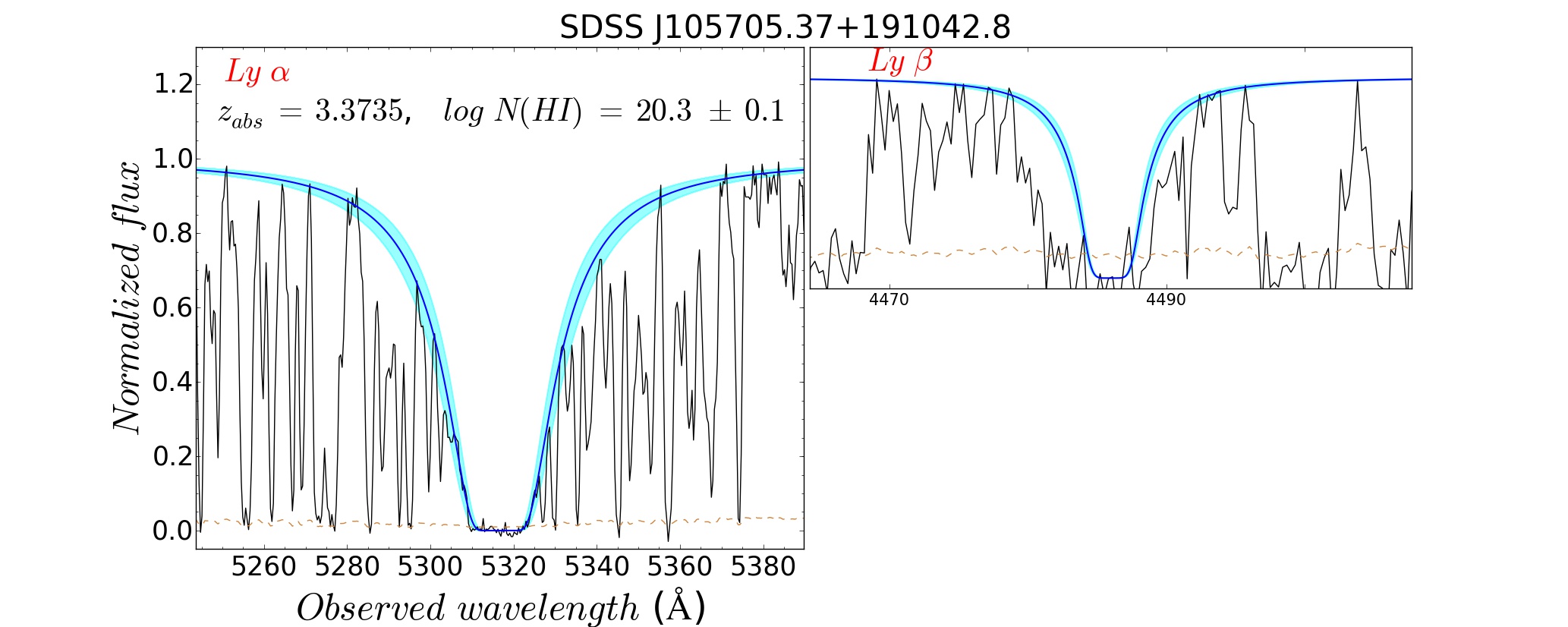} \\ 
\includegraphics[width=0.5\textwidth]{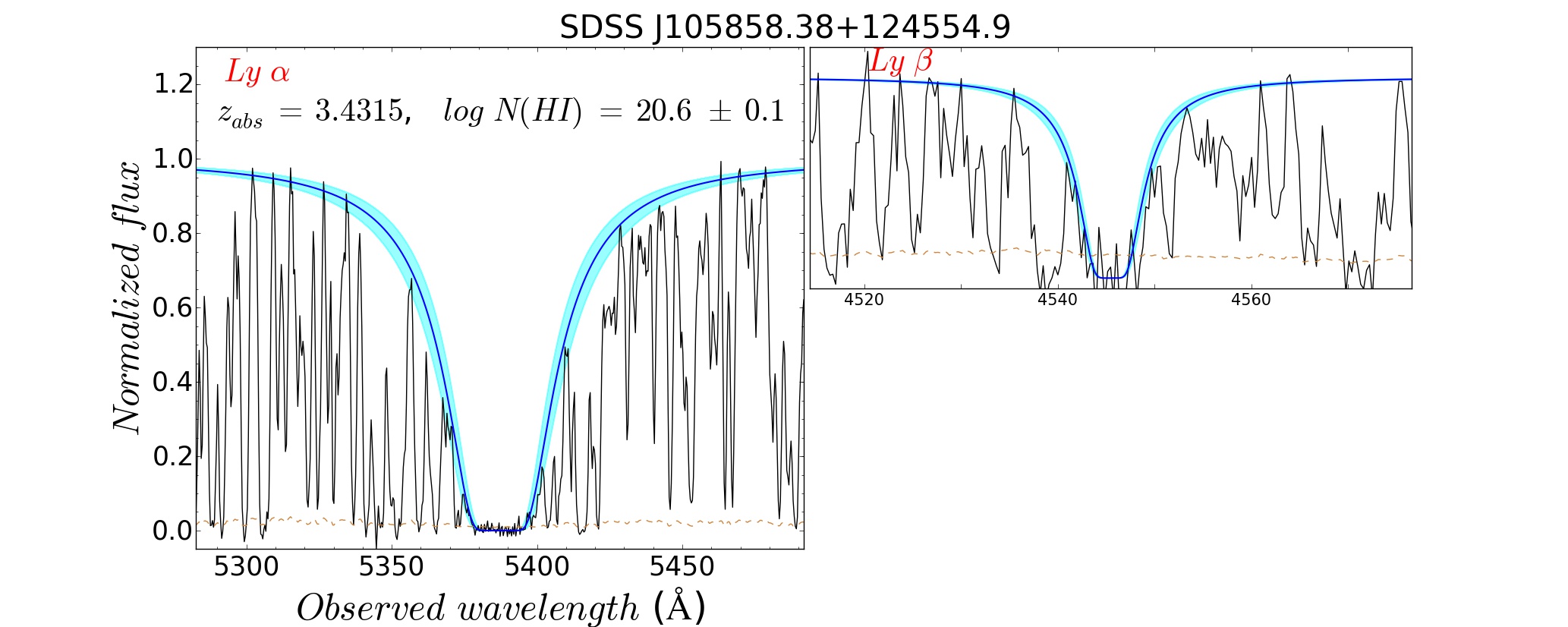} \\ 
\includegraphics[width=0.5\textwidth]{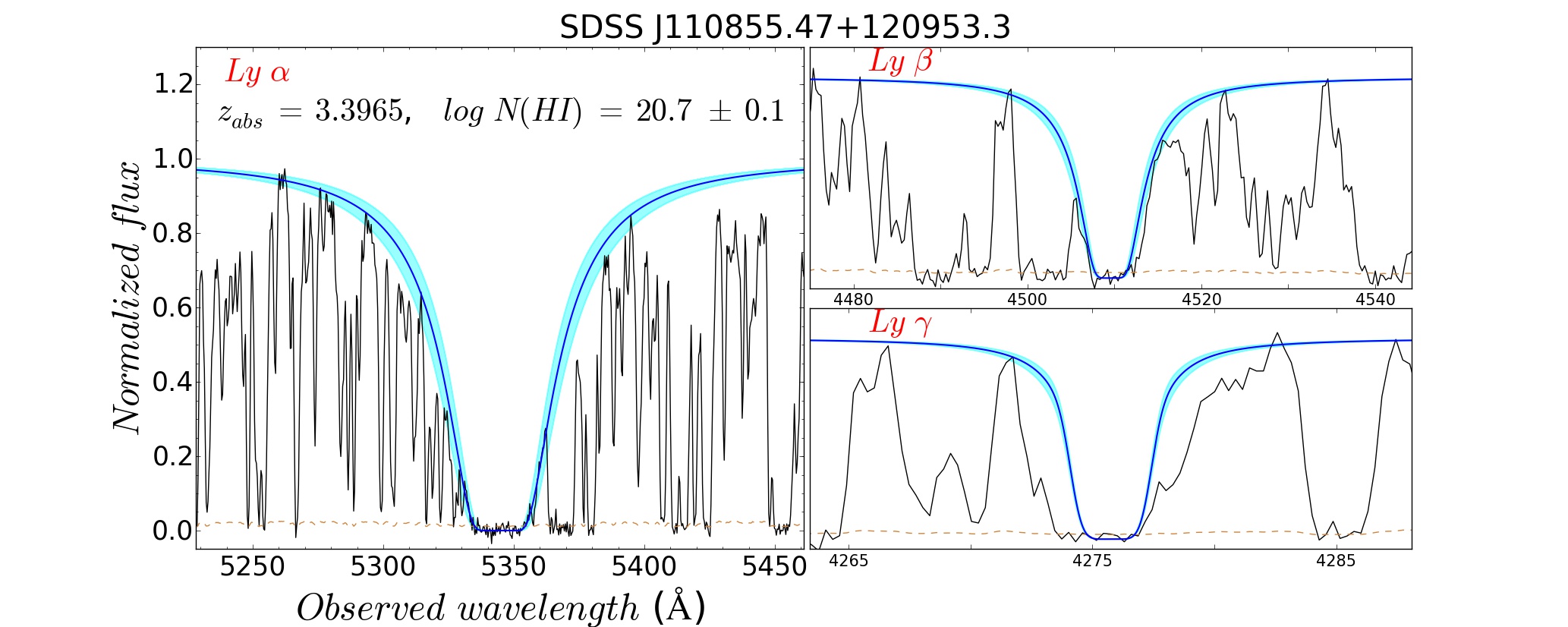} \\ 
\includegraphics[width=0.5\textwidth]{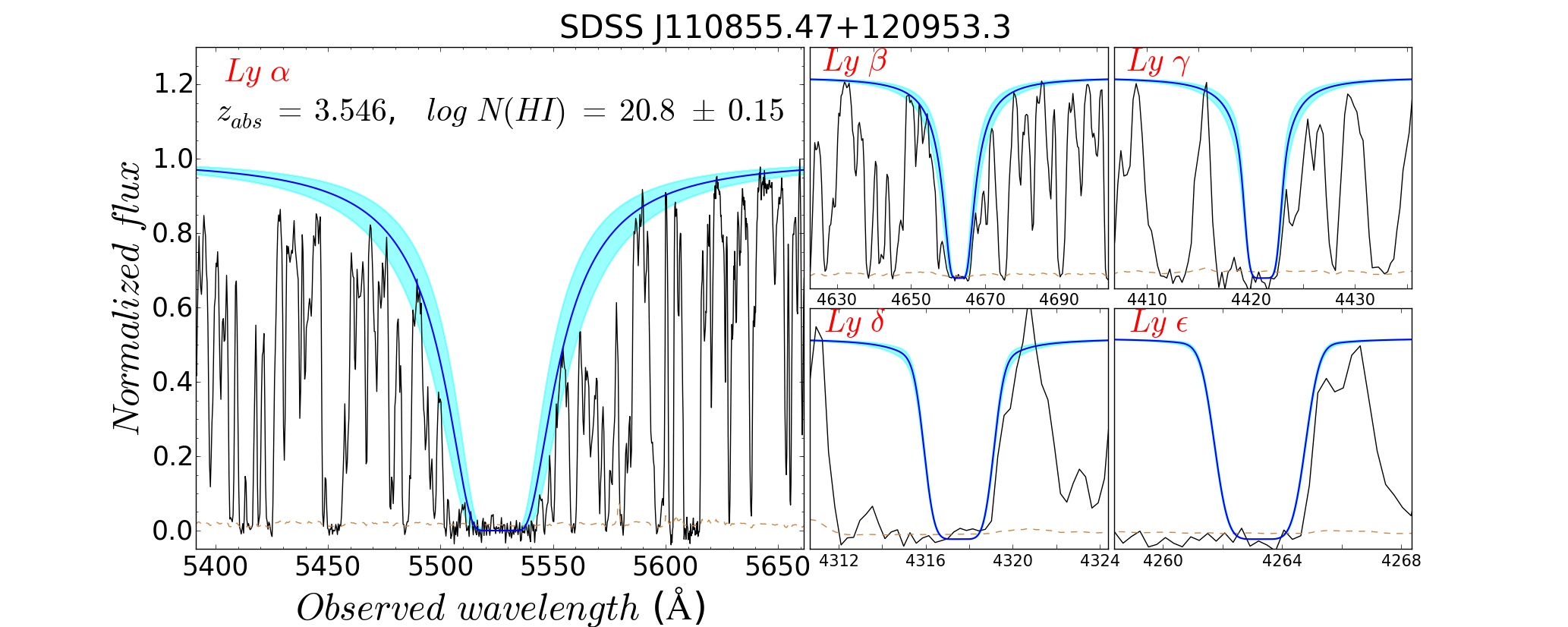} \\ 
\includegraphics[width=0.5\textwidth]{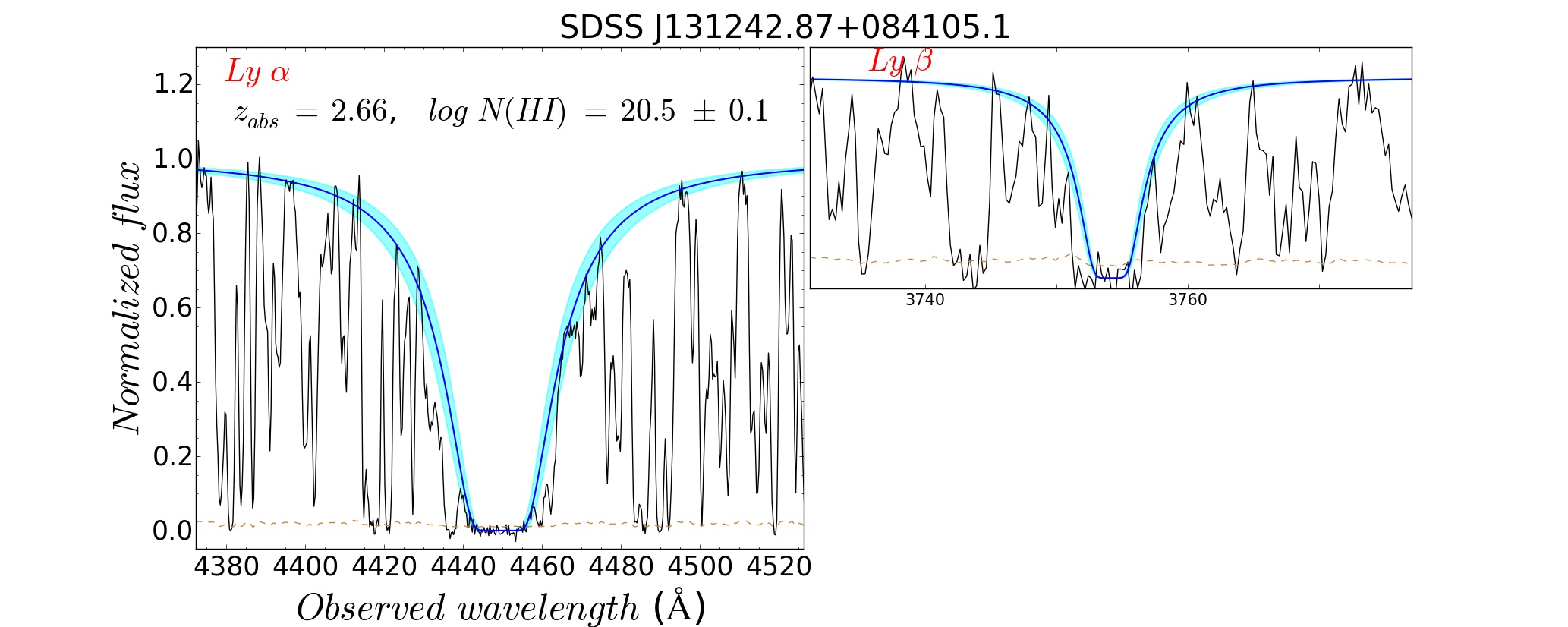} \\ 
\includegraphics[width=0.5\textwidth]{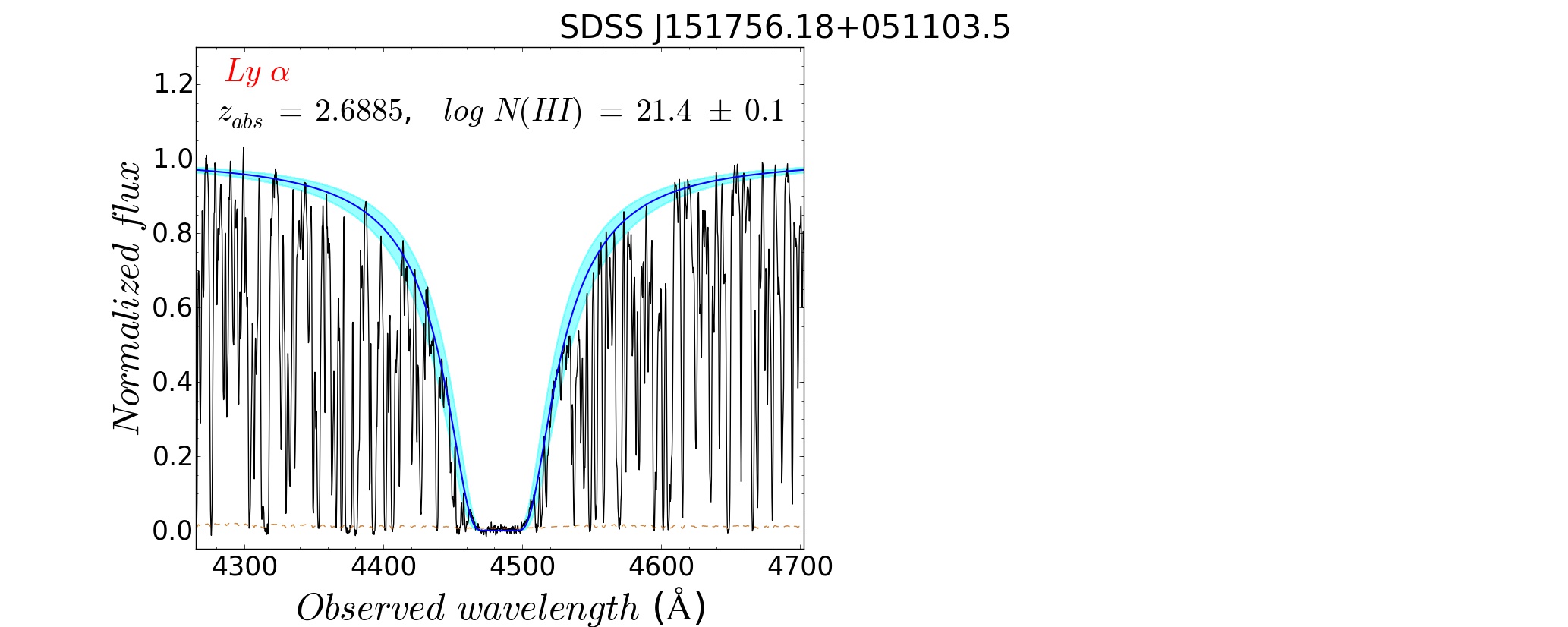} \\ 
\includegraphics[width=0.5\textwidth]{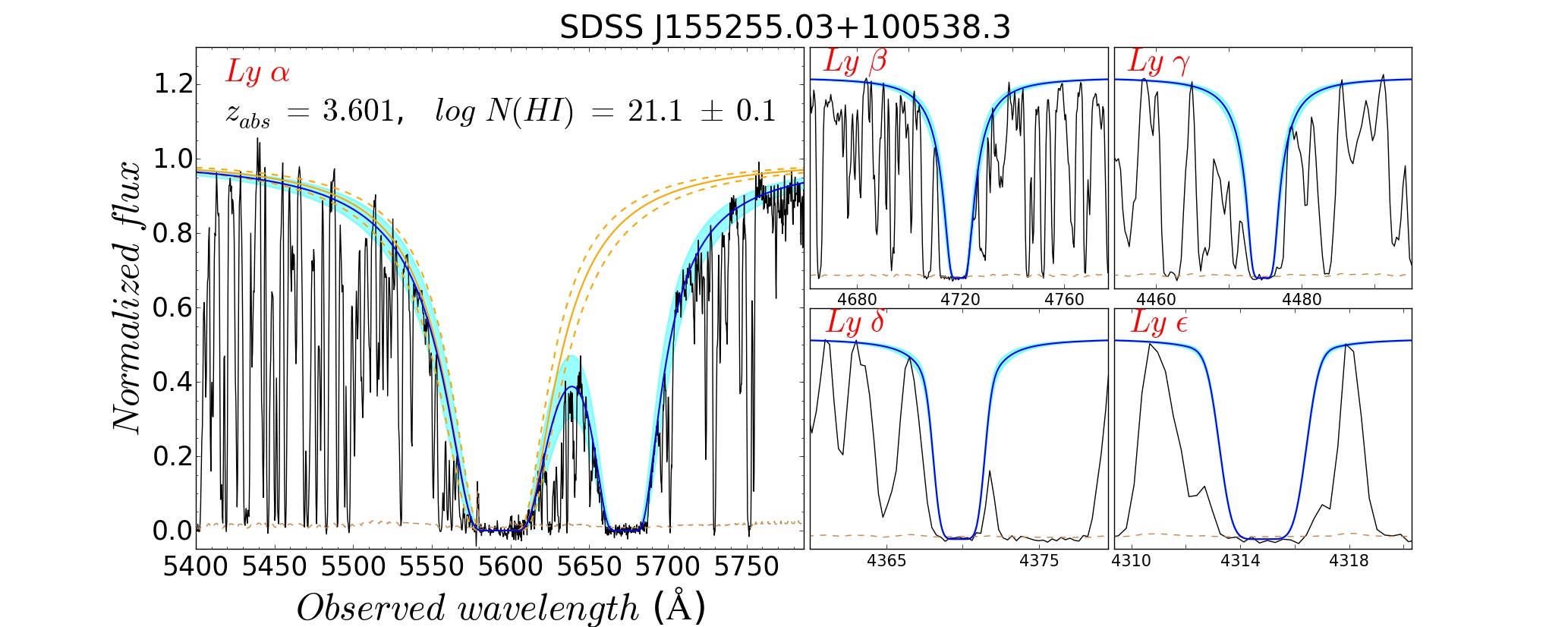} \\ 
\includegraphics[width=0.5\textwidth]{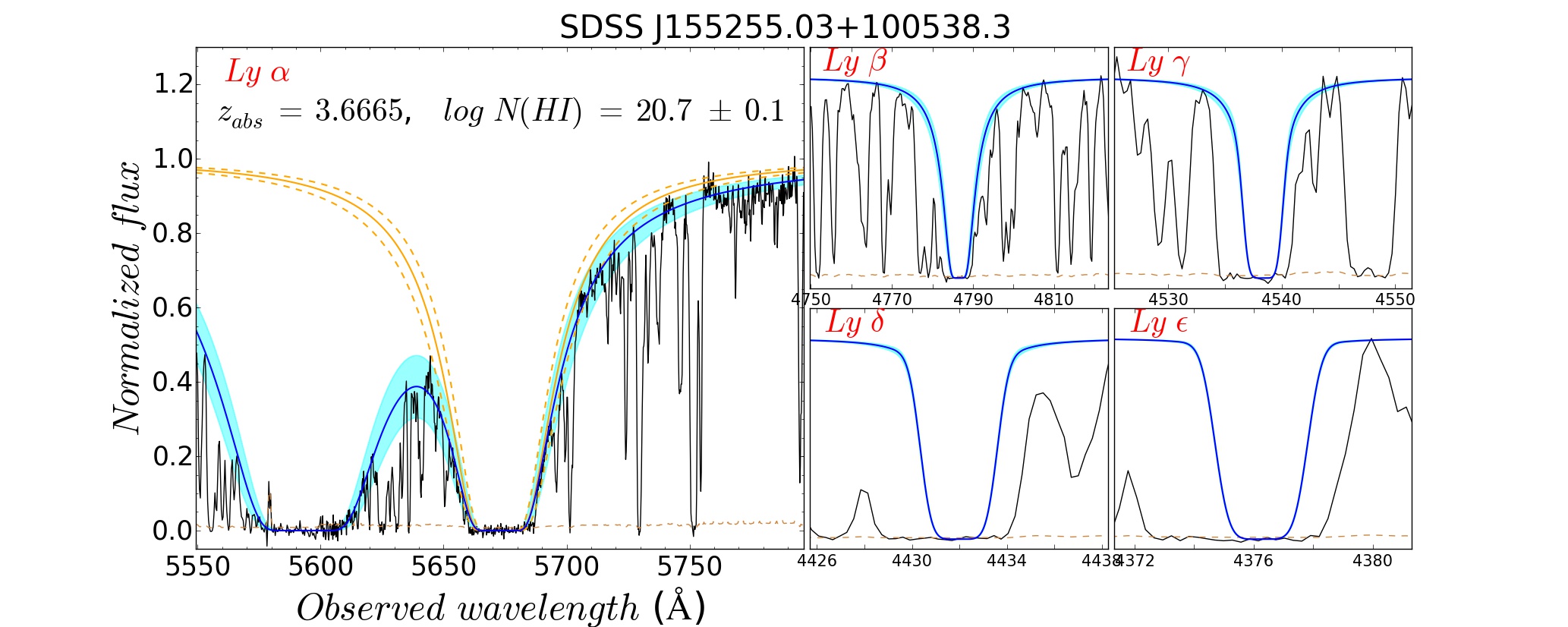} \\ 
\includegraphics[width=0.5\textwidth]{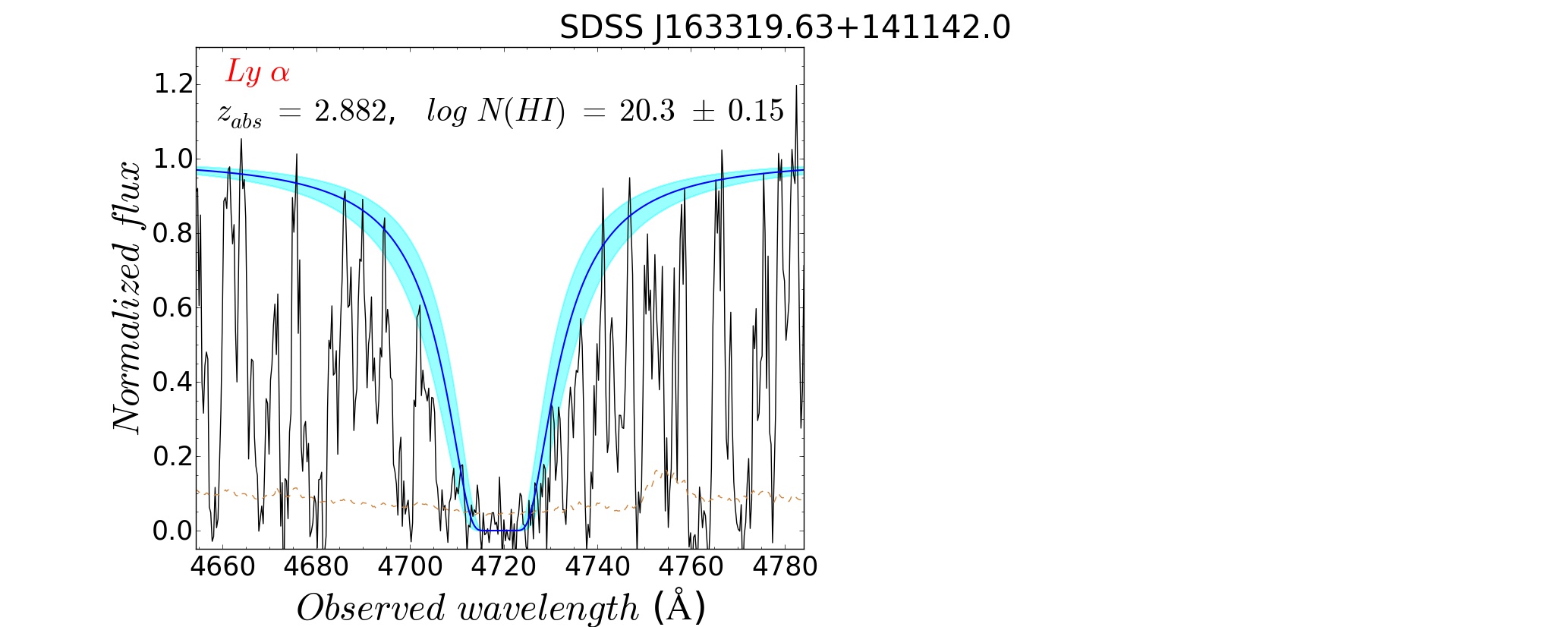} \\ 
\includegraphics[width=0.5\textwidth]{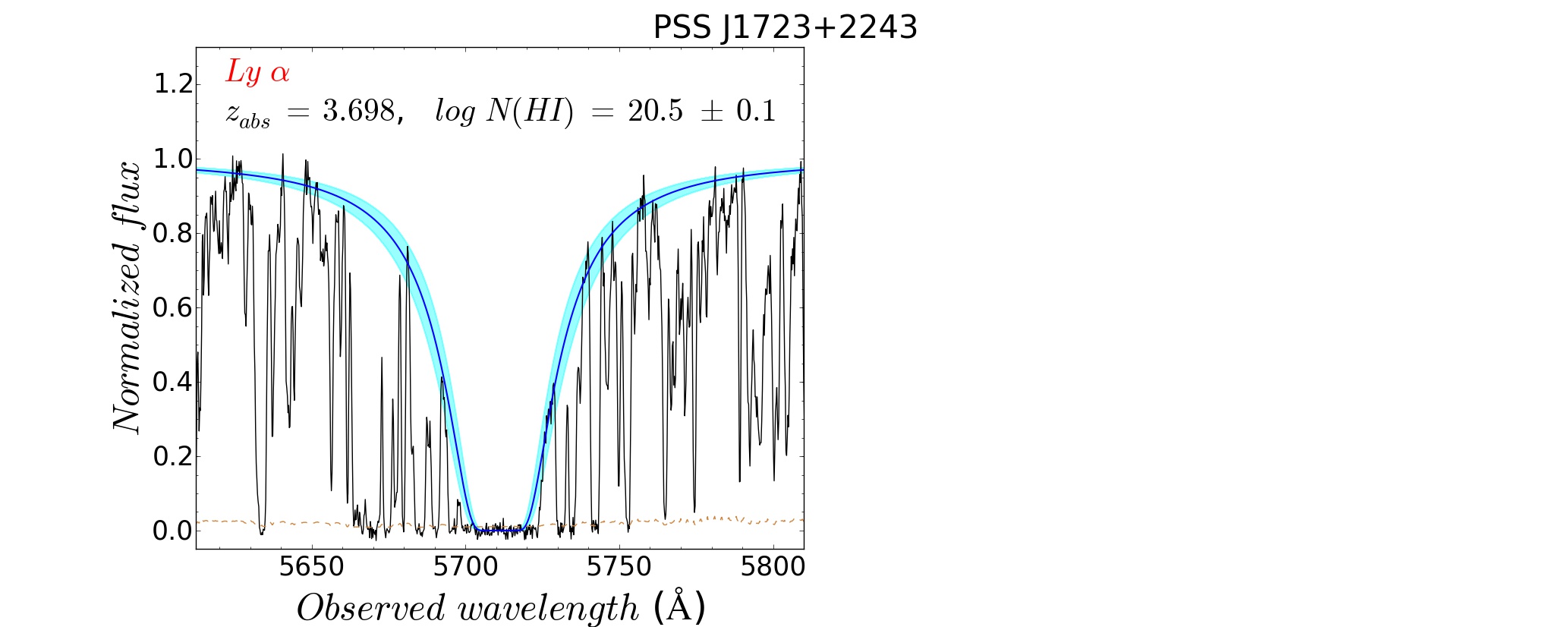} \\ 
\includegraphics[width=0.5\textwidth]{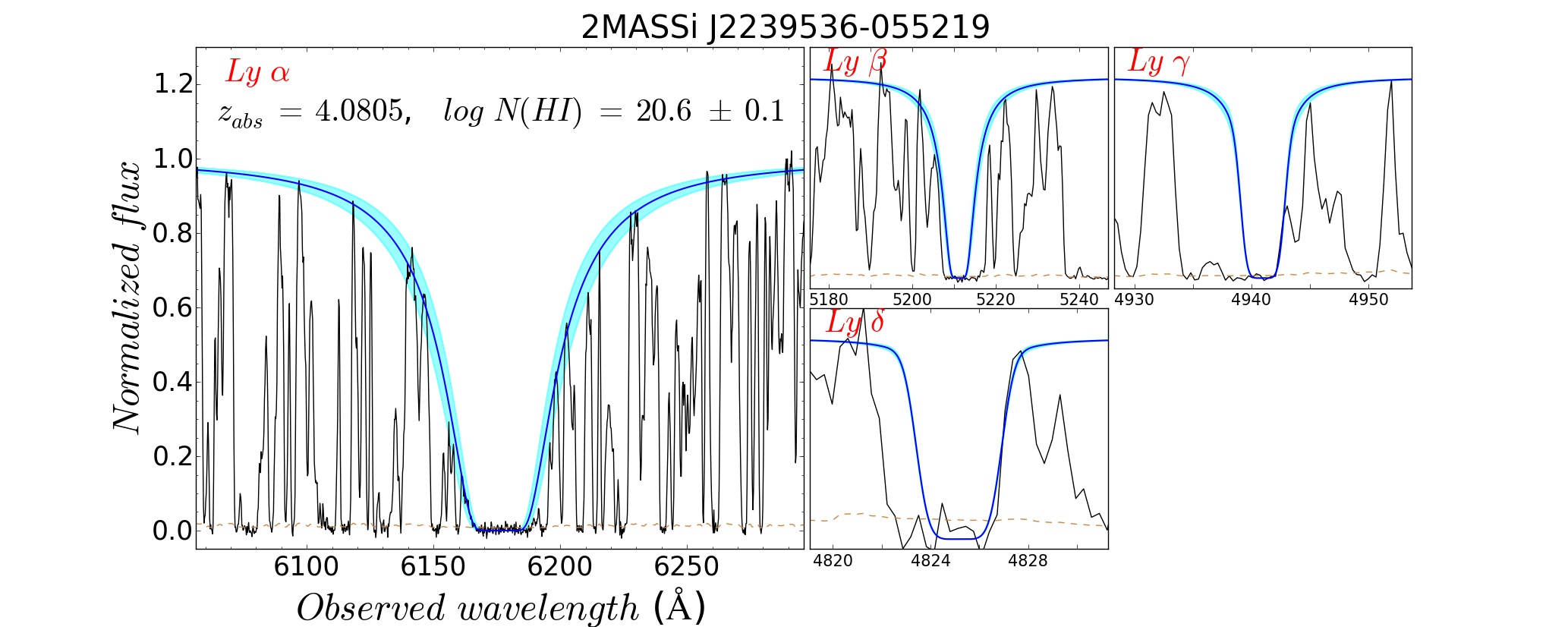} \\ 
\includegraphics[width=0.5\textwidth]{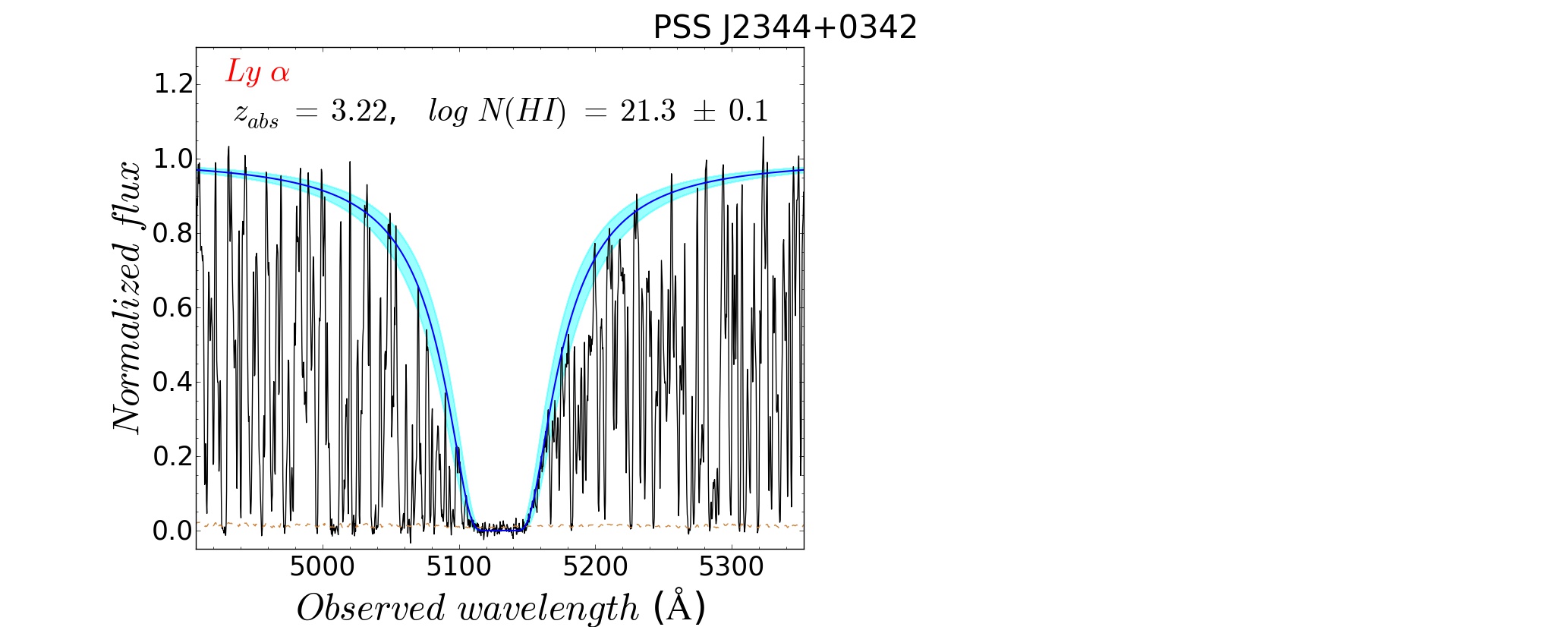} \\ 

\bsp

\label{lastpage}

\end{document}